\documentstyle[11pt,aaspp4,flushrt,psfig]{article}

\slugcomment{accepted by ApJ, September 9, 1998}

\def\Siir{[{\ion{S}{2}}]~$\lambda$6717/$\lambda$6731\/}

\lefthead{Allen et al.}  \righthead{}

\begin{document}

\title{PHYSICAL CONDITIONS IN THE SEYFERT GALAXY NGC~2992}

\author{\sc Mark G. Allen\altaffilmark{1}, 
            Michael A. Dopita\altaffilmark{1},  
            Zlatan I. Tsvetanov\altaffilmark{2} and
            Ralph S. Sutherland\altaffilmark{1}  }

\altaffiltext{1}{Mount Stromlo and Siding Spring Observatories, The
Australian National University, Private Bag Weston Creek P.O., ACT 2611,
Australia; mga, mad, ralph@mso.anu.edu.au}

\altaffiltext{2}{Department of Physics and Astronomy, Johns Hopkins
University, Baltimore MD 21218, USA; zlatan@pha.jhu.edu}

\begin{abstract}
This paper presents long slit spectral maps of the bi-cone shaped
extended narrow line region (ENLR) in the Seyfert galaxy NGC~2992. We
investigate the physical properties of the ENLR via emission line
diagnostics, and compare the observations to shock and photoionization
models for the excitation mechanism of the gas. The line ratios vary
as a function of position in the ENLR, and the loci of the observed
points on line ratio diagrams are shown to be most consistent with
shock+precursor model grids.  We consider the energetics of a nuclear
ionizing source for the ENLR, and perform the $q$-test in which the
rate of ionizing photons from the nucleus is inferred from
measurements of the density and ionization parameter. The $q$-test is
shown to be invalid in the case of NGC~2992 because of the limitations
of the \Siir\ density diagnostic.  The excitation of the gas is shown
to be broadly consistent with the kinematics, with higher
[\ion{N}{2}]~$\lambda$6583/H$\alpha$ present in the more dynamically
active region. We also show that the pressure associated with the
X-ray emitting plasma may provide a large fraction of the pressure
required to power the ENLR via shocks.
\end{abstract}


\section{INTRODUCTION}

The narrow line regions (NLR) of active galaxies are the largest structures
which directly derive their energy input from the active galactic nucleus
(AGN). In many nearby Seyfert galaxies the NLR are spatially resolved, and
often extend over distances of several kpc, so that the energy input
processes may be studied in detail. Such spatially-resolved NLR are often
referred to as extended narrow line regions (ENLR) (\cite{ung87} 1987). In
both Seyfert 1 and Seyfert~2 galaxies the ENLR appears as a filamentary
system of high excitation gas and in Seyfert~2 galaxies this gas is often
confined to a bi-cone structure centred on the nucleus, the so called
``ionization cones''. In other cases, the morphology is more complex, and
bent jet-like structures are sometimes observed.

According to the model developed some twenty years ago (\cite{kos78} 1978;
\cite{fer83} 1983; \cite{sta84} 1984), and which remains the dominant
paradigm (\cite{rob87} 1987, \cite{vie92} 1992), the NLR are excited by
photoionization by a powerful (approximately power-law) non-thermal UV
continuum produced by the AGN. This scenario appeals to the unified schemes
for AGN in which the nucleus and the broad-line region are surrounded by an
optically thick torus (\cite{ant85} 1985). In this picture the differences
between Seyfert types 1 and 2 are ascribed to an orientation effect in which
type 1 Seyferts are oriented so that the line of sight falls within the
opening angle of the cones, while in type 2 Seyferts the line of sight to
the nuclear region is obscured by the torus. This scenario explains the
conical ENLR observed in some type 2 Seyferts as due to shadowing of the
ionizing radiation from the nucleus by the absorbing torus.

Shocks may also be important for the morphology, kinematics and excitation
of the NLR in Seyfert galaxies. In some Seyferts the morphology of the
emission line gas is directly related to that of the radio emission,
providing evidence for dynamical interaction between the radio plasma and
the ambient interstellar medium of the host galaxy. For example, high
resolution narrow band imaging with HST has revealed bow-shock shaped
emission line regions around radio lobes, and jet-like emission line regions
in sources with jet-like radio structures (\cite{axo98} 1998, \cite{pog96}
1996, and references therein).

Shocks probably dominate the physics of the large scale outflows
observed in Seyfert galaxies. In a series of papers, (\cite{col96a}
1996a, \cite{col96b} 1996b, \cite{col98} 1998) Colbert and
collaborators show that optical emission suggestive of large scale
outflows is present in $\gtrsim \frac{1}{4}$ of their sample of 22
edge-on Seyfert galaxies, $\gtrsim \frac{1}{2}$ of the sample have
kiloparsec scale radio structures, and 3 of their sample, including
NGC~2992, have similarly extended soft X-ray emission. \cite{col98}
(1998) favour an interpretation in which these large scale outflows
originate as AGN driven non-thermal jets that entrain the surrounding
thermal material, and heat it by shocks over kpc spatial scales. An
alternative possibility is that the X-ray gas results from starburst
driven winds from the circumnuclear region.

Seyfert galaxies are radio quiet, and simple equipartition arguments
have long suggested that the pressure in the relativistic plasma
responsible for the radio wavelength synchrotron emission is
inadequate to drive high velocity shocks into the interstellar medium
(\cite{wil88} 1988). However, this picture may be wrong if mass
entrainment into the jet occurs close enough to the source. Recently
\cite{bic98} (1998) has shown that although the radio jets of Seyferts
start off mildly relativistic, over most of their length the internal
energies of Seyfert jets must be dominated by thermal plasma. The
combined pressure of the thermal and relativistic gas is then
sufficient for the jets to power the optical emission of the NLR. This
idea finds observational support from the work of \cite{col98} (1998)
which suggests that the large scale outflows in Seyfert galaxies are
predominantly winds of thermal X-ray emitting gas. This thermal phase
will dominate the pressure, consistent with the requirements of the
theory described in \cite{bic98} (1998).

Models of shocks for Seyfert galaxies have considered various
scenarios for the excitation of the gas. Bow-shock models developed by
\cite{tay92} (1992) (which can account for the wide range of line
profile shapes observed in the NLR) consider the excitation mechanism
to be photoionization, whereby the shocks simply serve to compress the
gas which is then illuminated by UV continuum from the nucleus. More
recent models show that shocks themselves can provide an alternative
mechanism for the ionization of the NLR based on the input of
mechanical energy. These shock models advanced by \cite{sut93} (1993)
and developed in detail by \cite{dop96} (1996) show that high velocity
shocks generate strong UV radiation in the high temperature cooling
zone behind the shock front, the hardness of which depends on the
shock velocity. This photon field is then available to produce a
highly ionized precursor \ion{H}{2} region, which emits a spectrum
like that observed in the NLR. \cite{dop95} (1995) find that their
shock+precursor models can account for the observed optical emission
line ratios of a sample of Seyfert galaxies. This raised the
possibility that such models may be generally applicable to narrow
emission line regions.

In order to find observational support in favour of either the shock
or the photoionization model of the NLR, we have undertaken a detailed
spectroscopic and dynamical study of the ENLR of a number of nearby
Seyferts. Amongst these, the subject of this paper, NGC~2992 provides
a particularly good test case to compare the predictions of the models
with observations. NGC~2992 is a gas-rich (\cite{hut82} 1982;
\cite{san85} 1985) Sa type galaxy classified as having a Seyfert 1.9
active nucleus. It is inclined at $\sim$70\arcdeg\ to our line of sight
and is crossed by a disturbed dust lane. It is interacting with
NGC~2993 which is connected to the southern part of NGC~2992 by a
tidal tail with a projected length of 2.9\arcmin. The nuclear redshift
is z=0.0077$\pm 0.0002$ which gives angular scale 150 pc arcsec$^{-1}$
assuming H$_{0}$=75 km~s$^{-1}$ Mpc$^{-1}$.

It provides an interesting test case because it displays a striking
and unusually extensive bi-conical ENLR which emerges almost
perpendicularly from the plane of the galaxy. Narrow band images show
the spatial extent of bright bi-cones is $\sim$4 kpc, with the SE cone
brighter than the NW cone.  Fainter emission is extended up to 8
kpc. Figures 1a and 1b show narrow band filter images obtained at the
ESO New Technology Telescope (NTT). These are respectively
[\ion{O}{3}]~$\lambda$5007, and H$\alpha$+[\ion{N}{2}] and are
presented on a logarithmic stretch so as to bring out the fainter
regions. Figure~1c shows a blue continuum image of NGC~2992.

The radio structure of NGC~2992 is described in detail in a paper by
\cite{weh88} (1988) which includes a synopsis of previous radio
observations.  In summary NGC~2992 has 20 cm radio emission which
extends approximately 25\arcsec\ along PA$\sim $160\arcdeg, and has a one
sided extension along PA$\sim $130\arcdeg. At a wavelength of 6 cm there
is visible a smaller scale figure-8-shaped structure, with its long
axis extending approximately 8\arcsec\ along PA$\sim $-26\arcdeg, and
centred on the nucleus. The figure-8 may represent edge brightened
bubbles of relativistic plasma or else magnetically-compressed arches,
either of which are indicative of an expanding region of
over-pressured relativistic plasma.  However, comparison of the radio
images with optical narrow band images shows no obvious
correlation between the structures - see Figures 1d, 1e and 1f where
these radio maps are overlaid on our [\ion{O}{3}]~$\lambda$5007, and
continuum images.

NGC~2992 was included in Durret and Bergeron's investigation of the
physical properties of the ionized gas in emission line galaxies
(\cite{dur88} 1988, \cite{dur90} 1990). In the earlier paper they
present long slit observations of NGC~2992. The second paper describes
the physical properties of the ionized gas and other general
characteristics derived from their spectra.  Here we briefly summarise
their results.

First, in the nuclear spectrum a broad component of H$\alpha$ is
detectable, but there is no corresponding broad H$\beta$ component.
NGC~2992 is therefore classified as an intermediate type Seyfert
(1.9). This classification is confirmed from the shape of the far-IR
to far-UV continuum (\cite{boi86} 1986). Second, \cite{dur88} (1988)
found the extranuclear spectra along PA = 197\arcdeg\ have similar line
ratios to the nuclear region, and are characteristic of those found in
Seyfert~2 galaxies.  This excludes the possibility that the ENLR is
excited by OB stars. The excitation of the gas appears to decrease
from the nucleus to the outer regions. The reddening of the inner
regions is high, ranging up to A$_{v}\sim$3, and a similar reddening
was estimated by \cite{mal83} (1983) from the
[\ion{S}{2}]~$\lambda\lambda$(4069+4076)/(6717+6731) ratio. Third,
using the H$\alpha$/H$\beta$ to correct line ratios for reddening
\cite{dur88} (1988) found a temperature of T$_{e}$=19900K from the
[\ion{O}{3}] lines in the nuclear regions.

The velocity field of NGC~2992 is complex. A kinematical study by
\cite{col87} (1987) revealed blue asymmetries in the [\ion{O}{3}]
lines at the center of the galaxy, and showed that the nuclear region
is dynamically decoupled from the off-nuclear regions. \cite{col87}
(1987) find some indication of radial outflow within a plane
highly inclined to the galaxy plane, and that the kinematics of the 
SE region may be accounted for by tidal interaction with NGC~2993.
Recently \cite{mar98} (1998) have
reported a kinematical study of NGC~2992 with the H$\alpha$ and
[\ion{O}{3}] lines measured along nine position angles. They note
that outflow of the gas within a conical envelope or on the surface of
a hollow cone as one possible picture to account for the double peaked
profiles and assymetries detected in several regions.  However they
use a simple kinematical model of circular rotation with a constant
radial outflow, where the out-flowing velocities are considered as
radial motions in a plane very close to that of the gas disk. While
this model provides a good approximation to the observed kinematics we
caution that such a model is not unique, and prefer to interpret the
kinematics as a general outflow or wind, not necessarily constrained
to a particular plane.  This was reported briefly by us in
\cite{tsv95} (1995) and details will appear in a future paper. In
summary, our higher spectral resolution study of the NGC~2992 NLR
shows double component line profiles at most positions.  One of the
kinematical components follows a galactic rotation curve, while the
other component is identified as a wind which is expanding out of the
plane of the galaxy with velocities up to 200 km~s$^{-1}$.

These kinematical studies generally assume the disk of NGC~2992 to be
oriented with NW edge closest, in concordance with trailing spiral
arms. In this orientation, with the axis of the bi-cone 
perpendicular to the disk, the SE cone is the closer and is directed
towards us, consistent with the outflow interpretation. If on the
other hand the spiral arms lead, then the SE edge of the galaxy
and the NW cone would be closer.  In this orientation the kinematics
would indicate inflow instead of outflow. Other possibilities also
exist if the bi-cone axis is highly inclined with respect to the 
galaxy disk.

The X-ray spectral properties of NGC~2992 are described by
\cite{wea96} (1996) (and references therein) to have a significant
absorbing column and a prominent Fe K$\alpha$ fluorescence emission
line. The X-ray flux varies by a factor of 2$-$3 on timescales of days
to months. There is an excess of soft X-ray (0.1$-$2.0 keV) flux above
that expected from the extrapolation of the 2$-$10 keV power law
($\Gamma$=1.7), and there is also some evidence for extended soft
X-ray emission on a scale of $\sim$2 kpc. X-ray images of NGC~2992
taken with the {\em ROSAT} Position Sensitive Proportional Counter
(PSPC), and High-Resolution Imager (HRI) are presented by \cite{col98}
(1998). Soft X-ray (0.2 $-$ 2.4 keV) emission extends out to
$\sim$1\arcmin\ along PA$\sim$100\arcdeg, and is roughly co-spatial with
the 20~cm radio emission.

Recently \cite{gla97} (1997) have reported that the near infrared
light-curve of NGC~2992 (monitored $\sim$3 times a year over the past
15 years, with an 9\arcsec\ aperture) shows an outburst in 1988. The
amplitude of the event was greatest in the L(3.5 $\mu$m) band, which
increased 102 mJy above the base level in less than 254 days, then
decayed exponentially with an e-folding time of $\sim$900
days. \cite{gla97} (1997) compare the total energy output of the event
(6$\times10^{43}$ J) to the explosion energies of types I and II
supernovae. They find that while a conventional un-obscured supernova
does not provide a direct fit, the outburst may be explicable in terms
of a model similar to that of \cite{ter92} (1992), which seeks to
explain the characteristics of Seyfert nuclei in terms of multiple
supernova explosions in dense \ion{H}{2}\ regions.

This paper presents the results of mapping the ENLR of NGC~2992 with
long slit spectroscopy in order to investigate the detailed spatial
dependence of line ratios and the physical properties with the
objective of understanding the mode of excitation of the ENLR. The
observations and data reductions are described in section 2, including
the line flux measurements and representative spectra. Three distinct
spectral regions of the ENLR are identified in section 3 in which we
show the spatial variation of the line ratios. In section 4 we derive
the physical properties of the emission line gas. The observations are
compared to shock and photoionization model predictions with line
ratio diagnostic diagrams in section 5. We discuss the $q$-test for
photoionization from a central source in section 6. In section 7 we
compare the excitation of the gas with the kinematics, and consider
the possibility of a thermal pressure energy source for the excitation
of the ENLR.

\section{OBSERVATIONS AND DATA REDUCTION}

NGC~2992 was observed with the Double Beam Spectrograph (DBS) on the
ANU 2.3m telescope at Siding Spring, Australia on March 5th and 9th
1994. Loral CCD detectors were used on both the red and blue arms of
the spectrograph.  The 158 line~mm$^{-1}$ grating was used in the red
arm giving a dispersion of 4.1~\AA~pixel$^{-1}$, a spectral
resolution of 11.9~\AA\ and covering a wavelength range 5840$-$9725 \AA. 
The 300 line~mm$^{-1}$ grating was used in the blue arm, providing a
dispersion of 2.2 \AA~pixel$^{-1}$, and yielding a 4.9 \AA\ spectral
resolution over the wavelength range 3440$-$5570 \AA . The slit width was
fixed at 2\arcsec\ for all observations of NGC~2992, but was increased
to 5\arcsec\ for observations of standard stars so as to ensure that
slit losses were negligible. The spatial scale along the slit was the
same in both red and blue arms; 0\farcs89 pixel$^{-1}$.

Spectra were obtained for a total of nine slit positions all
orientated at PA 30\arcdeg. These were placed $\sim $2\arcsec\ apart so
as to efficiently cover the ENLR and provide a set of spatially
independent spectra. A total of three 1000~s exposures were taken at
each slit position. The FWHM of the standard star profiles along the
wide slit provides an estimate of the seeing plus the guiding
errors. At the beginning of the first night the median seeing was
$\sim $2\arcsec\ but improved to $\sim $1\farcs8 by the end of the
night. The seeing in the middle of the second night was
$\sim$2\farcs3.

Data reduction was performed using the standard IRAF routines. Pixel
to pixel variations in the CCD response were removed by appropriately
normalised spectral flat fields derived from Quartz-Iodine lamp
spectra, while a correction for the slit function was made by a low
order fit to the illumination of the slit by the twilight sky. The DBS
has very low vignetting, so this correction is not very important. The
wavelength scale was established by fitting 5th order Chebychev
polynomials to lines identified in Helium-Argon comparison lamp
spectra obtained throughout the observations on both nights.

Long-slit observations are usually affected by misalignments and
optical distortions which cause the dispersion direction not to be
exactly aligned along the lines of the CCD. Spectra of standard stars
through a narrow slit revealed this effect to be of order 2.5 pixels
over the 1000-pixel length of the CCD, and the arc lines are tilted by
approximately 2 pixels. The distortion in each object and standard
star frame was found by tracing the continuum along the dispersion
direction. The distortion in the slit direction was similarly found by
tracing the lines in the arc frames. Two dimensional fits to the
wavelength and geometrical correction as a function of position was
done with the IRAF task {\it fitcoords}. The two transformations were
inverted to map the wavelength as a linear function along lines of the
frame, and the position along the slit as a linear function along the
columns. The subtraction of the background sky contribution could then
be performed, having first normalised the spectra by fitting a 2nd
order polynomial to the background in each column.

All frames were divided by a smoothed spectrum of a featureless DC
white dwarf star in order to remove the effects of the dichroic
transmission variability, which because of interference effects is a
complex oscillatory function of wavelength. The sensitivity function
derived from the measured standard stars can then be fit by a low
order polynomial. The spectrophotometric standards L745-46A and W485A
were used to calibrate the absolute fluxes of the spectra on the first
night, while only L754-46A was observed on the second night. The
nights appeared to be photometric, with atmospheric transmission
varying less than 10\%. No attempt was made to remove the atmospheric
molecular absorption bands.

An error frame was calculated for each 2-dimensional spectrum by
taking the square root of the counts after the instrumental
corrections and propagating the error through the reductions.

Since the observations were made without the aid of an auto-guider,
and the slit positions were set up by offsetting from the nuclear peak
of light as it appeared in the TV display, the slit positions were not
very accurately controlled. More accurate slit positions were found by
matching the intensity profile of the brightest line in the spectrum
[\ion{O}{3}]~$\lambda$5007 to the intensity profile generated from an
narrow band image of the galaxy in this same emission line. This
technique involves firstly rotating and binning the image to the same
pixel scale and spatial orientation as the 2-d spectra (0\farcs89
pixel$^{-1}$, PA 30\arcdeg). Simulated intensity profiles for slits
positioned at 0\farcs3 steps across the ENLR were then generated from
the image, and were compared to the observed slit profiles. The best
match in each case was found by cross-correlating the observed
intensity profile with the set of simulated slit profiles. The
IRAF task {\it fxcor} was used in its pixel correlation mode for this
purpose. The correlation function calculated for each simulated slit
with the observed profile is a measure of how well the simulated
profile matches the observed profile, and gives the relative shift in
pixels between the two profiles. The height of the correlation
functions provide three or four best match candidates, which are then
inspected manually and a best match chosen.  For all but slits -5 and
3 the best match was obvious upon inspection so in general we regard
the slit positions as known to within 0\farcs3. The derived slit
positions are shown overplotted on the [\ion{O}{3}]~$\lambda5007$
image contours in Figure~2. In the remainder of the paper we refer all
spatial positions in the ENLR of NGC~2992 to the coordinate system
shown in this figure.

The relative and absolute flux calibrations of all slits was assessed
by extracting the flux of [\ion{O}{3}]~$\lambda$5007 from the
calibrated narrow band image at the $x,y$ coordinates of the spectra
and comparing with the line flux measured from the spectra. The
spectra and image fluxes match within 20\% at most $x,y$ positions.

One dimensional spectra were extracted from the long slit frames by
summing every three pixels along the slit. The $y$ coordinate of the
extraction is known from the $y$ position of the best match simulated
slit, and the $x$ coordinates are determined from the relative shift
given by the cross-correlation. This results in a grid of spectra
covering the ENLR of NGC~2992 with a spacing of approximately
2\arcsec\ in the $y$ direction, and closer spacing in the $x$
direction. To improve the signal-to-noise ratio in the faint, outer regions 
of the ENLR we have co-added up to 8 spectra into a more coarse grid.

Average spectra of the NW cone, SE cone and inter-cone regions of the
ENLR are shown in Figures 3a, 3b and 3c. A sky feature is seen at
6860 \AA\, and the red end of the spectra are not shown as that region
is contaminated with sky features.

\subsection{Emission line measurements}

The fluxes, wavelengths and widths of the bright emission lines were
measured in each grid point spectrum by fitting Gaussians to the line
profiles with a linear fit to the continuum in the wavelength region
of the line. We find that approximating the continuum as a linear
function does not significantly affect the measurement of the bright
lines. The bright emission line measurements were done with the IRAF
STSDAS task $specfit$. For the [\ion{S}{2}], [\ion{N}{2}] and [\ion{O}{3}]
doublets the FWHM and wavelength ratio of the lines were fixed in the
fitting procedure. $1\sigma$ uncertainties in the flux, wavelength and
FWHM measurements were calculated from the error spectrum derived
from the error frames described above.

\subsection{Subtraction of the stellar population and faint line measurements}

At all locations in the ENLR the emission line spectrum is
superimposed on an underlying galaxy stellar spectrum. Figure~3b shows
that the galaxy contribution is much stronger on the SE side of the
galaxy. The main stellar features are readily identified as the Ca H
and K bands, the G band of C-H at 4290$-$4313~\AA, H$\beta$ absorption,
Mg~b at around 5170~\AA\ and \ion{Na}{1} 5890$-$5896~\AA. We find that
these stellar features and the general shape of the underlying galaxy
spectrum can be matched by the spectrum of the E4 peculiar galaxy
NGC~1316 (\cite{pic85} 1985) with some correction for reddening.

Subtraction of the stellar population is very important for the
detection and measurement of faint emission lines. \cite{sto96} (1996)
emphasize the importance of subtracting stellar features to obtain
accurate measurements of the H$\gamma$, [\ion{O}{3}]~$\lambda$4363, and
\ion{He}{2}~$\lambda$4686 lines in ENLR spectra. Using \cite{sto96}
(1996) as a guide for matching the stellar features we have fitted the
observed spectra of the NW and SE cones, using the NGC~1316 galaxy
template to account for the underlying galaxy continuum. The fit was
done over the wavelength range 4200$-$5300~\AA, with the IRAF $specfit$
package. In the fitting procedure we constrain the redshift of the
template to a range close to that determined for the emission lines in
the spectrum. A reddening correction is applied in the fitting
procedure using the \cite{car89} (1989) reddening curve and allowing
E(B-V) to be a free parameter. The emission lines H$\gamma$,
[\ion{O}{3}]~$\lambda$4363, \ion{He}{2}~$\lambda$4686, H$\beta$, the
[\ion{O}{3}]~$\lambda\lambda$4959,5007 doublet and
[\ion{N}{1}]~$\lambda$5200 were included in the fit as Gaussian
components. The widths of the H$\gamma$ and H$\beta$ lines were linked
to have the same value in the fit, as were the widths of the
[\ion{O}{3}]~$\lambda\lambda$4959,5007 doublet
lines. [\ion{O}{3}]~$\lambda$4363 was not linked to the other
[\ion{O}{3}] lines because the asymmetry observed in the brighter
[\ion{O}{3}] lines is not seen in the the fainter
[\ion{O}{3}]~$\lambda$4363 line. An error spectrum derived from the
original photon statistics of the CCD observations was used in the
fitting procedure and allows the calculation of uncertainties in all
the fit parameters. The 1$\sigma$ uncertainties in the NW and SE cone
spectra are approximately 0.15 and 0.4
$\times10^{-16}$~ergs~s$^{-1}$~cm$^{-2}$~\AA$^{-1}$ respectively over
the wavelength range of the fit.

The NW cone and SE cone spectra are both well fit by this procedure,
with the template galaxy component matching the wavelengths and depths
of the stellar absorption features. The underlying galaxy spectra in
the NW cone required an E(B-V) of 0.16 and the SE cone required an 
E(B-V) of 0.17. Figures 4a and 4b
show the result of the fits for the NW and SE cones respectively where
we show the galaxy template component overplotted on the observed
spectra. For each case we have also subtracted the fitted galaxy
template spectrum to reveal the spectrum of the emission line gas. A
small amount of residual structure is seen the spectrum after template
subtraction, however this structure is within the 1$\sigma $ noise
level. The template subtracted spectrum of the NW cone clearly shows
the H$\gamma$, [\ion{O}{3}]~$\lambda$4363 and \ion{He}{2}~$\lambda$4686 
lines. In the SE cone H$\gamma$ and \ion{He}{2}~$\lambda$4686
are clearly seen, but [\ion{O}{3}]~$\lambda$4363 is only just above
the noise level resulting in a relatively uncertain measurement of the
[\ion{O}{3}]~$\lambda$4363 flux for the SE cone.

\subsection{Reddening correction}

NGC~2992 has a prominent dust lane and the observed spectra show that
the reddening varies as a function of position in the ENLR, and that
the reddening derived from the emission lines is higher than for the
underlying galaxy spectrum. The most severely reddened spectra occur
in the regions associated with the dust lane, with the
H$\alpha$/H$\beta$ ratio observed as high as 10. For each grid point
spectrum where both H$\alpha$ and H$\beta$ are detected we derive a
value of the reddening from the H$\alpha$/H$\beta$ ratio, using the
reddening curve of \cite{car89} (1989) and assuming an intrinsic
H$\alpha$/H$\beta$ ratio of 3.1. The reddening within the dust lane
thus corresponds to A$_{V}$=3.6 and decreases outwards along the $y$
direction away from the dust lane.  The average reddening values in
the NW and SE cones are A$_{V}$=0.86 and 1.8 respectively. With the
orientation of NGC~2992 described in section~1, we might expect the
NW cone, seen through the disk of the galaxy to have the greater
reddening. The disturbed nature of the dust lane may account for this
discrepancy.

The emission line measurements for the average spectra of the NW and
SE cone, and inter-cone regions are given in Table 1 along with their
reddening corrected values. In each case the line fluxes are given
relative to H$\beta$=100.

\section{RESULTS}

The spectra of the ENLR of NGC~2992 are typical of Seyfert~2 type
galaxies with strong lines of [\ion{O}{3}]~$\lambda$5007,
[\ion{O}{2}]~$\lambda$3727, H$\alpha$ and also the low ionization lines
[\ion{S}{2}]~$\lambda\lambda$6717,6731, [\ion{N}{2}]~$\lambda$6583 and
[\ion{O}{1}]~$\lambda$6300. Highly ionized Iron lines (\ion{Fe}{7},
\ion{Fe}{9}, \ion{Fe}{10}) sometimes seen in Seyfert galaxies are not
detected in our spectra. [\ion{Ne}{5}]~$\lambda$3426 is detected at the
far blue end of the wavelength range, and the
[\ion{S}{3}]~$\lambda\lambda$9069,9532 lines are detected but are
strongly affected by night sky features. We also detect a blue
asymmetry in the line profiles of the
[\ion{O}{3}]~$\lambda\lambda$4959,5007 lines in the SE cone. The
fainter lines of [\ion{O}{3}]~$\lambda$4363, H$\gamma$ and
\ion{He}{2}~$\lambda$4686 are also detected in co-added spectra of the
NW and SE cones.

\subsection{Spatial variation of the spectra}

A striking feature of this set of spectra is the strong variation of
the line ratios as a function of position in the ENLR. Figure~5a shows
how the [\ion{O}{3}]~$\lambda$5007/H$\beta$ and
[\ion{N}{2}]~$\lambda$6583/H$\alpha$ line ratios separate the spectra
of the ENLR into three groups that correspond to different regions
within the ENLR. The open circles indicate spectra extracted from the
NW cone of the ENLR, the filled squares are for points within the SE
cone, and the open triangles correspond to spectra extracted from
regions between the two cones, and from the outer edges of the two
cones.

This effect is shown even more clearly in Figure~5b where we have used
the ratio of [\ion{O}{3}]~$\lambda$5007 to H$\alpha$ instead of
H$\beta$.  Using the brighter H$\alpha$ line allows us to include
spectra down to a fainter brightness limit, thus covering more of the
ENLR, especially those fainter regions between the cones. The
disadvantage of using the [\ion{O}{3}]~$\lambda$5007/H$\alpha$ ratio is
that the reddening correction is greater than the correction to
[\ion{O}{3}]~$\lambda$5007/H$\beta$. This larger reddening correction
with its greater uncertainty causes more scatter of the points in
Figure~5b. For the faintest regions where H$\beta$ is not detected, we
have applied a value for the reddening which has been extrapolated
from adjacent points at which H$\beta$ is measured. This procedure is
most uncertain for the spectra extracted from points between the cones
and within the dust lane of the galaxy. For these points we apply the
average reddening of the dust lane points which is equivalent to
assuming an H$\alpha$/H$\beta$ ratio of 6 before reddening correction.
Despite these uncertainties in the reddening correction for the
[\ion{O}{3}]~$\lambda$5007/H$\alpha$ ratio, the inferred spatial
variation of the line ratios remains similar to those found in Figure~5a.

Figure~6 shows the spatial locations of the spectra used in Figures
5a and 5b. We shall maintain the distinction between these three
spatial regions by using the same plotting point styles throughout the
rest of this paper.

\subsection{[\ion{N}{2}]/H$\alpha$ variation}

Figures 5a and 5b show that the the NW and SE cones have significantly
different [\ion{N}{2}]~$\lambda6583$/H$\alpha$ ratios. The
[\ion{N}{2}]~$\lambda 6583$/H$\alpha$ ratio of the average spectrum of
the NW cone is 0.46 compared to 0.76 for the SE cone.

Since the SE cone has a stronger underlying galaxy continuum we have
investigated the possibility that the variation in
[\ion{N}{2}]~$\lambda6583$/H$\alpha$ between the two cones is due to a
greater amount of H$\alpha$ absorption in the SE cone. To test this
possibility we use the fit to the blue region of the underlying
stellar distribution that was discussed in the previous section, and
apply it to the spectral range around the H$\alpha$ and [\ion{N}{2}]
lines.  In the spectral region around H$\alpha$ it is more difficult
to make an independent fit for the underlying galaxy continuum. We
therefore fit the galaxy template in the 6450$-$6700~\AA\ spectral
range using the same multiplicative scale and reddening
parameters derived for the blue region. The H$\alpha$ line and each of
the [\ion{N}{2}]~$\lambda\lambda$6548,6583 emission lines are fitted
with single component Gaussian profiles with the fit constrained so
that the [\ion{N}{2}] doublet line wavelengths were at the known
ratio. For the fit the flux ratio
[\ion{N}{2}]~$\lambda$6583/$\lambda$6548 was fixed at 3.0, and both
[\ion{N}{2}] components were constrained to have the same line widths.

The results of the fits to the NW cone and SE cone and dust-lane
region spectra over the 6450$-$6700~\AA\ range are shown in Figure~7. 
In each panel we show the total fit overlaid on the data and also
the result of subtracting the template from the data. The lighter
weight curves in Figure~7 show the fitted galaxy template and the
individual Gaussian component fits to the H$\alpha$ and [\ion{N}{2}]
emission lines.

We obtain an excellent fit to the spectrum in the NW cone. Since the
stellar contribution is faint in this cone there is only a small
amount of H$\alpha$ absorption which is barely discernable in the
plot, and is insignificant relative to the emission line fluxes. The
fit to the SE cone (middle panel of Figure~7) demonstrates that
stellar H$\alpha$ absorption is more important on the SE side of the
ENLR. This has the effect of increasing the H$\alpha$ flux measurement
by $<5$\% and decreases the [\ion{N}{2}]~$\lambda$6583/H$\alpha$ ratio
by $<7$\% compared to the values obtained by using a linear continuum
level in the measurement of these lines. This is not large enough to
make a significant difference to the
[\ion{N}{2}]~$\lambda6583$/H$\alpha$ ratio. We have thus established
that the variation of the ratio between the NW and SE cones cannot be
attributed to H$\alpha$ absorption, and must therefore be due to real
physical differences between the excitation conditions in the cones.

The slight excess shown by the fit at the wavelength of the
[\ion{N}{2}]~$\lambda$6548 line is most likely due to a blue asymmetry
in the H$\alpha$ line similar to the asymmetry in the
[\ion{O}{3}]~$\lambda$5007 line seen at the same location. While it is
possible to add another component of H$\alpha$ to obtain a better fit
to the spectrum, we would also need to include similarly blue shifted
[\ion{N}{2}] components, and this data simply do not have the
resolution or signal to noise required to properly constrain the
relative fluxes or line widths of such additional components.

The lower panel of Figure~7 shows the fit to the brightest point in
the dust-lane region of the ENLR. Here H$\alpha$ and the [\ion{N}{2}]
lines are well fitted by single component Gaussians, and yield a
[\ion{N}{2}] $\lambda6583$/H$\alpha$ ratio of 0.59 $\pm0.01$ which is
intermediate to the values measured in the NW and SE cones.

\section{Physical properties of the emission line gas}

\subsection{Temperature}

The [\ion{O}{3}]~$\lambda$4363/[\ion{O}{3}]~$\lambda$5007 line ratio is
the well known temperature diagnostic often referred to as
R$_{OIII}$. We estimate the temperature from R$_{OIII}$ using the IRAF
STSDAS $nebular$ package which draws on the atomic data of
\cite{men83} (1983) for the O$^{+2}$ ion. An assumed density of 100
cm$^{-3}$ was used in the temperature calculation, and we note that
the calculation is relatively insensitive to density. The average
spectrum of the NW cone has R$_{OIII}$=0.019 $\pm 0.004$ corresponding
to T$_{e}$=14780$_{-980}^{+1550}$. The average spectrum of the SE cone
has R$_{OIII}$=0.025 $\pm 0.009$ corresponding to
T$_{e}$=16990$_{-2300}^{+5600}$, the greater uncertainties in the SE
cone are due to the large uncertainty in the [\ion{O}{3}]~$\lambda$4363
measurement. These temperatures are lower than the value of 19900 K
obtained by \cite{dur88} (1988) for the nuclear region. It is well
known that temperature fluctuations in the ionized region tend to
drive the value of R$_{OIII}$ up, so the measured temperatures are
probably somewhat higher than the mean temperature characterising the
O$^{+2}$ zone.

\subsection{Density}

The ratio of [\ion{S}{2}]~$\lambda$6717/[\ion{S}{2}]~$\lambda$6731 is a
density diagnostic sensitive in the range 100$-$10000~cm$^{-3}$. Density
estimates from the [\ion{S}{2}] ratios were calculated with the IRAF
STSDAS $nebular$ package which draws on the atomic data of
\cite{cai93} (1993) for the S$^{+1}$ ion. Figure~8 shows these
densities plotted against the projected radius of the spatial
locations within the ENLR. The average error bar on the density
measurements is approximately 80 cm$^{-3}$. We find that the
[\ion{S}{2}] ratios are close to the lower limit of the diagnostic
range, and that some points have [\ion{S}{2}] ratios which can only be
used as an upper limit for the density. The central regions have the
higher density values. Although Figure~8 shows a general trend of
decreasing density with distance from the nucleus, the densities in
the cones (when these can be measured) remain relatively constant with
distance.

\subsection{Ionization Parameter}

\cite{pen90} (1990) demonstrated that the [\ion{O}{3}]/[\ion{O}{2}]
line ratio is proportional to the ionization parameter $U$ for
radiation-bounded photoionized clouds irrespective of the shape of the
ionizing continuum. Following \cite{pen90} (1990) we estimate the 
ionization parameter by the relationship:

\begin{equation}
U = 1.82\times10^{-3}\;[{\mathrm O\;III}]\;\lambda5007\;/\;[{\mathrm O\;II}]\; \lambda 3727
\end{equation}

where we use the reddening corrected ratio of
[\ion{O}{3}]~$\lambda$5007 / [\ion{O}{2}]~$\lambda$3727. Figure~9 shows
the values of $U$ plotted against the projected radius. Interpreted
literally, this shows that the NW cone has slightly higher mean $U$
than the SE cone, while the dust lane and outer regions of the cone
have the lowest ionization parameter.

\section{LINE RATIO DIAGRAMS}

The utility of line diagnostic diagrams consisting of selected pairs
of emission line ratios is well established. With such diagrams,
different excitation mechanisms can be distinguished, chemical
abundances may be estimated (in special circumstances), and physical
parameters may be derived. The most commonly used diagrams involve
ratios of bright, easily measured lines (\cite{bal81} 1981), and the
best of these restrict the choice of line ratios to emission lines
that are close together in wavelength, avoiding the necessity to
determine reddening corrections with great accuracy (\cite{vei87}
1987).

Our main objective in this section is to analyse the excitation
mechanisms responsible for the emission lines. To do this we compare
the observed line ratios with grids of shock and photoionization model
predictions.

The models used here are:

\begin{itemize}
\item The MAPPINGS~II shock and shock+precursor models of \cite{dop96}
(1996). These models are parameterised by the shock velocity
$V_{shock}$=200$-$500 km~s$^{-1}$, and the magnetic parameter
$B/n^{1/2}$ = 0$-$4 $\mu$G~cm$^{3/2}$. In these models, EUV photons
produced within the cooling region of the shock, diffuse both upstream and
downstream. Those that go downstream are trapped by the shocked gas
and produce a low-ionization parameter photoionized region near the
point where hydrogen recombines. The combined spectrum of the cooling
and photoionized gas we refer to as the ``shock only'' model. The
(undoubtedly important) effect of thermal instabilities are not
accounted for in this model. The photons which pass upstream produce a
photoionized region of high ionization parameter and excitation in the
pre-shock medium.  The importance of this region depends upon whether
this medium is able to absorb the ionizing photons. Shock models which
include the contribution of both the shocked gas and its ionized
precursor are referred to as ``shock+precursor''.

\item The power-law photoionization models presented in \cite{all98}
(1998). These are simplified photoionization models calculated with a
ionizing spectrum of the form F$_\nu\propto\nu^\alpha$, with
$\alpha$=-1, -1.4 and -2 and a range of ionization parameters
($10^{-4}<U<1$), and densities (100$-$1000 cm$^{-3}$). These models
were calculated with the same MAPPINGS~II code as the shock models
described above.

\item The $A_{M/I}$ sequence of \cite{bin96} (1996). This is a
sequence of photoionization models which include both matter bounded
and ionization bounded clouds, such that the ionizing spectrum
incident on the ionization bounded clouds is filtered by the matter
bounded component. \cite{bin96} (1996) find that this two component
model can account for some of the discrepancies of single component
photoionization models. This sequence is parametrised by $A_{M/I}$
which is the solid angle ratio of the matter bounded clouds to the
ionization bounded clouds.

\end{itemize}

In plotting the observed points on the resulting line ratio diagrams,
we maintain the same plotting point styles as used in section 3 to
distinguish between the NW cone (open circles), SE cone (closed
squares) and inter-cone (open triangles) regions. In each diagram we
show the average error bar of the observed line ratio, and also
provide an arrow indicating the direction of reddening correction. The
shock model grids are labelled with the shock velocity and magnetic
parameter. The power-law photoionization $U$ sequences are plotted for
$n=100$~cm$^{-3}$ and $n=1000$~cm$^{-3}$ with grid lines of constant
$U$ every 0.25~dex.  The $A_{M/I}$ sequence is labelled with $A_{M/I}$
and tick marks at intervals of 0.2~dex.

\subsection{The Bright-line Diagnostic Plots}

In Figures 10 through 16 we show the bright-line diagnostic plots
which are generally used for determining excitation conditions in
AGN. On these diagrams, we have sufficient observational points to
analyse the detailed spatial structure of the ionization cones and the
inter-cone region. In all of these diagrams the line ratios fall
within limits that characterise Seyfert galaxies, in fact the points
from this single object cover almost completely the regions of this
diagram filled by samples of Seyfert galaxies (\cite{vei87} 1987),
indicating that the NLR of NGC~2992 is characterized by a
wide range of physical parameters. This is also reflected in the way
that the points fall over a large range of the parameter space covered
by the various models.

In section 3 we have already shown how the
[\ion{O}{3}]~$\lambda$5007/H$\beta$, and [\ion{O}{3}]~$\lambda$5007/H$\alpha$ 
versus [\ion{N}{2}]~$\lambda$6583/H$\alpha$ diagrams
separate the NLR of NGC~2992 into 3 distinct spatial regions. 
These diagrams are repeated here (Figures 10 and 11) with the model grids
overlaid. Similar behaviour is clearly shown in other excitation-dependent 
diagnostic plots such as [\ion{O}{3}]~$\lambda$5007/H$\beta$ versus
[\ion{S}{2}]~$\lambda\lambda$6717,6731/H$\alpha$ (Figure~12), or
[\ion{O}{3}]~$\lambda$5007/H$\beta$ versus
[\ion{O}{2}]~$\lambda$3727/[\ion{O}{3}]~$\lambda$5007 (Figure~13), but is not
apparent when the ratios of low-excitation lines with hydrogen are
compared, such as the [\ion{N}{2}]~$\lambda$6583/H$\alpha$ and
[\ion{S}{2}]~$\lambda\lambda$6717,6731/H$\alpha$ line ratios (Figure~14). 
In these diagrams the points from the different regions tend to cluster more
tightly, and are not clearly separated as they are in Figures 10 and 11. 

We will now consider how well the various theoretical sequences match
(or do not match) the observations.  First, the shock-only grid fails
to match any of the observational data, for any of these diagnostic
plots. Thus, if the ENLR gas is shock-excited, the shocks must be
occuring in a matter-rich environment where the EUV photons produced
by the shocks can be absorbed. We will not discuss the shock-only
grids further.

The shock+precursor grids do much better. In all of the diagrams, with
the exception of those which involve ratios of the [\ion{O}{1}] lines,
such as [\ion{O}{1}]~$\lambda$6300/[\ion{O}{3}]~$\lambda$5007 (Figure 15), 
the observations define shock velocities which are reasonably
consistent from one diagnostic diagram to another. For example, in
Figures 10a, 11a and 12a the points from the NW and SE cones fall
within the high shock velocity region (350$-$500 km~s$^{-1}$) of the
shock+precursor model grid. The inter-cone points show a large range
in [\ion{O}{3}]~$\lambda$5007/H$\beta$ which extends these points
along lines of decreasing shock velocity, down to as low as 
200~km~s$^{-1}$ with approximately constant magnetic parameter in the
shock+precursor model grid.  However, the different diagnostic plots
do not agree very well as to the value of the magnetic parameter,
other than broadly excluding the possibility that this is less than
about $B/n^{1/2}\sim1\mu$G~cm$^{3/2}$. However, such scatter as is
observed in this parameter between the different diagrams can 
probably be ascribed to differences between the abundance set used in the
models, and those that are applicable to NGC~2992, as well as to
collisional de-excitation effects in the [\ion{O}{2}] and [\ion{S}{2}]
lines, which are not modelled in these (low density limit) shock
models.

Turning now to the simple photoionization models. In each of the
diagnostic plots, the photoionization models overlap the same region
of parameter space in this diagnostic diagram as the shock+precursor
models, and so the observed points also fall within the
photoionization $U$-sequences. However, different regions tend to be
characterized by different ionization parameters and spectral
index. For example, on Figure~10b, the NW cone points are within the
$\alpha=-1.4$ models with ionization parameter
$10^{-2.5}<U<10^{-2.25}$.  The points from the SE cone show more
scatter, but indicate an average ionization parameter which is lower
than in the NW cone.  Most of the SE cone points fall between the $\alpha
=-1.4$ and $\alpha =-1.0$ sequences which may indicate a higher
spectral index in the NW cone than in the SE cone. There is however a
degree of degeneracy in the photoionization models such that increase
in density or an increase in spectral index have a similar effect on
the predicted line ratios.

The spread of the inter-cone points may therefore be partially
explained by a wider range in density than we computed in the
photoionization models. To investigate this possibility we extended
the photoionization grid by calculating models with densities down to
10$^{-2}$ cm$^{-3}$, for ionization parameters $U= 10^{-2}$ and
$10^{-2.75}$, and spectral index $\alpha =-1.4$. The low density limit
occurs for density $n_{e}\sim$1.0 cm$^{-3}$, below which
[\ion{O}{3}]~$\lambda$5007/H$\beta$ remains constant at $\sim10^{0.6}$
for $U=10^{-2}$ and $\sim10^{0.9}$ for $U=10^{-2.75}$. However, the
observed points extend over a wider range down to
[\ion{O}{3}]~$\lambda$5007/H$\beta\sim10^{0.3}$, making it difficult
to account for the observations with a photoionization models with a
single value of the spectral index $\alpha$ such as would be required
in a self-consistent photoionization model of the whole NLR.

Different diagnostic plots yield different estimates for both $\alpha$
and for $U$. For example, the NW cone points that fall within the
$\alpha =-1.4$ $U$-sequence in Figures 10b and 11b, are shifted closer
to the $\alpha =-1$ $U$-sequences in Figure~12b, with a lower value of
$U$. The shift of the points with respect to the photoionization
$U$-sequences is significant, being approximately 4 times the average
1$\sigma$ error bar of the
[\ion{S}{2}]~$\lambda\lambda$6717,6731/H$\alpha$ observations. For the
[\ion{S}{3}]~$\lambda\lambda$9069,9532/H$\alpha$ versus
[\ion{S}{2}]~$\lambda\lambda$6717,6731/H$\alpha$ plot (Figure~16),
$\alpha\sim-2$ is indicated.  In general, most of the plots are
consistent with $U\sim10^{-2.5}$ for the bright parts of the cones,
but those involving [\ion{N}{2}]~$\lambda$6583 /H$\alpha$ versus
[\ion{S}{2}]~$\lambda\lambda$6717,6731/H$\alpha$, and
[\ion{S}{3}]~$\lambda\lambda$9069,9532/H$\alpha$ versus
[\ion{S}{2}]~$\lambda\lambda$6717,6731/H$\alpha$ are more consistent
with $U\sim10^{-4}$.

We can therefore conclude that not only are the observations
inconsistent with a single value of $\alpha$ within individual line
ratio diagrams, but also that different diagnostic diagrams do not
furnish consistent values of $\alpha$ between one another. Since
photoionization models should be characterised by a unique value of
$\alpha$ determined by the intrinsic nuclear spectrum (although $U$
may change from one region to another), these observations provide a
serious challenge to photoionization modellers.

By contrast, the shock+precursor models indicate values of the shock
velocity which are consistent from one diagram to another; of order
350$-$500 km~s$^{-1}$ in the cones, falling to as low as
200~km~s$^{-1}$ in the inter-cone region. In the shock+precursor
picture, the apparent changes in both $\alpha$ and $U$ according to
the photoionization interpretation is the result of changes in the
spectral energy distribution and the total energy flux in the ionizing
radiation as a function of shock velocity.

These problems with photoionization models are not resolved by use of
the $A_{M/I}$ sequence. Indeed, the fact that $A_{M/I}$ sequences are
very similar to $U$-sequences, when plotted on the traditional line
ratio diagrams was shown in \cite{bin96} (1996). This is apparent in
Figures 10b, and 11b where the $A_{M/I}$ sequence is very close to the
higher density $\alpha$=-1.4 $U$-sequence. With respect to the
$A_{M/I}$ sequence, the observed points from the SE and NW cones shown
in Figures 10b and 11b cluster around the sequence for values of
$A_{M/I}$ in the range 1$\lesssim A_{M/I}\lesssim$6. The SE cone
points tend to have [\ion{O}{3}]~$\lambda$5007/H$\beta$ higher than
the $A_{M/I}$ sequence prediction, while the NW cone points have
[\ion{O}{3}]~$\lambda$5007/H$\beta$ slightly lower than the $A_{M/I}$
sequence, and favour higher values of $A_{M/I}$. As with the
$U$-sequences, the inter-cone points tend to diverge from the
$A_{M/I}$ sequence. In general the $A_{M/I}$ sequences can neither
explain the range or the direction of scatter of the observations in
NGC~2992. In addition, in some cases such as in the
[\ion{N}{2}]~$\lambda$6583/H$\alpha$ versus
[\ion{S}{2}]~$\lambda\lambda$6717,6731/H$\alpha$ and the
[\ion{S}{3}]~$\lambda\lambda$9069,9532/H$\alpha$ versus
[\ion{S}{2}]~$\lambda\lambda$6717,6731/H$\alpha$ diagrams (Figures 14
and 16 respectively), the $A_{M/I}$ sequence is offset from the
observations by an amount which is larger than could be explained by
abundance variations.

The result of comparing the ratios of the bright lines mapped over the
NLR of NGC~2992 can be summarized as follows: The ratios of
[\ion{O}{3}], [\ion{N}{2}], [\ion{S}{2}] and [\ion{S}{3}] generally
fall within the parameter space covered by both the shock+precursor,
and photoionization $U$-sequence models. The ratios of these lines
fall consistently in the same location of the diagnostic diagrams
relative to the shock+precursor grid, but do not provide consistency
with respect to the photoionization models. [\ion{O}{2}] appears to be
slightly overestimated in both shock+precursor and photoionization
models, and [\ion{O}{1}] is predicted to be too strong in all the models.

\subsection{Faint line diagnostic diagrams}

The lines of [\ion{Ne}{5}]~$\lambda$3426, \ion{He}{2}~$\lambda$4686 and
[\ion{O}{3}]~$\lambda$4363 also are important diagnostics, but are
relatively faint lines so we are not able to spatially map them over
the entire ENLR of NGC~2992. In the following three diagrams we compare
the model predictions with the observations of these lines in the
average spectra of the NW and SE cones.

Figure~17 shows [\ion{S}{2}]~$\lambda\lambda$6717,6731/H$\alpha$ versus
\ion{He}{2}~$\lambda$4686/H$\beta$. In this diagram the shock-only and
shock+precursor models overlap, so we show the shock-only model grid
with dashed lines. It is clear from this diagram that the
\ion{He}{2}~$\lambda$4686/H$\beta$ ratio is underestimated in the
shock+precursor models but the photoionization $U$-sequence shown in
Figure~17b can account for the large
\ion{He}{2}~$\lambda$4686/H$\beta$ ratio with $\alpha =-$1 and
ionization parameter $10^{-3}<U<10^{-2}$.

The line ratios used in Figure~17 are identified by \cite{bin96}
(1996) as one of the combinations for which $A_{M/I}$ and $U$-sequences
show markedly different behaviour. This is apparent in Figure~17b
where the $A_{M/I}$ sequence cuts across the $U$-sequences curves. The
data points for the NW and SE cones fall relatively close to the
$A_{M/I}$ sequence for A$_{M/I}\sim$1.

Figure~18 shows \ion{He}{2}~$\lambda$4686/H$\beta$ with
[\ion{O}{2}]~$\lambda$3727/[\ion{O}{3}]~$\lambda$5007, where we have
plotted the shock models, photoionization models and $A_{M/I}$
sequence on the one diagram. The relative positions of the various
grids in this diagram are similar to those Figure~17, and we find that
the points fall in similar positions with respect to the $A_{M/I}$
sequence and $U$-sequences, but fall well off the shock and
shock+precursor grids.

Figure~19 combines the ratio of the high excitation
[\ion{Ne}{5}]~$\lambda$3426 line to [\ion{Ne}{3}]~$\lambda$3869 versus
the [\ion{O}{3}]~$\lambda $5007/H$\beta$ ratio. The data points from
the SE and NW cones fall approximately 3$\sigma$ to the right of the
shock+precursor models, but do fall within the $\alpha=-1$
photoionization model. The inferred ionization parameter from the
position of the data points on the $U$-sequences is much higher than
inferred from the other diagrams, however the fact that the
U-sequences fold over themselves in this region of the diagram
(causing some regions to have double valued solutions within the
photoionization models) makes this inference very uncertain. The
$A_{M/I}$ sequence again shows different behaviour to the $U$-sequences
and the data points fall near the A$_{M/I}\sim$0.6 prediction.

Shock models of this simple class are known not to provide a good
description of the very high-excitation line spectrum. This includes
the \ion{He}{2}\ and [\ion{Ne}{5}] lines as well as higher excitation
species such as [\ion{Fe}{7}]. These lines are better predicted by
models which include a 
component of optically thin gas with high ionization parameter. 
The A$_{M/I}$ sequence is an example this. The two components used in the
A$_{M/I}$ models can be considered to be a simple approximation of
an inhomogeneous medium which may contain regions of different optical
thickness to ionizing radiation.
The success of the A$_{M/I}$
sequence in predicting the line ratios as shown in the faint-line
diagnostic diagrams may attest to the presence of inhomogeneities in
the ionized clouds. The ionized precursor gas in our shock+precursor
models is homogeneous and is thus not expected to reproduce these lines
accurately. Nevertheless, \cite{dop97} (1997) has shown that an optically 
thin, high ionization parameter component which emits high ionization lines,
can be produced in the context of more
complex shock models associated with shocked entrainment layers at the
boundary of ionization cones.
 
It may also be possible for combined shock and central source 
photoionization models to provide a better match to the observed 
line ratios. Such models are however difficult to constrain and
we find no obvious spatial or spectral separation of shock and
central source components in our data, so we have not attempted 
a combined model. Nevertheless it is clear that these diagnostics 
involving faint high excitation lines will be very useful in developing 
more sophisticated models of shocks, photoionization or combinations
of shocks and central source phtoionization.

\subsection{The temperature sensitive diagnostic diagram}

A long-standing problem in AGN modelling has been the so-called
temperature problem. Briefly the nature of this problem is that simple
photoionization models give low temperatures (less than 15000~K),
which disagrees with observations of
[\ion{O}{3}]~$\lambda$4363/[\ion{O}{3}]~$\lambda$5007 in NLRs of some
active galaxies, for which the inferred electron temperatures range up
to $\sim$22000~K (\cite{tad89} 1989). Shocks can better account for
the high temperatures because the [\ion{O}{3}]~$\lambda$4363 emission
from shocks is dominated by the 40000-20000~K region of the cooling
zone behind the shock front (\cite{dop95} 1995).  Recently
\cite{bin96} (1996) have shown the $A_{M/I}$ models can also predict
high values of
[\ion{O}{3}]~$\lambda$4363/[\ion{O}{3}]~$\lambda$5007. Also some of
the earlier [\ion{O}{3}]~$\lambda$4363 line measurements may be
overestimated due to poor subtraction of the underlying galaxy
spectrum.  We now compare the temperature sensitive line ratios
observed in NGC~2992 to the model predictions.

Figure~20 employs the temperature sensitive ratio of
[\ion{O}{3}]~$\lambda$4363/[\ion{O}{3}]~$\lambda$5007 on the y-axis,
versus \ion{He}{2} $\lambda$4686/H$\beta$ on the x-axis. 
Due to the
different temperature regimes of the shock and photoionization models,
the grids are widely spaced in this diagram. A temperature axis is
plotted on the left side of the diagram showing the inferred
temperatures calculated from the
[\ion{O}{3}]~$\lambda$4363/[\ion{O}{3}]~$\lambda$5007 ratio in the same
manner as discussed in section 4.1.

The large [\ion{O}{3}]~$\lambda $4363/[\ion{O}{3}]~$\lambda$5007 values
predicted by the shock-only model reflect the high temperatures that
occur in the cooling region behind the shock front.
The photoionization model grids display the well
known result that $U$-sequences generally have [\ion{O}{3}]
temperatures less than 15000~K. The highest [\ion{O}{3}] temperatures
obtained in photoionization models occur for the $\alpha=-1$ model
with ionization parameter $U=0.1$, with the $A_{M/I}$ able to produce
a similarly high [\ion{O}{3}] temperature. As shown in section 4.1 the
average spectra of the NW and SE cones have [\ion{O}{3}] temperatures
of 14780~K and 16990~K respectively. The location of the data points on
this temperature axis are consistent within the error bars with the
[\ion{O}{3}] temperatures predicted by the shock models, the $A_{M/I}$
sequence and also the $\alpha$=-1, $U$=0.1 photoionization models.

\cite{bin96} (1996) show that high temperatures are obtained in the
$A_{M/I}$ sequence because the MB component is chosen to be
sufficiently thin that the temperature is governed by the
photoionization of He$^{+}$, and that the O$^{+2}$ zone is truncated
before the internal temperature of that zone falls enough to decrease
the [\ion{O}{3}]~$\lambda$4363/[\ion{O}{3}]~$\lambda$5007 ratio. The
$A_{M/I}$ sequences described by \cite{bin96} (1996) (and displayed in
Figure~12 of \cite{bin96} 1996) can reach
[\ion{O}{3}]~$\lambda$4363/[\ion{O}{3}]~$\lambda$5007 $\sim$0.019 for an
incident $\alpha=-1.3$ ionizing spectrum (as plotted in Figure~20), up
to $\sim$0.025 for an incident ionizing spectrum with $\alpha=-1.1$,
and as high as $\sim$0.048 when the metallicity of the gas is
decreased to Z=0.2. As such these $A_{M/I}$ sequences are able to
reproduce the relatively high temperatures observed in the NW and SE
cones of NGC~2992.

\section{CENTRAL SOURCE TEST}

\subsection{Determination of $Q$ from [\ion{S}{2}] density.}

\cite{met96} (1996) describe a method to test whether the line
emitting gas in the NLRs of Seyfert galaxies is photoionized by a
central source. The test uses the definition of the ionization
parameter for radiation from a central source incident on an emission
line cloud:

\begin{equation}
U=\frac{q}{n_er^2c} > \frac{q}{n_e(r/\cos\theta)}  
\end{equation}

where $q$ is the number of ionizing photons per unit time per steradian,
$n_e$ is the electron density, and $r$ is the distance of the emission
line cloud from the source. $\theta$ is the angle between a line from
the source to the cloud and the plane of the sky. Using
[\ion{O}{3}]~$\lambda$5007/[\ion{O}{2}]~$\lambda$3727 to estimate $U$,
\Siir\ to measure the density $n_e$ and estimating $r$, allows $q$ to be
determined. If the assumption of a central source is correct, then $q$
should be independent of position in the NLR.  Using long slit
spectroscopy \cite{met96} (1996) applied this technique to the NLR of
three nearby Seyfert galaxies including NGC~2992. In the case of
NGC~2992 their slit was aligned along the radio structure (PA = 26\arcdeg).
They find that $q$ rises with increasing distance from the nucleus in
all three NLR.

Here we perform a similar test on NGC~2992 using our data which has
more complete spatial coverage of the NLR than the single slit in
\cite{met96} (1996).  We confirm the result that $q$ measured in this
way does rise with increasing distance from the nucleus. However we
find that the \Siir\ density diagnostic is not sensitive in the
density range required to measure $q$ in the NLR of NGC~2992, and thus
demonstrate that the test is invalid.

To perform the ``$q$-test'' with our spectra of NGC~2992 we have
considered only the regions within the bi-cone of opening angle
116\arcdeg\ centred on the nucleus in which have well determined
[\ion{S}{2}] density measurements.  $U$ is determined by using the
reddening corrected
[\ion{O}{3}]~$\lambda$5007/[\ion{O}{2}]~$\lambda$3727 ratio as described
in section 4.3.  The values of $n_e$ and $U$ for points within the cone
are shown in Figures 21a and 21b respectively.  To estimate the de-projected
distances of points in the bi-cone to the nucleus, we have considered
all the points to lie within the geometric volume of the
bi-cone. Narrow band images show that the bi-cone axis is aligned at
roughly right angles to the plane of the galaxy. Here we assume that
the bi-cone has the same inclination on the sky as the galaxy, and
that the SE cone is the near one.  Using this geometry we can then
calculate two values for the de-projected radius of each point
\footnote{\begin{equation}
r_{deprojected} = \frac{y(\cos(\psi + i) + \sin\psi\sin i) \pm \sqrt{y^2\sin^2\psi\sin^2i - x^2\sin^2i\cos^2(\psi + i) -2x^2\cos(\psi + i)\sin\psi\sin^3i}}{\cos^{2}(\psi + i) + 2\cos(\psi + i)\sin\psi\sin i} 
\end{equation}
where $i$ is the angle subtended by the cone axis to the plane of the sky,
and $\psi$ is the half angle of the cone. },
which correspond to the near and far surfaces of the cones. These two
values represent the extrema for the de-projected radii of points
within the cones, and we calculate $q$ using both values, and likewise with
the projected radius.

The diagrams on the left hand side of Figure~22 show the values of 
$Q=4\pi q$ derived for; (i) de-projection onto the near side for both 
cones (top panel), (ii) de-projection onto the far side of the cones 
(middle panel) and (iii) no de-projection (bottom panel). In all 
cases $Q$ is found to rise with increasing distance from the nucleus, 
and the result is relatively insensitive to the de-projection used.

A rising $Q$ is inconsistent with photoionization from the central nucleus,
because in this case $Q$ should have the same value wherever it is measured.
\cite{met96} (1996) rule out the possibility that this result is due to a
projection effect because it would require $\theta$ to increase towards the
nucleus, which is inconsistent with the observed sharp edged shape of the
cone. \cite{met96} (1996) also note that a two-component ionizing continuum
due to a mixture of ionization bounded and matter bounded clouds cannot
account for the observed increase in $Q$. Although photoionizing shocks 
(\cite{dop95} 1995) provide a local source of UV which might explain the
rising trend of $Q,$ it is more likely that the apparent increase in $Q$ is
spurious, and is the result of overestimating the average density
using the [\ion{S}{2}]~$\lambda$6717/$\lambda$6731 ratio.

The [\ion{S}{2}]~$\lambda$6717/$\lambda$6731 ratio is only sensitive
as a density diagnostic over the range $\sim$100$-$10000 cm$^{-3}$. The
ratio tends to a limit of 1.4 for lower densities, and 0.44 for higher
densities. In NGC~2992 we find that the [\ion{S}{2}] density for some
locations near the nucleus is in the range 100$-$400 cm$^{-3}$, however
for many locations the [\ion{S}{2}]~$\lambda$6717/$\lambda$6731 ratio
lies near or below the lower limit of the usable range of this density
diagnostic. If we take the value of $Q$ as measured near the nucleus
where the density is relatively well determined, we find $Q=10^{52.9}$, 
in units of photons s$^{-1}$. This value is probably
reliable, and reflects the true value for the central ionizing
source. Using this value of $Q$, and the measurements of the
ionization parameter and de-projected distances, we rearrange equation~2 
to calculate the density required at each location,
$n_{e,Q},$ to keep $Q$ constant:

\begin{equation}
\log(n_{e,Q}) = \log(52.9) - \log(4 \pi c) - \log U -2\log r
\end{equation}

The diagrams on the right hand side of Figure~22 show the results of 
this calculation, with $\log n_{e,Q}$ plotted against $r$ for each
of the de-projections. The dashed line indicates the
lower useful limit for the [\ion{S}{2}] density diagnostic. Most of
the points indicate that $n_{e,Q}$ is below this limit. Thus if $Q$ is
actually constant at $10^{52.9}$~photons~s$^{-1}$, we would not be
able to measure it at large $r$ because the density diagnostic is not
sensitive in the range required. The $q$-test is therefore invalid in
this case. However, if the local density is really as high as
200$-$300~cm$^{-3}$ in the regions at large $r$ where we can measure
it, then we would still require shocks if only to produce a local
compression.

\subsection{Determination of $Q$ from the H$\alpha$ luminosity}

In this section we compute the rate of hydrogen recombinations $Q_{rec}$
from the H$\alpha$ luminosity and compare this to the rate of ionizing
photons $Q$, derived from the emission line ratios above. In photoionization
equilibrium the rate of hydrogen recombinations must equal the rate of
ionizing photons. The rate of hydrogen recombinations $Q_{rec}$ can be
derived from the H$\alpha$ luminosity by comparing the effective
recombination coefficient to produce H$\alpha$ photons with the total (Case
B) recombination coefficient for hydrogen:

\begin{equation}
Q_{rec}\simeq 1.39 \times 10^{52}\, {\mathrm L}_{40}({\mathrm H}\alpha)\ {\mathrm photons\ s^{-1}}
\end{equation}
where L$_{40}({\mathrm H}\alpha)$ is the H$\alpha$ luminosity (in units of
10$^{40}$ ergs~s$^{-1}$).  In the case where a central source is 
illuminating the NLR:

\begin{equation}
Q_{rec}\simeq 1.39 \times 10^{52}\, {\mathrm L}_{40}({\mathrm H}\alpha)\,\frac{4\pi}{\epsilon \omega}\ {\mathrm photons\ s^{-1}}
\end{equation}

where $\epsilon$ is the fraction of ionizing photons that are
absorbed, and $\omega$ is the solid angle subtended by the H$\alpha$
emitting region as seen from the nucleus.

The reddening corrected H$\alpha$ luminosities of the NW and SE cones
are estimated from our H$\alpha$+[\ion{N}{2}] narrow band image,
corrected for reddening with a reddening map derived from the grid of
H$\alpha$/H$\beta$ ratios measured from the spectra. The reddening map
was constructed by re-sampling the semi-regularly spaced grid of
reddening constants from the spectra onto the image pixel grid with
the IDL {\em TRIGRID} routine.  The H$\alpha$+[\ion{N}{2}] flux is
summed over the specified region in the reddening corrected image, and
a correction applied for the [\ion{N}{2}] contribution using the
average [\ion{N}{2}]/H$\alpha$ ratio of the region as measured from
the spectra. This yields the total H$\alpha$ flux which is then
converted to a luminosity using the distance to NGC~2992 (30.8 Mpc).
For each of the SE and NW cones, we have used only the H$\alpha$
luminosity of the cones (corresponding to the same region over which
$Q$ is calculated in the previous section) and have specified
$\omega$= 2.26 steradian, the solid angle of the cone subtended at the
nucleus. The calculation of $Q_{rec}$ for the entire NLR region
ignores the cone geometry and uses the total H$\alpha$ luminosity and
$\omega=4\pi$. The dominant source of uncertainty in these
calculations are errors in the reddening correction, and the implicit
assumption that $\epsilon$ is unity. This means that the $Q_{rec}$
values derived represent a lower limit estimate of the number of
ionizing photons $Q$.

The values of L$_{40}$(H$\alpha$) and $Q_{rec}$ computed for the NW
and SE cones, and the NLR as a whole are shown in table
2\footnote{Table 2 also contains L$_{40}$([\ion{O}{3}]) calculated
from the reddening corrected [\ion{O}{3}] narrow band
image}. Comparing $Q_{rec}$ derived from the H$\alpha$ luminosity, and
the $Q$ derived from the forbidden emission line ratios we find that
$Q_{rec}$ lies within the computed range of the $Q$ values for all the
regions considered. If the whole region was photoionized by a single
central source then we should expect the values of $Q_{rec}$ to
correspond to the value of $Q$ as determined for the inner parts of
the NLR in the previous section. However, our lower limits for
$Q_{rec}$ are however consistently higher than the value of $Q$
derived at the inner edge of the NLR (10$^{52.9}$ photons s$^{-1}$).

In conclusion, although the $q$-test in its original form is invalid,
there does appear to be real evidence for a local source of ionizing
photons with a hard spectrum within the cone. This can only come from
photoionizing shocks. We note that a method along 
the lines
of the $q$-test as described here, but devoid of the density
diagnostic limitations, to quantify the relative contributions to
the ionizing field from local and central sources. This would aid the
development of combined shock and central source photoionization
models.

\section{SHOCK ENERGETICS}

The emission line properties of the ENLR of NGC~2992 appear to be
understandable in the context of shock excitation of the gas. In
this section we compare the excitation of the gas to the kinematics of
the outflow/wind, and also compare the energetics of the radio and 
X-ray plasma to the emission line gas.

\subsection{Excitation and Kinematics}

As described in section~1, the kinematics of the emission line gas
shows a outflow velocity field superimposed on the underlying galaxian
rotation.  Figure~23a shows a map of the the velocity difference of
the wind component to the galaxy rotation from our high resolution
spectra of the [\ion{O}{3}] line. The wind component is systematically
blueshifted in the SE cone, and redshifted in the NW cone with
velocities of up to $\sim$200 km~s$^{-1}$.  A map of the FWHM of the
wind component shown in Figure~23b reveals that the SE cone has the
broader FWHM, and we consider this to be evidence of greater dynamical
activity in the SE cone.

If shocks are important for the excitation of the gas, we may expect
the kinematical properties of the gas to in some way correlate with
excitation.  To compare the excitation of the gas with the kinematics
we re-binned the high and low resolution spectra onto a common spatial
grid covering the ENLR, and extracted the emission line ratios, and
kinematical properties at each grid point. We now compare the
[\ion{O}{3}]~$\lambda$5007/H$\beta$, and
[\ion{N}{2}]~$\lambda$6583/H$\alpha$ ratios to the kinematics.

In Figure~24a we plot the [\ion{O}{3}]~$\lambda$5007/H$\beta$ ratio,
which is a measure of the excitation of the gas, versus the FWHM of
the kinematical components. The kinematical component identified with
the galaxy rotation are plotted as the solid circles, the main outflow
components (as shown in Figure~23) are plotted as open squares and
where the extra component is detected (at a few locations), this is
plotted as open triangles. The uncertainty in the FWHM is generally
less than 100~km~s$^{-1}$.

The outflow component points show a wide range in FWHM from 100 to
670~km~s$^{-1}$, whereas the FWHM of the rotational component is much
more quiescent, and is generally less than $\sim$200 km~s$^{-1}$. The
excitation of the rotational component covers the full range in
[\ion{O}{3}]~$\lambda$5007/H$\beta$, while the wind components tend to
show higher [\ion{O}{3}]~$\lambda$5007/H$\beta$ ratios in locations
where the lines are wider. No regions with line FWHM$>$250 km~s$^{-1}$
display low excitation.

We also compare the data points on the excitation versus line width
diagram to the theoretical curves predicted by the shock+precursor
models. The solid curve assumes the FWHM is directly related to the
shock velocity, and the dashed curve assumes that the FWHM is
associated with the projection of randomly oriented shocks so that
$V_{shock}\simeq$FWHM/$\sqrt{3}$.  While the spread of the observed
points is very suggestive of shock activity, it is not what would be
expected for a simple shock moving into an undisturbed pre-shock
medium. In this case, the regions of very high
[\ion{O}{3}]~$\lambda$5007/H$\beta$ ratio would arise in the
precursor, which would be kinematically undisturbed, and therefore
there would be an inverse correlation of line width and
excitation. However, if the shocks arise in a mass-entraining flow,
the strong shocks are located near the outer boundary of the
entrainment flow, and the photoionized precursor medium is the
fast-moving material inside this (\cite{dop97} 1997). This geometry
would provide the direct correlation between line width and excitation
observed.

In Figure~24b we plot the [\ion{N}{2}]~$\lambda$6583/H$\alpha$ ratio
versus the FWHM of the outflow components. As in Figure~24a the square
points indicate the main outflow component, and the triangles indicate
the extra component. Here we make the distinction between the cones by
using filled symbols for the SE cone and open symbols for the NW
cone. The cones are clearly separated in Figure~24b because the NW
cone has lower [\ion{N}{2}]~$\lambda$6583/H$\alpha$ and lower FWHM,
while the more dynamically active SE cone has higher
[\ion{N}{2}]~$\lambda$6583/H$\alpha$. The shock+precursor curves
plotted on Figure~24b show that the models predict higher
[\ion{N}{2}]~$\lambda$6583/H$\alpha$ for faster shock velocities.  This
suggests that the higher [\ion{N}{2}]~$\lambda$6583/H$\alpha$ in the
SE cone is directly related to the greater dynamical activity in that
cone. 

The difference in shock velocity between the cones inferred from these
curves is however greater than that indicated by the
[\ion{O}{3}]~$\lambda$5007/H$\beta$ ratio which indicates only
slightly faster average shock velocity in the SE cone (see
Figure~10a).  As such, the increase in
[\ion{N}{2}]~$\lambda$6583/H$\alpha$ is likely due to a combination of
faster shock velocity and increased magnetic parameter, both of which
may be expected in the more turbulent conditions within the SE cone.

\subsection{Thermal pressure energy source for the ENLR}

Another aspect of shock energetics is related to the question of
whether there is sufficient energy in the radio plasma to power the
optical emission. For Seyferts, classical equipartition arguments
suggest that the pressure in the relativistic plasma is inadequate to
drive high velocity shocks into the interstellar medium and produce
the observed emission line flux (\cite{wil88} 1988). Recently
\cite{bic98} (1998) have argued that the radio plasma may only trace a
fraction of the available energy which exists in an entrained thermal
phase, which is not detected in synchrotron emission, and that the NLR
may well be powered by the energy and momentum flux originally carried
by the radio-emitting jets. 

Briefly summarising the physics detailed in \cite{bic98} (1998): 
The total energy of a jet-fed lobe is $f_{ad}F_Et$ where $F_E$
is the jet energy flux, $t$ is the age of the lobe and $f_{ad}\sim0.5$
is a factor which accounts for energy lost in adiabatic expansion. 
Expansion of the lobe gives rise to narrow line emission via shocks
and the jet energy flux $F_E$ can be estimated from the
[\ion{O}{3}]~$\lambda$5007 luminosity, L([\ion{O}{3}])$\sim0.01F_E$. 
The relativistic electrons in
the lobe emit radio wavelength synchrotron radiation, and the ratio of
the monochromatic radio power of the lobe (in synchrotron radiation)
to the jet energy flux can be expressed as:

\begin{equation}
\kappa_{\nu}\simeq C(a) B^{(a+1)/2} \nu^{-(a-1)/2} f_e f_{ad} t
\end{equation}

where $a$ is the power-law exponent of the electron energy distribution,
$C(a)$ depends on synchrotron parameters, and
$B$ is the randomly oriented magnetic field. $f_e$ is the fraction of
the total energy in relativistic electrons and positrons, and is thus
a measure of non-thermal to thermal composition of the plasma.
We now use the relationships detailed in \cite{bic98} (1998) to
investigate the possibility of a thermal energy source for
the ENLR of NGC~2992.

Our reddening corrected [\ion{O}{3}]~$\lambda$5007 luminosity gives an
estimate of $F_E=9.6\times10^{43}$~ergs~s$^{-1}$, and the 335 mJy flux
density detected in the Fleurs map of \cite{war80} (1980) gives a radio
power at 1.4 GHz of P$_{1.4}$=3.8 $\times10^{29}$~ergs~s$^{-1}$~Hz$^{-1}$. 
This yields $\kappa_{1.4}=4\times10^{-15}$~Hz$^{-1}$, which 
is typical of Seyfert galaxies (eg. \cite{bic98} 1998).

The total pressure can be expressed as:
\begin{equation}
P_{tot} = P_{particles} + B^2/8\pi
\end{equation}
where 
\begin{equation}
P_{particles} = P_{e} + P_{th} = (1 + k)\,P_{e}
\end{equation}
is the total pressure in relativistic electrons $P_{e}$ plus the 
pressure in thermal particles $P_{th}$. 

The minimum pressure magnetic field is obtained by minimising the
total pressure subject to the constraint of the observed radio surface
brightness.  Normally when estimating the pressure in the plasma of
radio galaxies via minimum pressure arguments, one takes $k\sim1$.
For NGC~2992, using $k=1$, an assumed spectral index of 0.7 and an
electron index $a=2.4$ we obtain
$P_{particles}=1.1\times10^{-11}$~dyn~cm$^{-2}$, and a minimum
pressure magnetic field of $B=1.5\times10^{-5}$~G. \cite{col98} (1998)
infer the pressure of the X-ray gas to be
$1.9\times10^{-11}$~dyn~cm$^{-2}$ for T=0.1~keV, and 
$7.9\times10^{-11}$~dyn~cm$^{-2}$ for T=1~keV.
This is larger than the value for $P_{particles}$ obtained in
the minimum pressure calculation with $k=1$, although only
marginally so for T=0.1~keV. Therefore it is possible
that the conventional estimate of the minimum pressure associated with
the non-thermal plasma is an underestimate. Moreover, we note that in
order to drive a shock with velocity $\sim350$~km~s$^{-1}$ into gas
with a pre-shock number density of $\sim1$~cm$^{-3}$, we require a
pressure of $\sim10^{-9}$~dyn~cm$^{-2}$. This is well above the upper
X-ray estimate, although the latter is based upon the total extended
X-ray luminosity and is therefore an average. Nevertheless the point
remains that the pressure of the X-ray emitting gas is possibly well
above the conventionally calculated non-thermal value.

Therefore we consider
values of the parameter $k \gg 1$. We find that the
thermal pressure, $\frac{k}{1+k}P_{particles}$, is equal to upper
value for the pressure in the X-ray gas for $k=75$. The minimum
pressure magnetic field in this case is $B=4.1\times10^{-5}$~G. If the
lower X-ray pressure is relevant, we obtain values of $k=6$ and
$B=2.2\times10^{-5}$~G; that are close to the minimum pressure values
for the $k=1$ case.

Using these values of $k$ we now estimate $f_e$:

\begin{equation}
f_e \equiv \frac{\epsilon_e}{\epsilon_e + \epsilon_{th}} = [1 + (P_{th}/2P_e)]^{-1} = [1 + k/2]^{-1} 
\end{equation}

where $\epsilon_e$ and $\epsilon_{th}$ are the energy densities of the
relativistic and thermal components. This gives $f_e$=0.026 for
$k=75$, and $f_e$=0.25 for $k=6$.  These low values of $f_e$, in
particular $f_e$=0.026, indicate that the relativistic gas which is
responsible for the radio emission may constitute only a small
fraction of the total energy.

Estimating the age to be $t=10^{6}-10^{7}$~yrs from the expansion
time-scale, and $f_{ad}\sim0.5$, we obtain $\tau = f_e f_{ad} t =
10^{4.1}-10^{5.1}$~yrs for $f_e$=0.026, and
$\tau=10^{5.1}-10^{6.1}$~yrs for $f_e$=0.25.  Using these values of
$\tau$ and the corresponding minimum pressure magnetic fields we can
read off $\kappa_{1.4}$ from Figure~2 in \cite{bic98} (1998).  This
gives $\kappa_{1.4}$ in the range
$2\times10^{-14}-7\times10^{-13}$~Hz$^{-1}$. These values are larger
than the observed value of $\kappa_{1.4}$ described above, implying
that the energy associated with the radio and X-ray gas is less than
the amount required to power the total optical emission of the ENLR.
Given the uncertainties in the X-ray images and the reddening
correction to the [\ion{O}{3}]~$\lambda$5007 luminosity, we may view
the results presented here as accurate to an order of magnitude.  Even
so, this result represents a significant difference compared to the
traditional value of $\kappa_{1.4}\approx10^{-11}$ which is implicitly
used when the radio emission is considered as a scaled down version of
a radio galaxy lobe.

Another possibility is that minimum pressure conditions are not
appropriate. In this case, with $k=1$, the magnetic field required for
the total particle pressure to equal the X-ray gas pressure is
$B=3.3\times10^{-6}$~G for the upper value of the X-ray gas pressure,
and $B=7.5\times10^{-6}$~G for the lower value of the X-ray
pressure. Using the same estimates for $t$ and $f_{ad}$ as above leads
to estimates of $\kappa_{1.4}=5\times10^{-15} -
6\times10^{-14}$~Hz$^{-1}$ for the upper value of the X-ray pressure
and $\kappa_{1.4}=3\times10^{-14} - 3\times10^{-13}$~Hz$^{-1}$ for the
lower value of the X-ray pressure. The observed value of $\kappa_{1.4}$
is outside of these ranges, but only marginally so for the upper
value of the X-ray pressure.

In the context of the theoretical work of \cite{bic98} (1998) the
properties of the optical, radio and X-ray emitting gas in NGC~2992
indicate that the energy associated with the radio plasma and the X-ray 
emitting gas may provide a larger fraction of the energy required to 
power the ENLR than anticipated by previous calculations for Seyferts
which assume  $\kappa_{1.4}=3\times10^{-11}$. The uncertainties in the above
calculations could be reduced if an accurate estimate of the
temperature of the X-ray emitting gas was available.

\section{CONCLUSIONS}

The physical properties of NGC~2992 derived from the emission line
spectra show that the ENLR is characterized by relatively high
temperature, [\ion{S}{2}] densities up to 400 cm$^{-3}$ and a fairly
wide range of excitation.

In comparing the spatially mapped line ratios with an extensive shock
and photoionization model grids, we have shown that the observations
generally fall within the parameter space defined by either class of
model.  However, in terms of the consistency of the model predictions
from one diagnostic diagram to another, the shock+precursor models
provide a better outcome. The ionization cones are characterised by a
shock velocity of 350$-$500~km~s$^{-1}$, consistent with the line
widths found in high resolution observations. In the inter-cone
region, and about the edges of the ionization cones, the velocities
inferred are lower, ranging down to about 200 km~s$^{-1}$. Single
component photoionization models would require a range of spectral
indices in the different regions, which is not allowed in a model
where the photons are provided by a central source. In addition,
different line ratios demand different values of the ionization
parameter in the same region, which is again inconsistent. The two
component $A_{M/I}$ sequence model of \cite{bin96} (1996) is generally
better able to account for the observed line ratios than the simple
single component photoionization models. Indeed, the fit provided by
both the shock and photoionization hypotheses, in particular for the
higher ionization lines, would be improved by more sophisticated 
multi-component models.

We have applied the $q$-test for a central ionizing source and have
shown that the apparent increase in the rate of ionizing photons, $q$,
with distance from the nucleus is most likely due to overestimation of
the average density with the [\ion{S}{2}]~$\lambda$6717/$\lambda$6731
ratio. The most reliable values of $q$ determined at the inner edge of
the NLR, are however below the lower limits for the hydrogen
recombination rate $Q_{rec}$ suggesting that there is a source of
locally produced ionizing photons which may be attributed to the
ionizing field generated by fast shocks.

Despite the lack of any direct morphological indication of interaction
between the radio or X-ray plasma with the emission line gas in
NGC~2992, shocks must play an important role in this object. The
outflow kinematics attest to this, because strong shocks with Mach
numbers of order 40 simply must occur in such dynamically active
gas. Furthermore, the observation of co-spatial X-ray and radio
emitting plasma support the entrainment theory of \cite{bic98} (1998)
in which the Seyfert jets are dominated by the entrained thermal
plasma. In this object we find that the energy associated with the
radio and X-ray plasma may account for a large fraction of the energy
required to power the total optical emission of the ENLR.

While the emission line ratios alone cannot uniquely determine shocks
as the dominant excitation mechanism, the observed ratios and the
velocity field and the kinematical line widths are all broadly
consistent with shocks, and with the energetics as determined from the
radio and X-ray plasma.

We find no spectral evidence for significant contribution to the
ionizing field or outflow from a starburst component, and favour the
interpretation of \cite{col98} (1998) that the source of the outflow
is an initially relativistic jet originating at the AGN.

\acknowledgments

We thank Geoff Bicknell for assistance with the energy budget
calculations in this paper, and for the use of his minimum pressure
program.  MA acknowledges the support of an APRA scholarship. We are
grateful for an Australian DIST International Science and Technology
Major Grant which facilitated the cooperation between MSSSO and The
Johns Hopkins University.  This research has made use of the NASA/IPAC
Extragalactic Database (NED), which is operated by the Jet Propulsion
Laboratory, Caltech, under contract with NASA.

\clearpage

\begin{deluxetable}{lccccccc}
\footnotesize
\tablecaption{Line intensities of NGC~2992}
\tablewidth{0pt}
\tablehead{
\colhead{} &
\multicolumn{2}{c}{NW cone} &
\multicolumn{2}{c}{SE cone} &
\multicolumn{2}{c}{inter-cone}  \\
\colhead{Line} & 
\colhead{Flux} &
\colhead{Flux} &
\colhead{Flux} &
\colhead{Flux} &
\colhead{Flux} &
\colhead{Flux}  \\
\colhead{} & 
\colhead{} &
\colhead{(r.c)} &
\colhead{} &
\colhead{(r.c)} &
\colhead{} &
\colhead{(r.c)}  \\
\colhead{(1)} &
\colhead{(2)} &
\colhead{(3)} &
\colhead{(4)} &
\colhead{(5)} &
\colhead{(6)} &
\colhead{(7)} }
\startdata
[\ion{Ne}{5}] $\lambda$3426            &    560  $\pm$  12  &   81 $\pm$  17  &  61  $\pm$  20  &  130 $\pm$  40   &   \nodata     &  \nodata        \nl  
[\ion{O}{2}] $\lambda$3727             &    440  $\pm$  20  &  610 $\pm$  30  &  423 $\pm$  15  &  802 $\pm$  28   &  62 $\pm$  3  &  176 $\pm$  9   \nl  
[\ion{Ne}{3}] $\lambda$3869            &    95   $\pm$  35  &  125 $\pm$  46  &  100 $\pm$   8  &  180 $\pm$  10   &  50 $\pm$ 20  &  130 $\pm$ 70   \nl  
H$_{8}$ + \ion{He}{1} $\lambda$3888    &    22   $\pm$   9  &   29 $\pm$  12  &  \nodata        &  \nodata         &   \nodata     &  \nodata        \nl  
[\ion{Ne}{3}] $\lambda$3967            &    31   $\pm$  11  &   40 $\pm$  15  &  \nodata        &  \nodata         &   \nodata     &  \nodata        \nl  
[\ion{S}{2}] $\lambda$4072             &    \nodata         &   \nodata       &  9   $\pm$   5  &  14  $\pm$   8   &   \nodata     &  \nodata        \nl  
H$\delta$                              &    22   $\pm$   8  &   27 $\pm$   9  &  16  $\pm$   5  &  26  $\pm$   8   &  24 $\pm$ 11  &  49  $\pm$ 24   \nl  
H$\gamma$                              &    42   $\pm$   4  &   49 $\pm$   4  &  37  $\pm$   6  &  51  $\pm$   8   &  55 $\pm$ 13  &  84  $\pm$ 21   \nl  
[\ion{O}{3}] $\lambda$4363             &    15   $\pm$   3  &   17 $\pm$   3  &  16  $\pm$   6  &  21  $\pm$   7   &   \nodata     &  \nodata        \nl  
\ion{He}{2} $\lambda$4686              &    28   $\pm$   4  &   29 $\pm$   4  &  32  $\pm$   7  &  35  $\pm$   7   &   \nodata     &  \nodata        \nl  
H$\beta$                               &    100  $\pm$   6  &  100 $\pm$   6  &  100 $\pm$   6  &  100 $\pm$   6   & 100 $\pm$ 10  &  100 $\pm$ 10   \nl  
[\ion{O}{3}] $\lambda$4959             &    320  $\pm$  20  &  310 $\pm$  20  &  290 $\pm$  40  &  280 $\pm$  40   &  70 $\pm$  5  &  65  $\pm$  5   \nl  
[\ion{O}{3}] $\lambda$5007             &    950  $\pm$  60  &  920 $\pm$  60  &  910 $\pm$ 120  &  840 $\pm$ 100   & 220 $\pm$ 18  &  194 $\pm$ 16   \nl  
[\ion{N}{1}] $\lambda$5200             &    7    $\pm$   2  &    7 $\pm$   2  &  \nodata        &  \nodata         &   \nodata     &  \nodata        \nl  
\ion{He}{1} $\lambda$5876              &    31   $\pm$  12  &   26 $\pm$  10  &  63  $\pm$  15  &  42  $\pm$  10   &  52 $\pm$ 17  &  27  $\pm$  9   \nl  
[\ion{O}{1}] $\lambda$6300             &    40   $\pm$   3  &   32 $\pm$   3  &  90  $\pm$  14  &  54  $\pm$   8   &  56 $\pm$ 11  &  24  $\pm$  5   \nl  
[\ion{O}{1}] $\lambda$6363             &    19   $\pm$   3  &   14 $\pm$   2  &  37  $\pm$  17  &  20  $\pm$  10   &   \nodata     &  \nodata        \nl  
[\ion{N}{2}] $\lambda$6548             &    63   $\pm$   1  &   48 $\pm$   1  &  143 $\pm$   4  &  79  $\pm$   2   & 163 $\pm$  2  &  63  $\pm$  1   \nl  
H$\alpha$                              &    410  $\pm$   5  &  310 $\pm$   3  &  560 $\pm$  10  &  310 $\pm$   8   & 810 $\pm$ 10  &  310 $\pm$  4   \nl  
[\ion{N}{2}] $\lambda$6583             &    189  $\pm$   4  &  142 $\pm$   3  &  430 $\pm$  10  &  237 $\pm$   7   & 490 $\pm$  7  &  185 $\pm$  3   \nl  
[\ion{S}{2}] $\lambda$6716             &    123  $\pm$   6  &   91 $\pm$   5  &  230 $\pm$  20  &  120 $\pm$  10   & 220 $\pm$ 32  &  79  $\pm$ 11   \nl  
[\ion{S}{2}] $\lambda$6731             &    95   $\pm$   5  &   71 $\pm$   4  &  180 $\pm$  20  &  98  $\pm$   9   & 174 $\pm$ 25  &  62  $\pm$  9   \nl  
[\ion{S}{3}] $\lambda$9069             &    67   $\pm$   5  &   38 $\pm$   3  &  83  $\pm$  42  &  25  $\pm$  13   & 251 $\pm$ 27  &  37  $\pm$  4   \nl  
[\ion{S}{3}] $\lambda$9532             &    197  $\pm$  11  &  109 $\pm$   6  &  300 $\pm$  70  &  90  $\pm$  20   & 512 $\pm$ 44  &  68  $\pm$  6   \nl  
\enddata
\tablecomments{ All intensities are given relative to H$\beta$=100.
              (r.c) indicates reddening corrected intensities. }
\end{deluxetable} 

\clearpage

\begin{deluxetable}{lccc}
\tablecaption{}
\tablewidth{0pt}
\tablehead{
\colhead{Region} & 
\colhead{L$_{40}$([\ion{O}{3}])} &
\colhead{L$_{40}$(H$\alpha$)} &
\colhead{log$_{10}$Q$_{rec}$} }
\startdata
NW cone                 &       5.01  & 1.9   &    53.2  \nl
SE cone                 &       5.63  & 3.7   &    53.5  \nl
NW cone (red corrected) &       14.4  & 4.2   &    53.5  \nl
SE cone (red corrected) &       76.9  & 29.9  &    54.4  \nl
Entire EELR             &       11.7  & 6.70  &    53.0  \nl
Entire EELR (red corrected) &   96.0  & 40.1  &    53.8  \nl
\enddata
\tablecomments{Luminosities of the NW, SE and inter-cone regions
               of NGC~2992 in units of $10^{40}$~ergs~s$^{-1}$, 
               and the inferred rate of hydrogen recombinations 
               $Q_{rec}$ (s$^{-1}$).}
\end{deluxetable}

\clearpage 

\figcaption[fig1a.ps,fig1b.ps,fig1c.ps,fig1d.ps,fig1e.ps,fig1f.ps]{ Optical 
and radio images of NGC~2992: a) [\ion{O}{3}]~$\lambda$5007 narrow
band image, b) H$\alpha$+[\ion{N}{2}] narrow band image, c) blue
continuum image, d) blue continuum image overlaid with the 20cm radio map
from \cite{hum83} (1983) 
(contours: -2, 2, 4, 7, 10, 20, 30, 50, 70, 90, 110 mJy per beam; 
beam=6\arcsec) e) [\ion{O}{3}] narrow band image overlaid with
20cm radio contours, f) [\ion{O}{3}] overlaid with the 6cm radio map
from \cite{ulv84} (1984) (contours: 3, 6, 9, 15, 30, 50, 70, 90\% of the 
peak of 7.2 mJy per beam; beam=0.5\arcsec). }

\figcaption[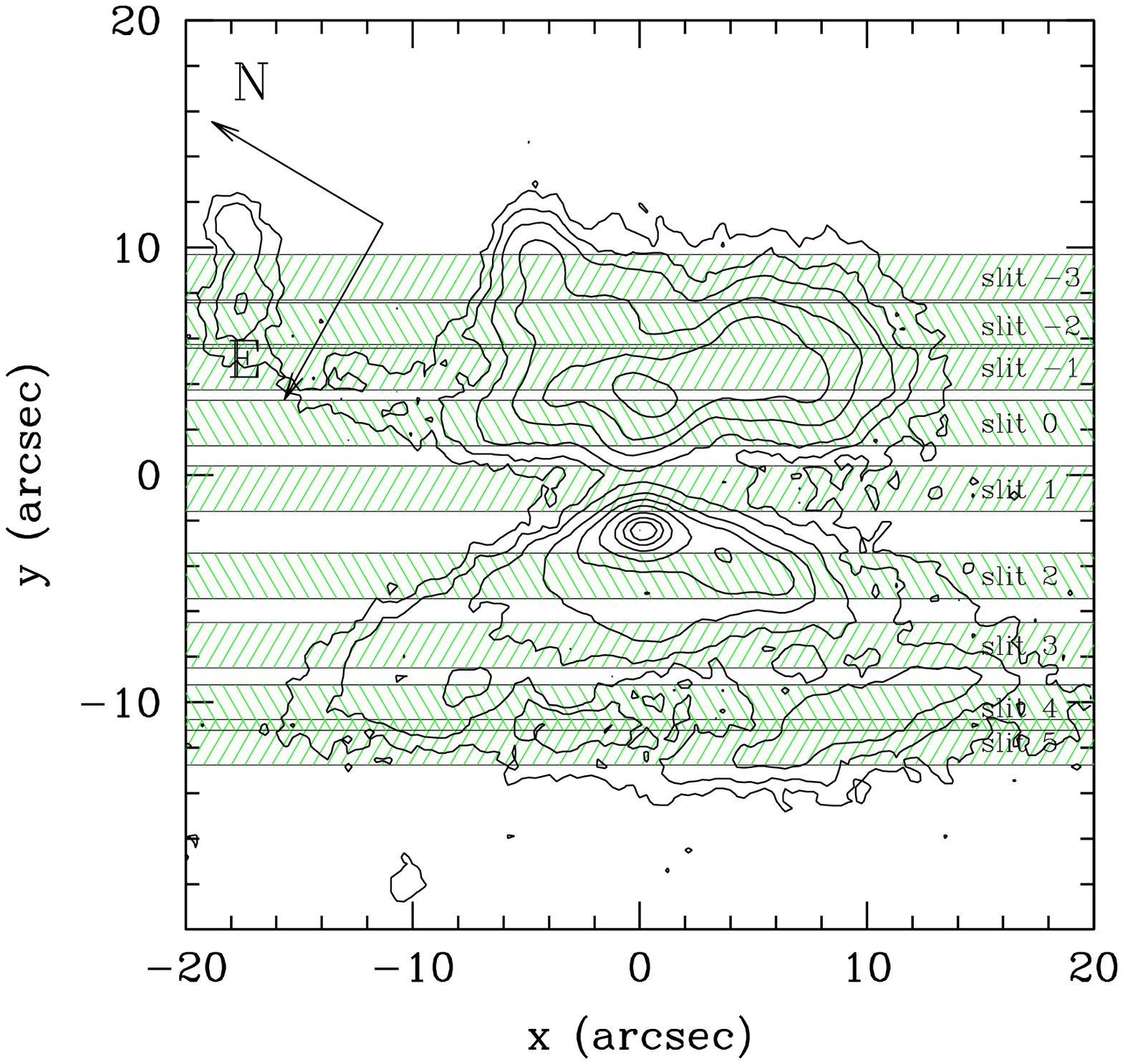]{Slit positions used for the long-slit observations
of NGC~2992 overlaid on the [\ion{O}{3}]~$\lambda$5007 contour map
with contour levels shown at  0.5, 1, 2, 5, 10, 20, 40, 60, 80, 100\%
of the peak value.  \label{fig2}}

\figcaption[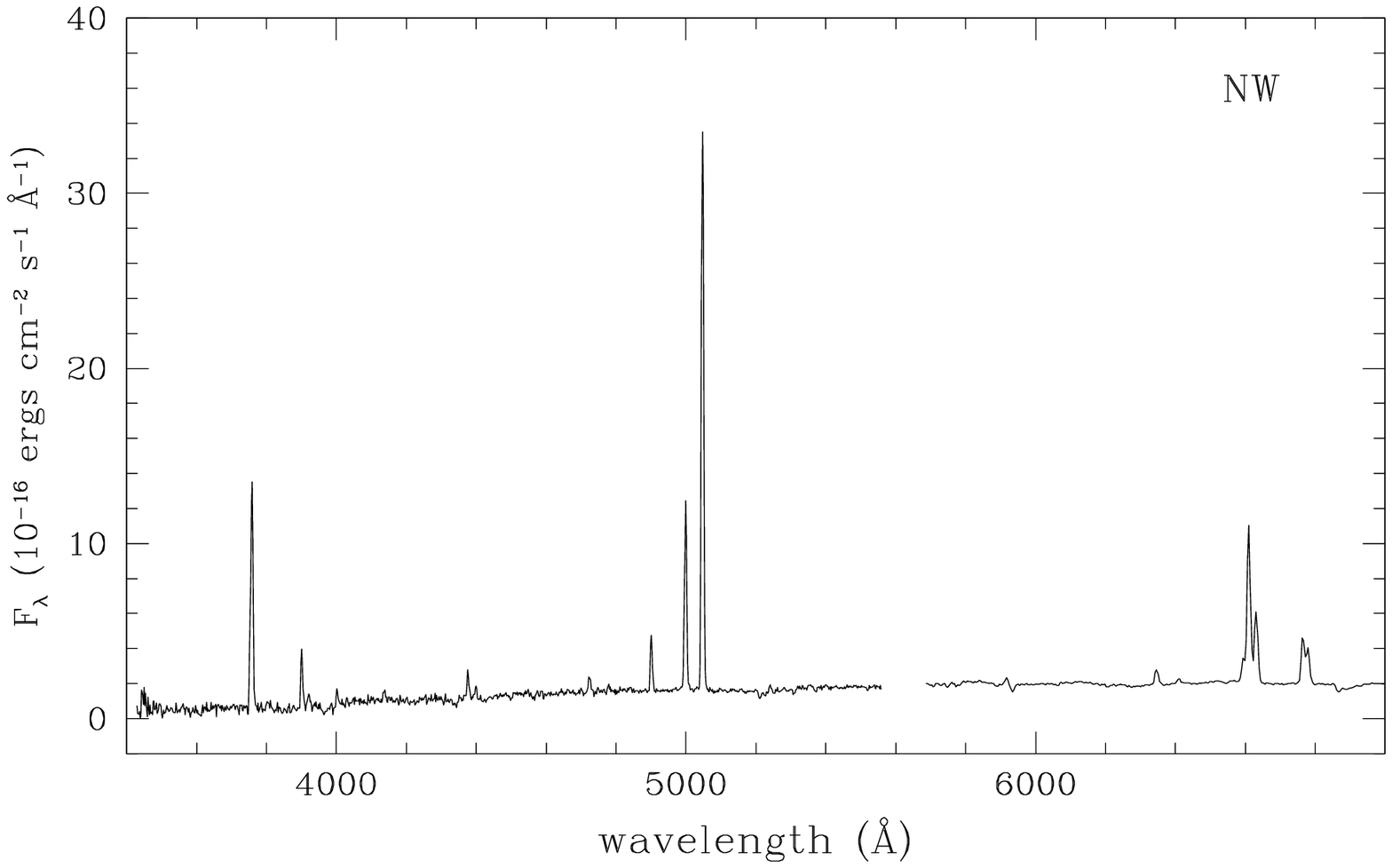,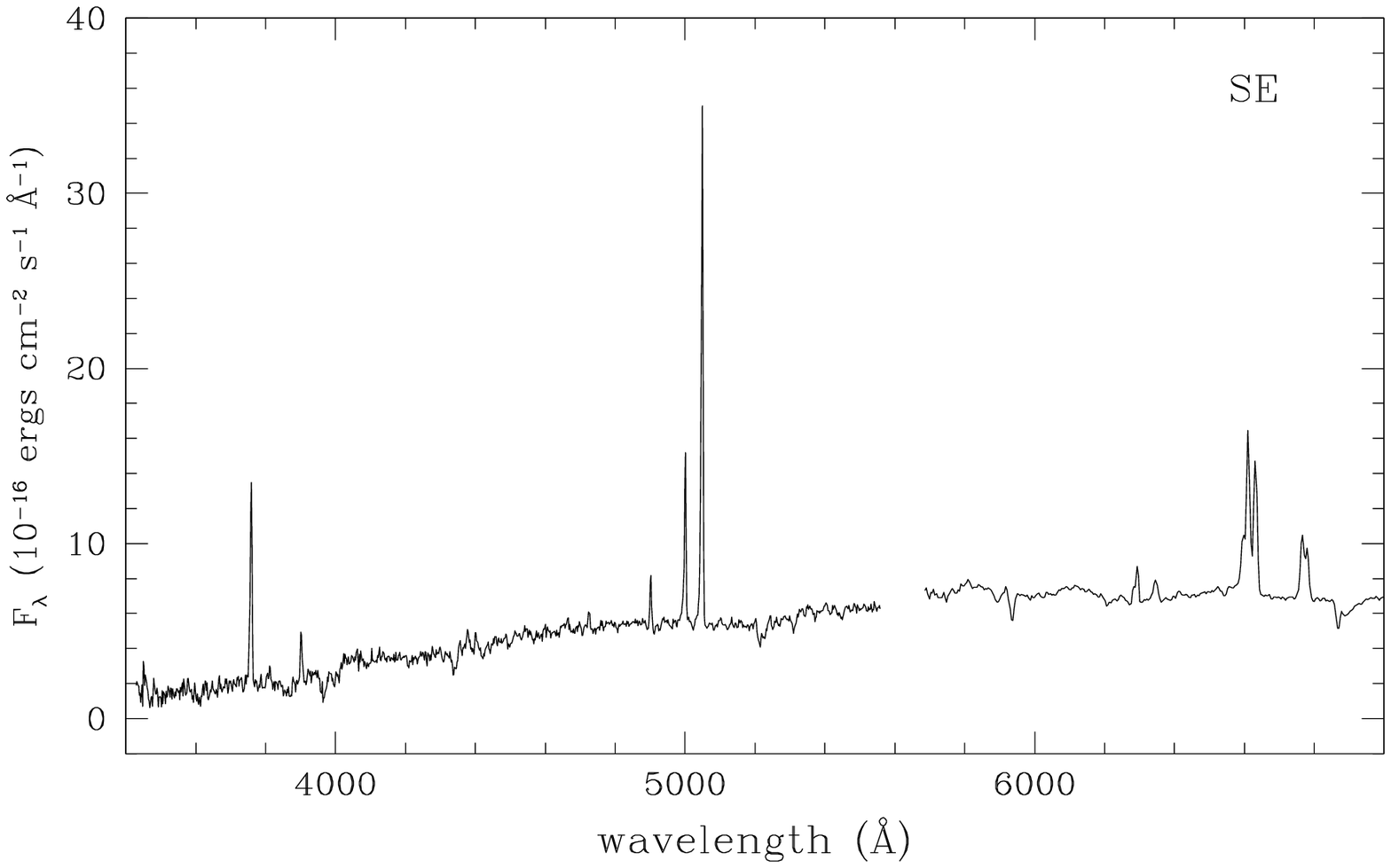,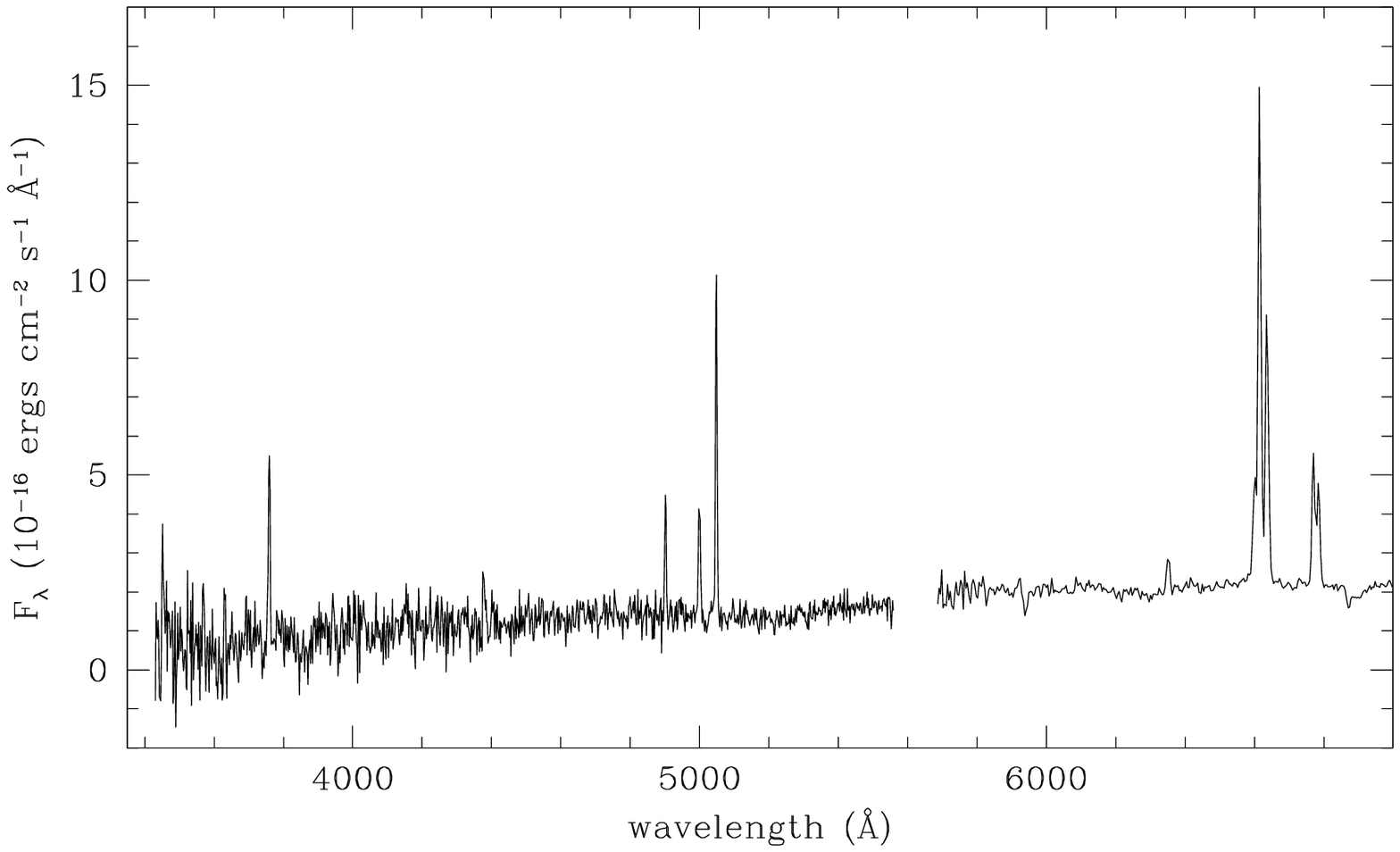]{ a) Average spectrum of the NW cone of 
NGC~2992, b) Average spectrum of the SE cone of NGC~2992, c) Spectrum of 
the inter-cone region of NGC~2992.  }

\figcaption[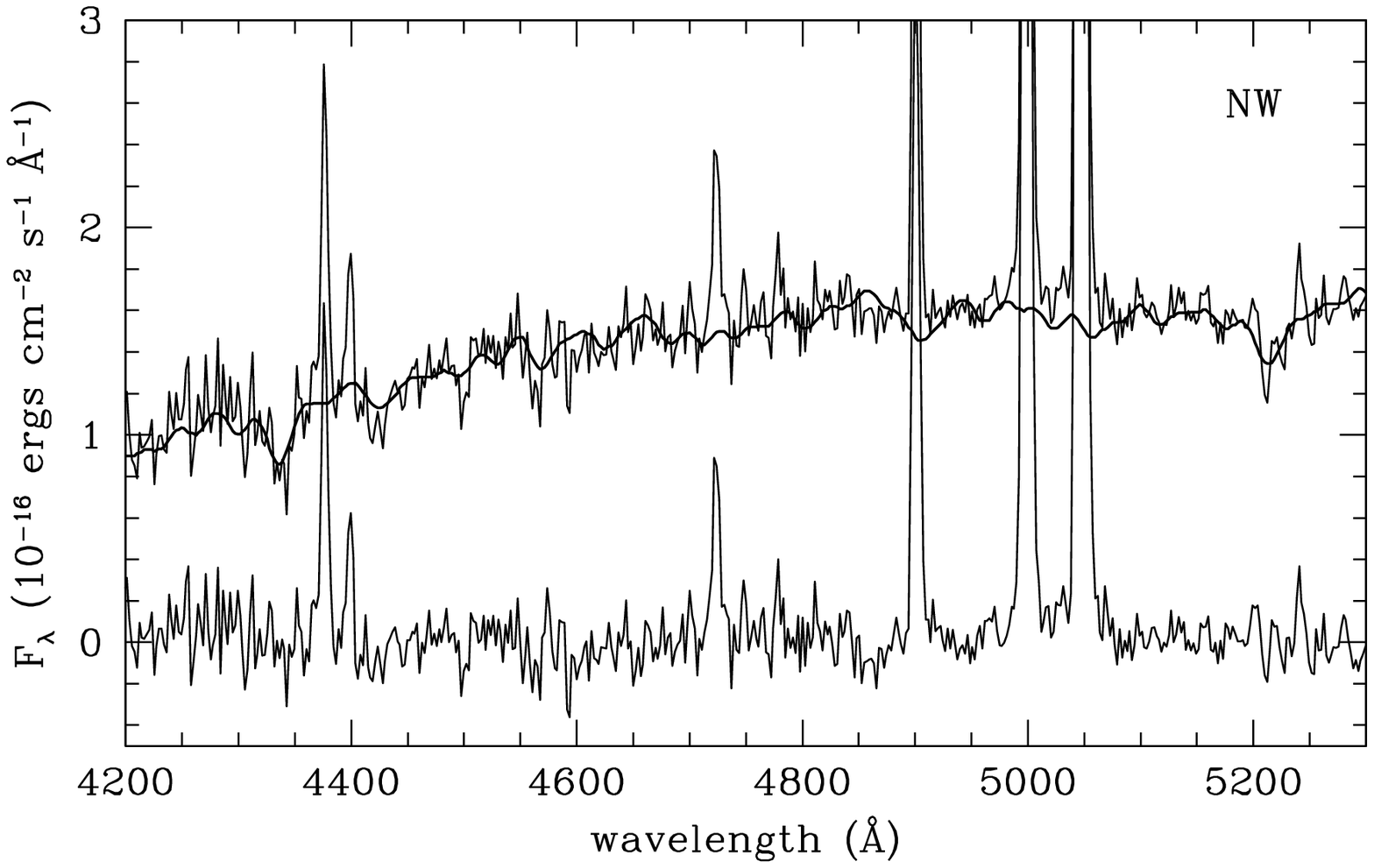,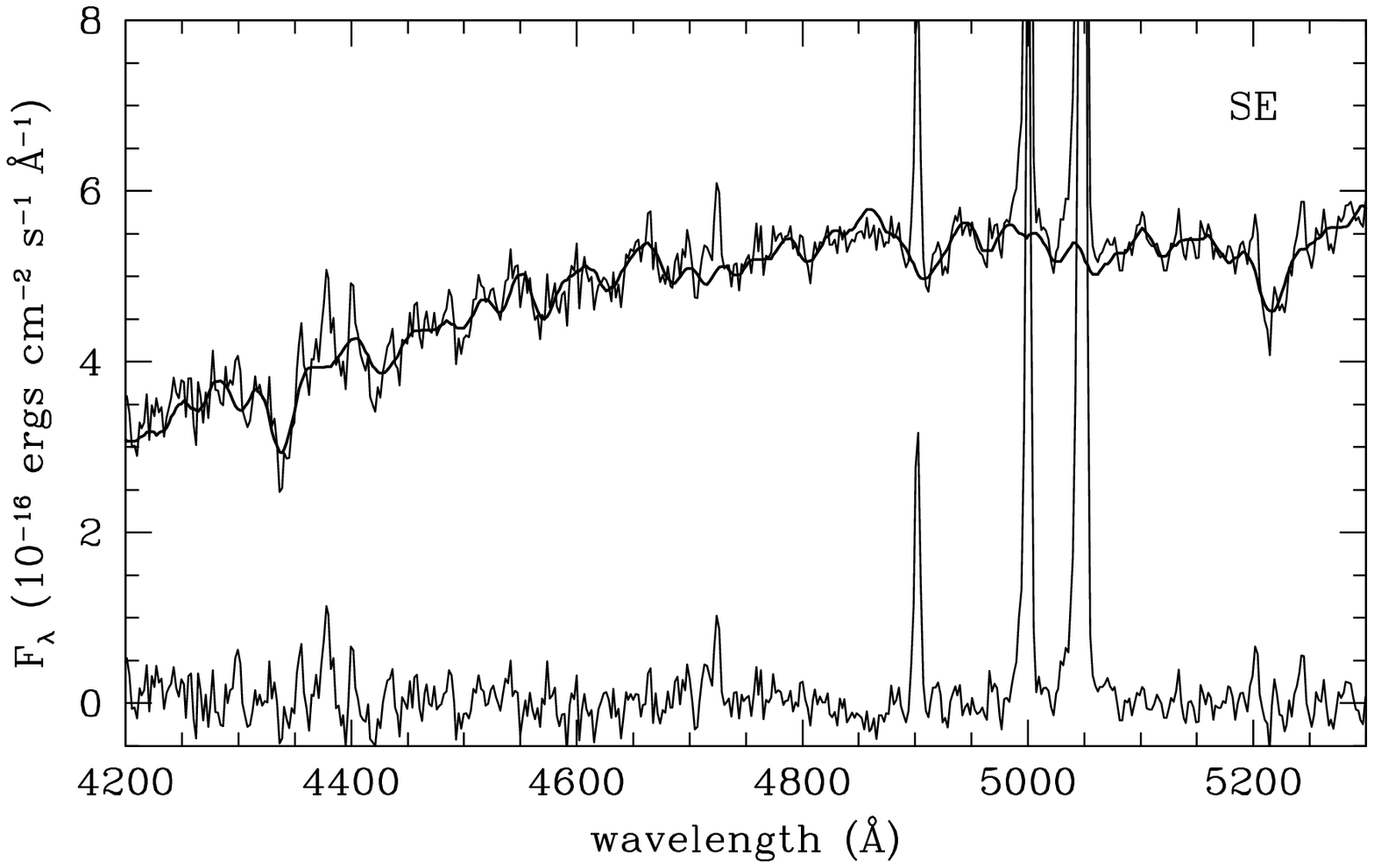]{ Blue spectra of the NW and SE cones of 
NGC~2992. 
The top curve shows the observed spectrum overlaid with the
template used to subtract the underlying stellar contribution.
The lower curve shows the template subtracted spectrum.}

\figcaption[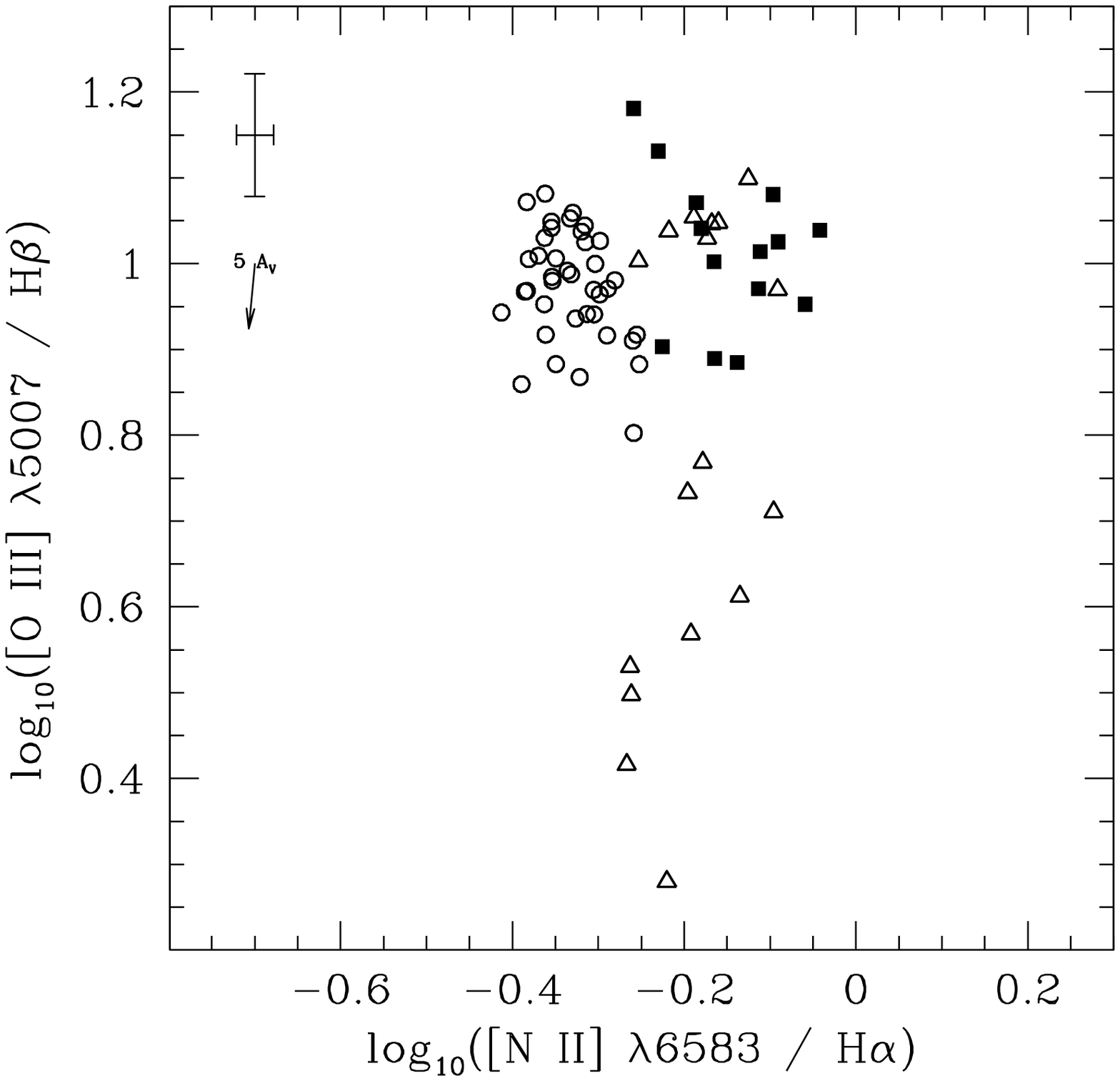,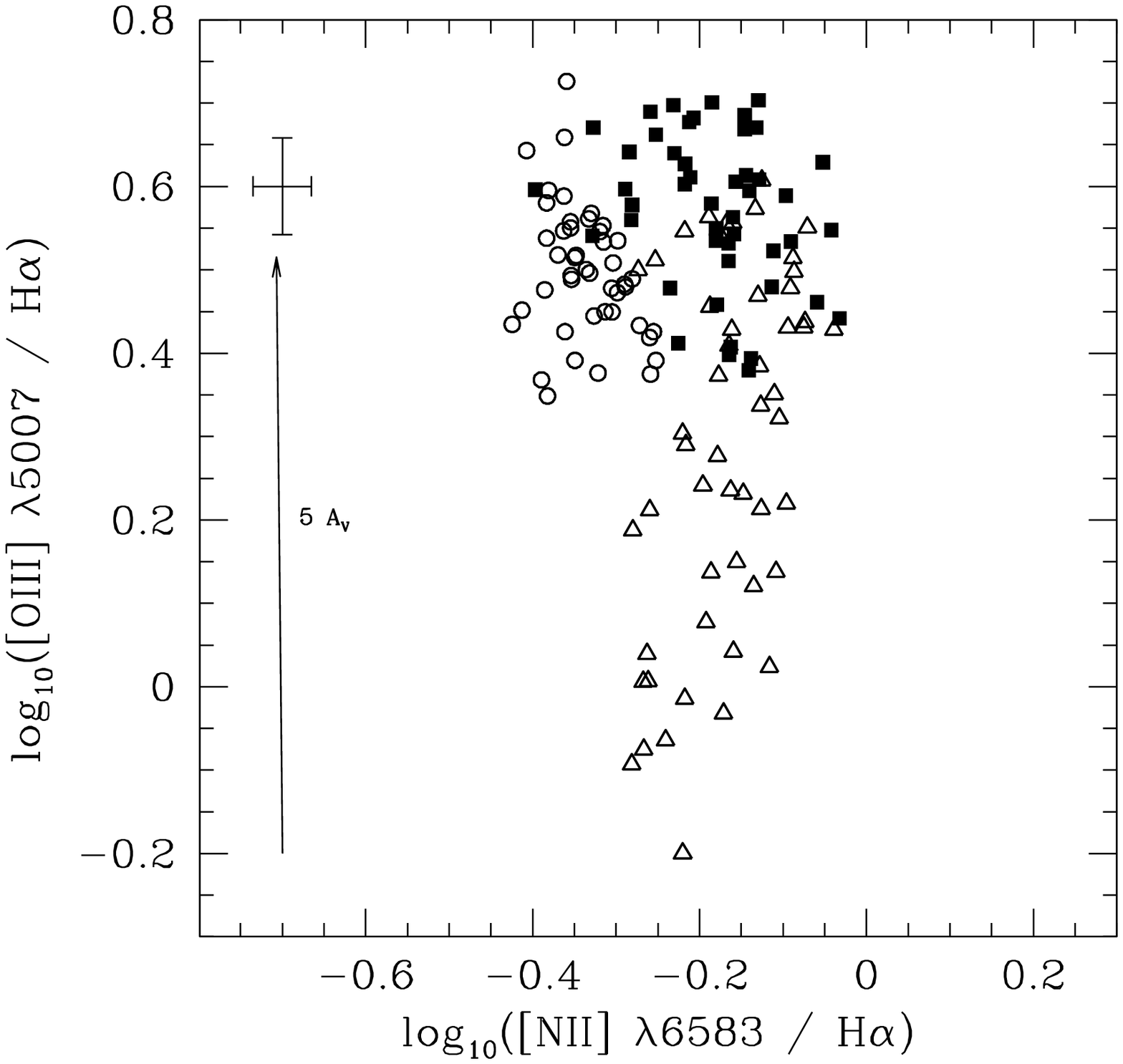]{Spatial variation of line ratios in NGC~2992.
The open circles indicate line ratios from positions in the NW cone, 
solid squares indicate points in the SE cone, and open 
triangles indicate the inter-cone region as shown in Figure~6.}

\figcaption[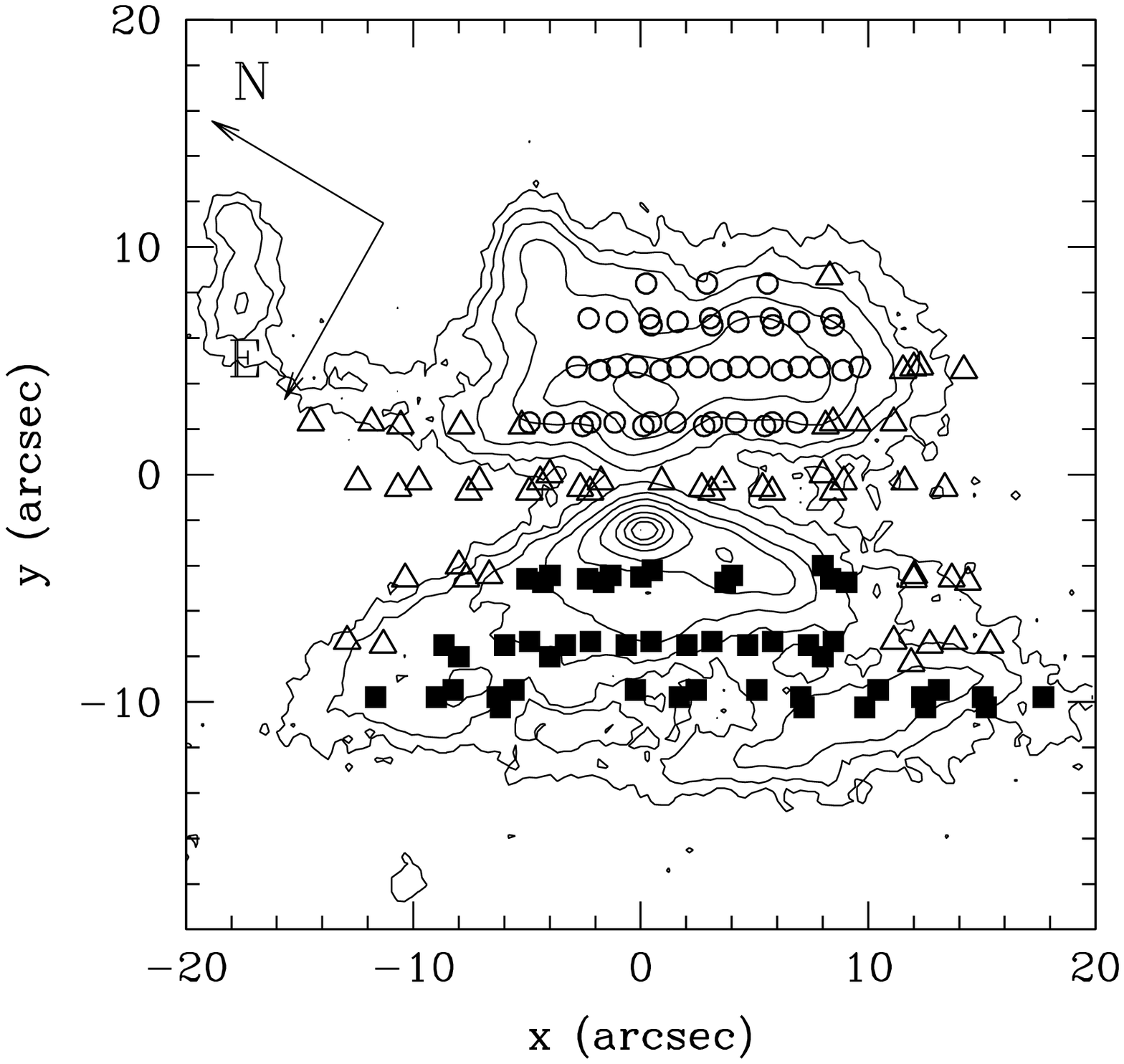]{Spatial locations of the extracted spectra 
overplotted on the 
[\ion{O}{3}]~$\lambda$5007 contour map. The contour levels are
as in Figure 2. The open circles indicate positions in the NW cone, 
solid squares indicate points in the SE cone, and open 
triangles indicate the inter-cone region.}

\figcaption[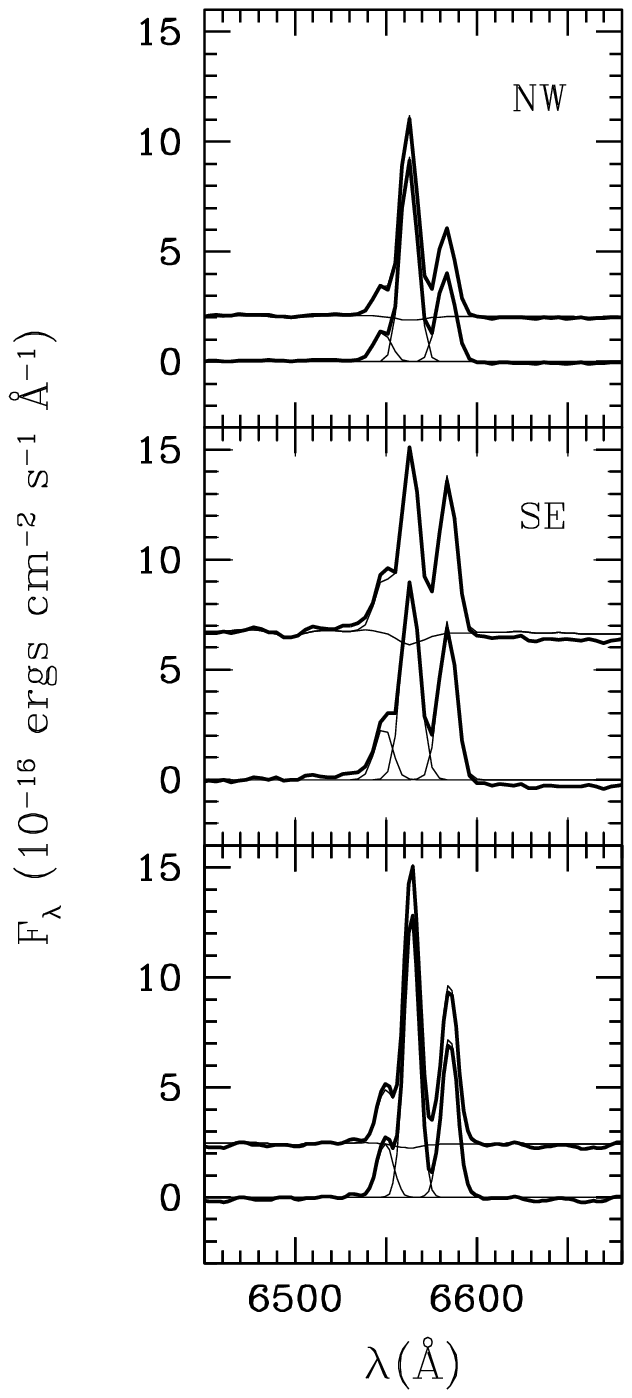]{H$\alpha$ + [\ion{N}{2}] $\lambda$6548,6583
profiles in the spectra of the NW cone (top), SE cone (center),
and the inter-cone regions (lower). In each panel we show the total 
fit overlaid on the data template subtracted data. The lighter
weight curves show the fitted galaxy template and the
individual Gaussian component fits to the H$\alpha$ and [\ion{N}{2}]
emission lines.}

\figcaption[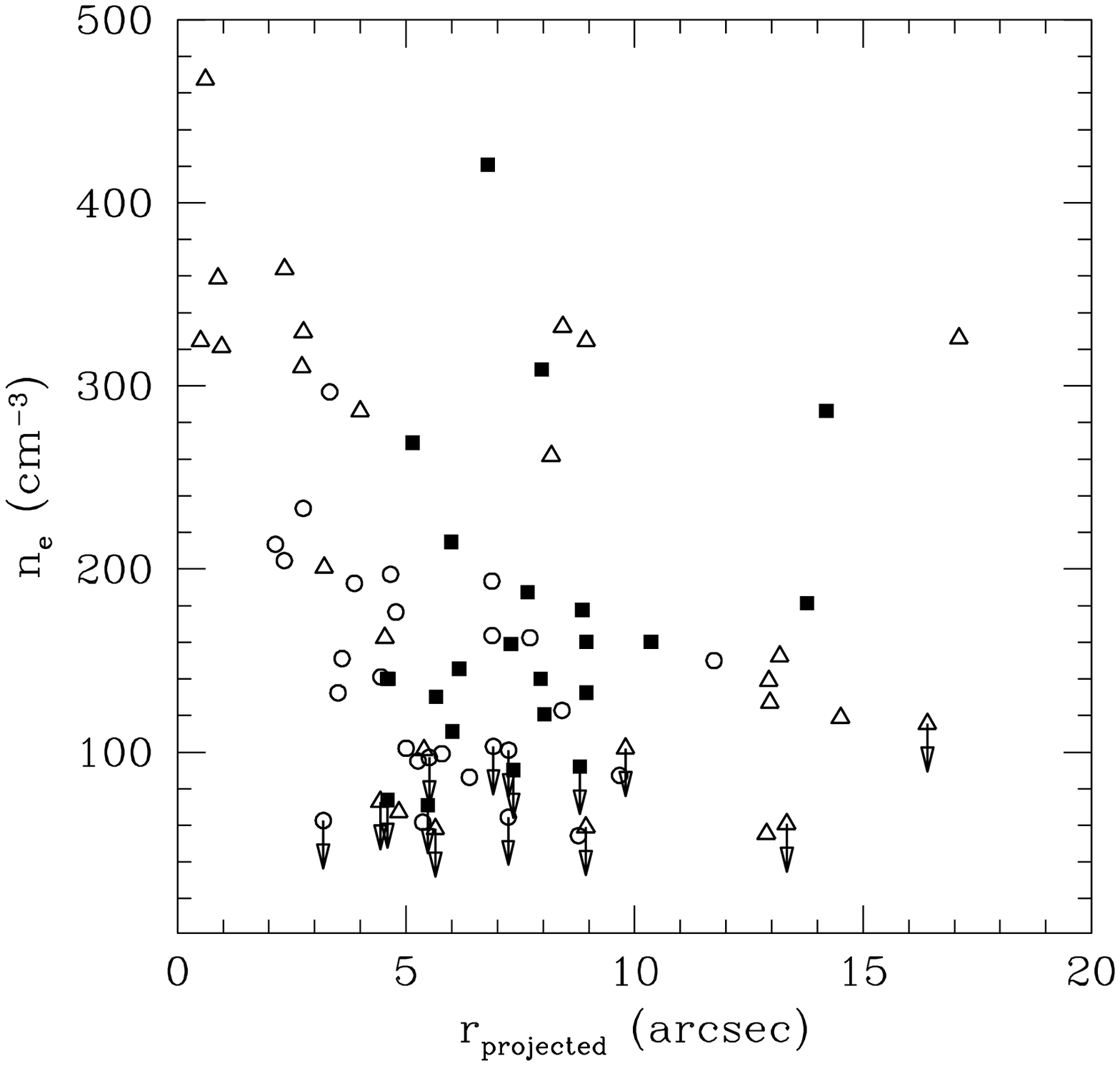]{Electron densities $n_e$ inferred from the
\Siir\ ratio in spectra extracted from various positions in the 
ENLR of NGC~2992. The plotting symbols represent the different 
regions of the ENLR as shown in Figure~6.}

\figcaption[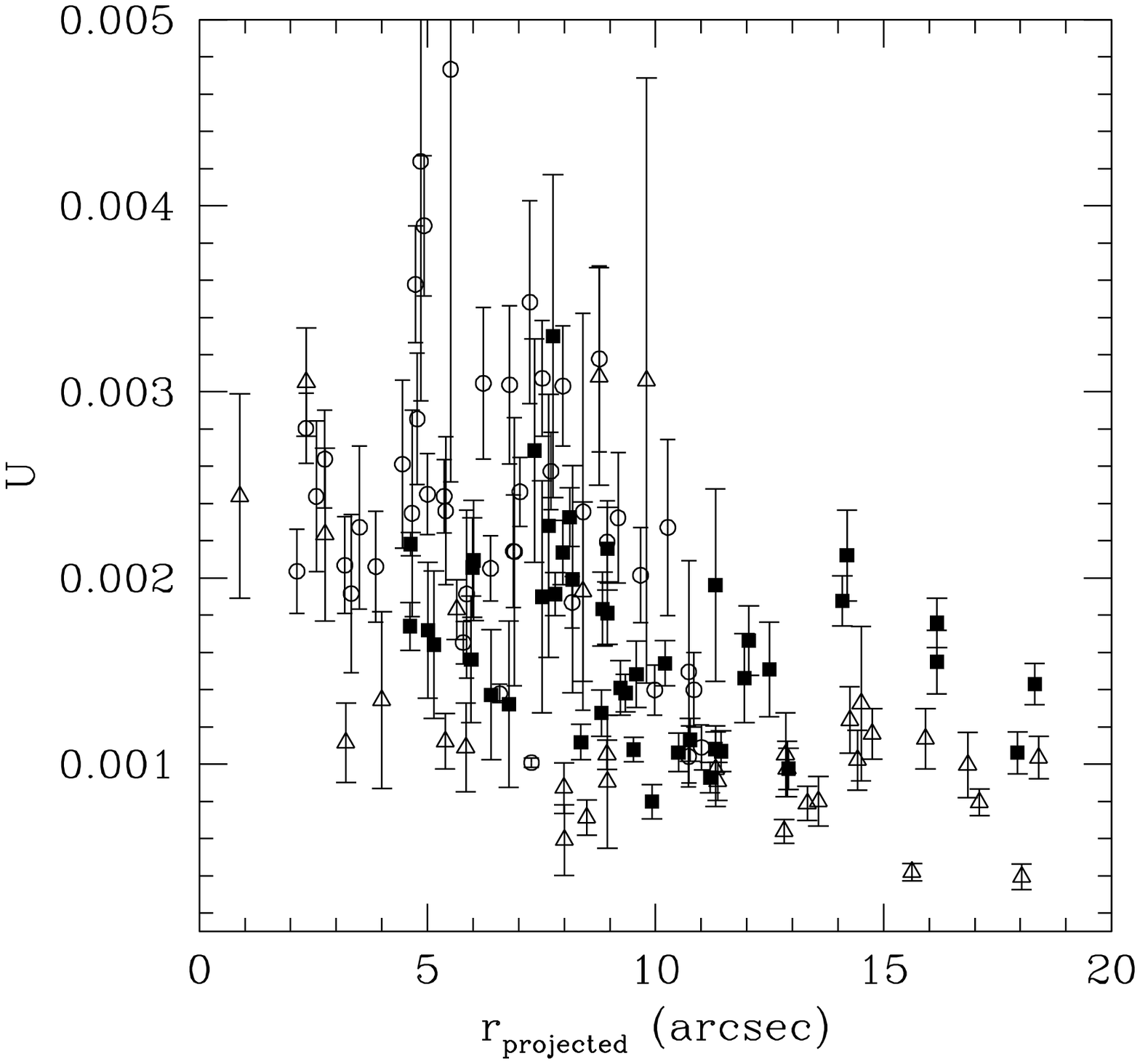]{Ionization parameter estimated from the 
[\ion{O}{3}]/[\ion{O}{2}] line ratio at each point in the spectral
map of NGC~2992. The plotting symbols represent the different 
regions of the ENLR as shown in Figure~6.}

\figcaption[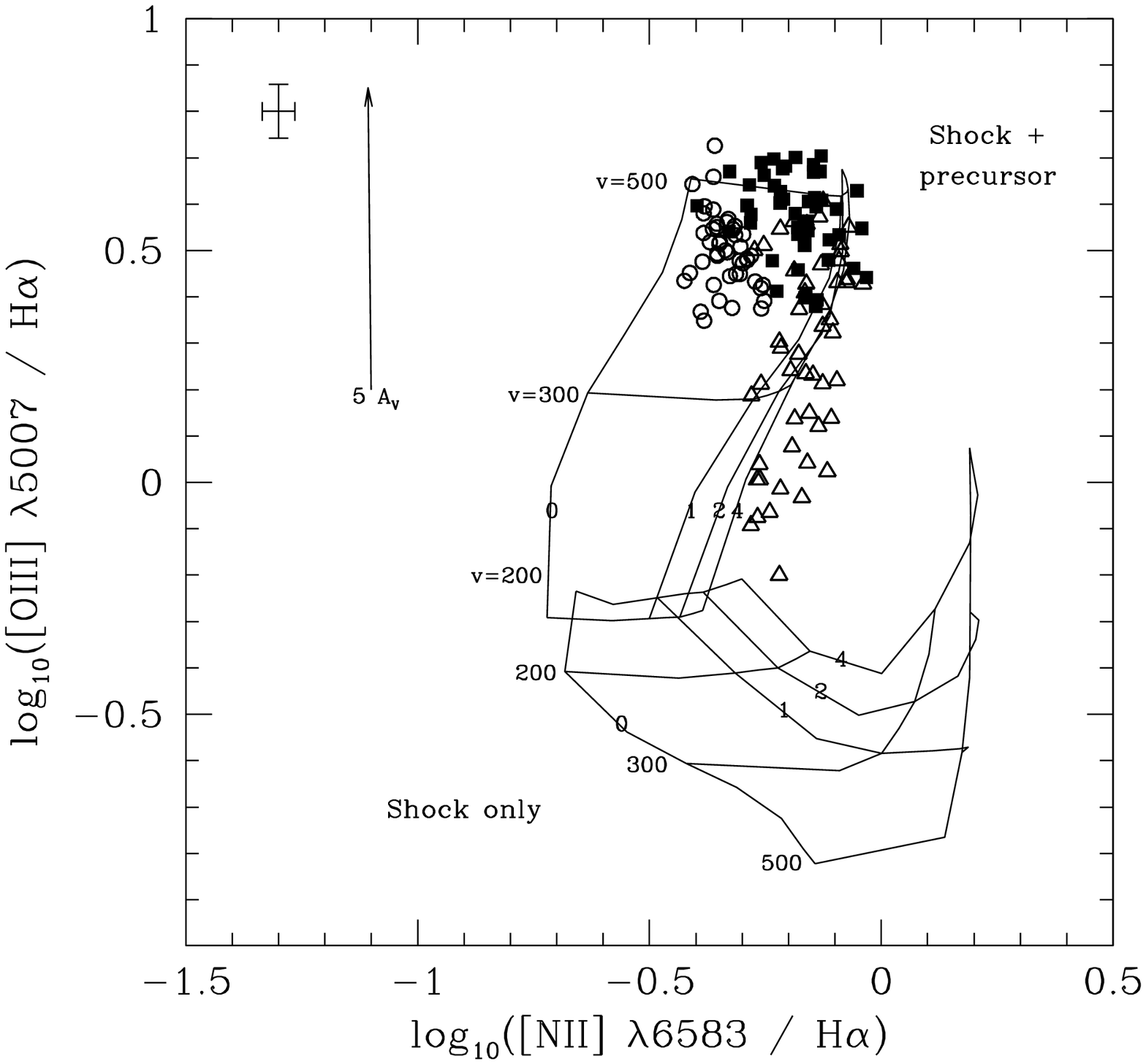,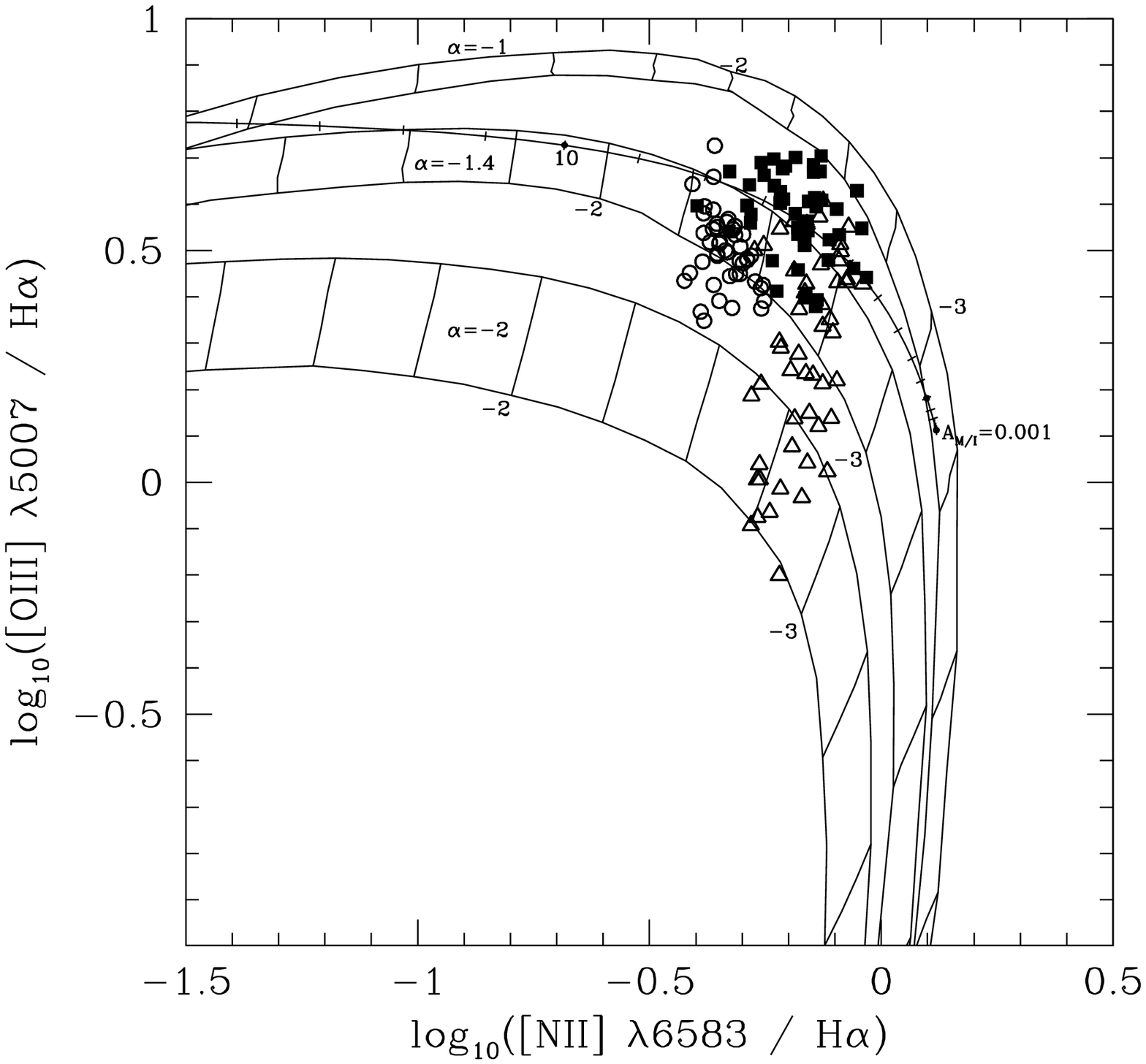]{[\ion{O}{3}]~$\lambda$5007/H$\alpha$ versus
[\ion{N}{2}]~$\lambda$6583/H$\alpha$. The observed line ratios are
shown for each point in the spectral map of the ENLR of NGC~2992.  The
plotting symbols represent the different regions of the ENLR as shown
in Figure~6; open circles denote points from the NW cone, solid
squares denote points from the SE cone, and open triangles indicate
the inter-cone region.  In panel (a) we over-plot the shock, and
shock+precursor model grids and also include the average 1$\sigma$
error bar and a vector which indicates the direction of reddening
correction. The shock and shock+precursor model grids are labelled
with shock velocity in units of km~s$^{-1}$, and with the magnetic
parameter $B/n^{1/2}$ = 0$-$4 in units of $\mu$G~cm$^{3/2}$. Panel (b)
shows the single component photoionization ($\alpha$=-1.0,-1.4 and -2.0)
$U$-sequences plotted for $n=100$~cm$^{-3}$ and
$n=1000$~cm$^{-3}$ with grid lines of constant $U$ every 0.25~dex, and
labelled with log$_{10}U$.  The $A_{M/I}$ sequence is also shown and
is labelled with $A_{M/I}$ and has tick marks at 0.2~dex intervals.}

\figcaption[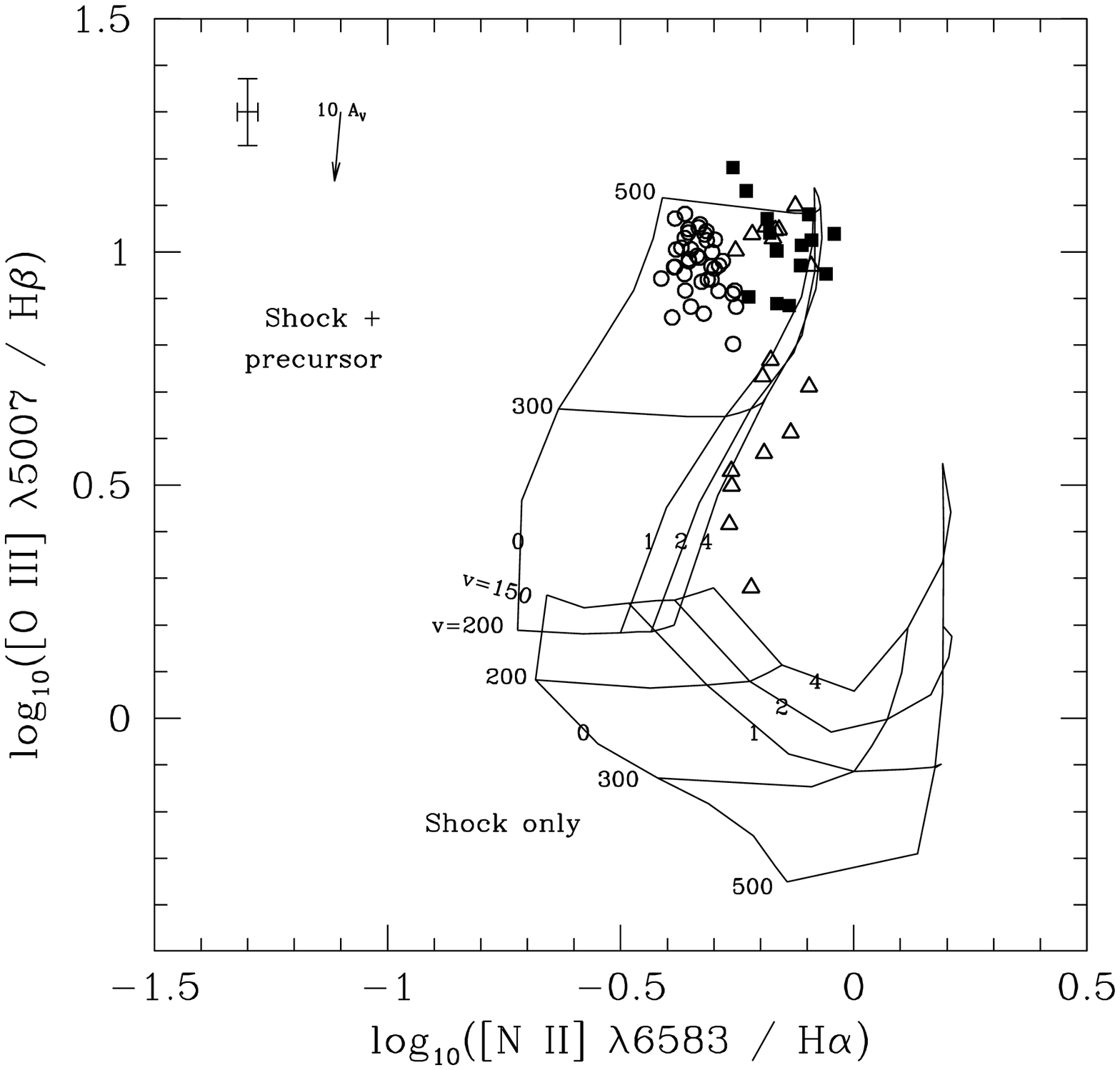,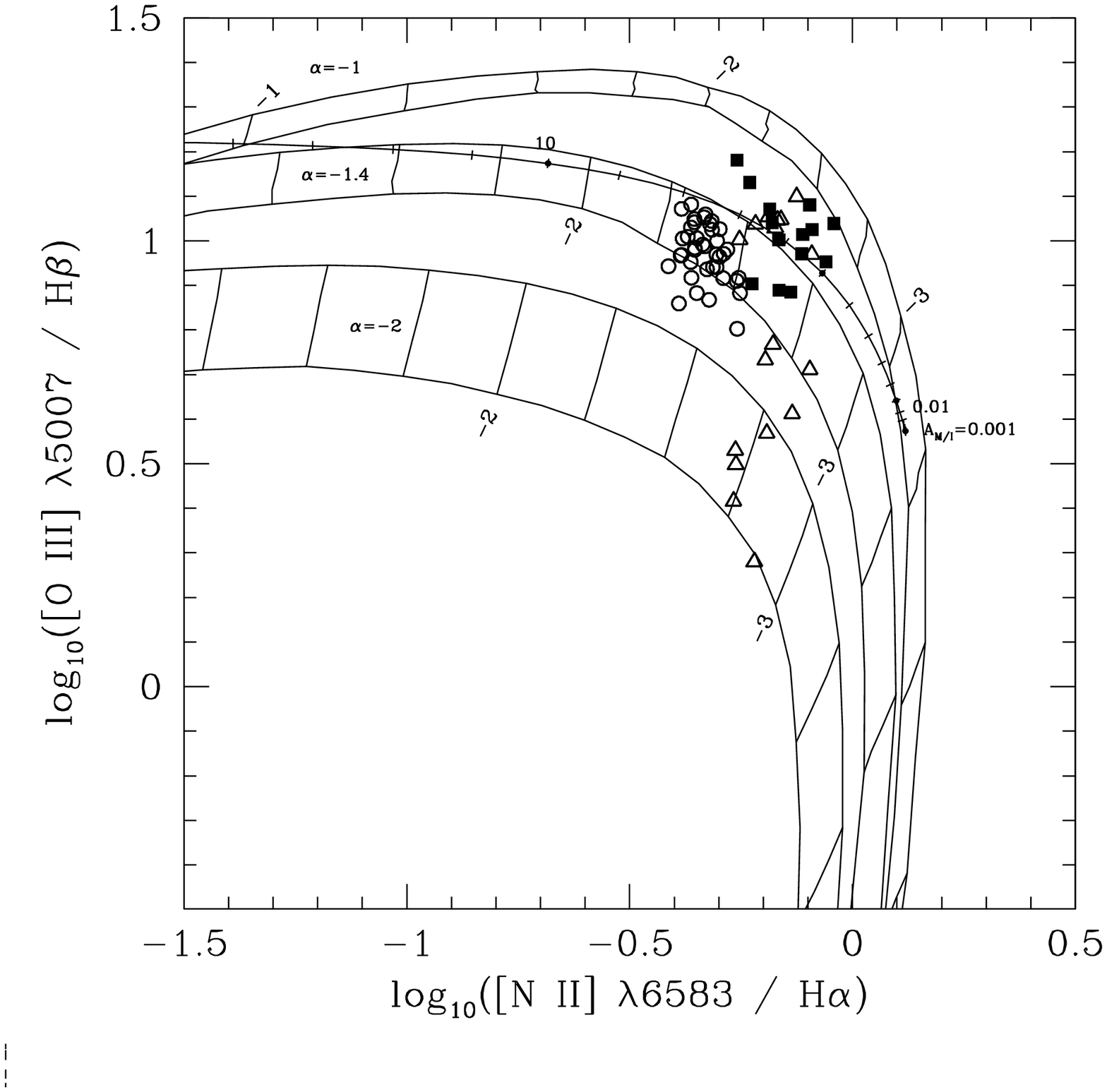]{[\ion{O}{3}]~$\lambda$5007/H$\alpha$ versus
[\ion{N}{2}]~$\lambda$6583/H$\alpha$. The data points and model grids 
are as described for Figure~10. }

\figcaption[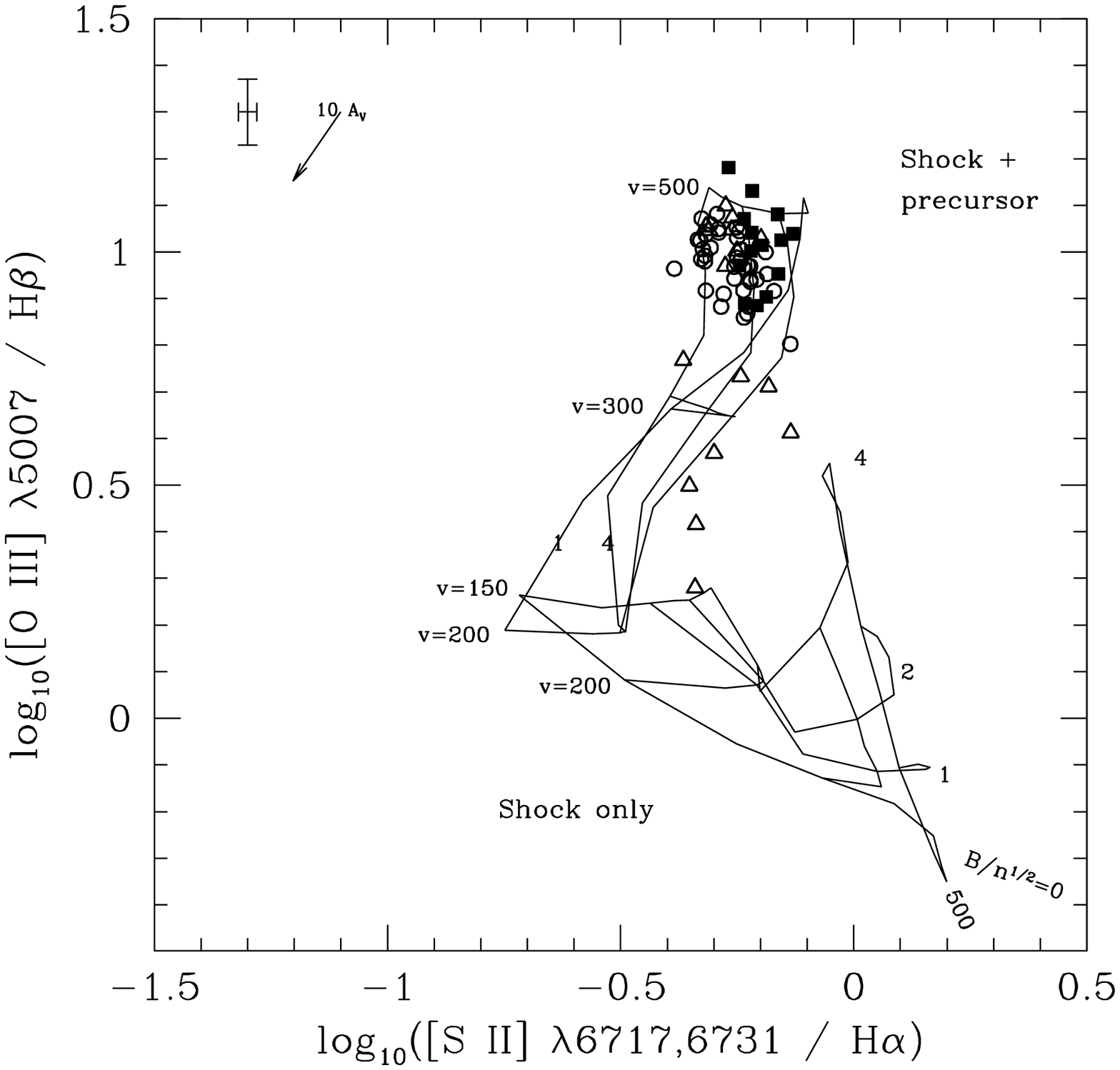,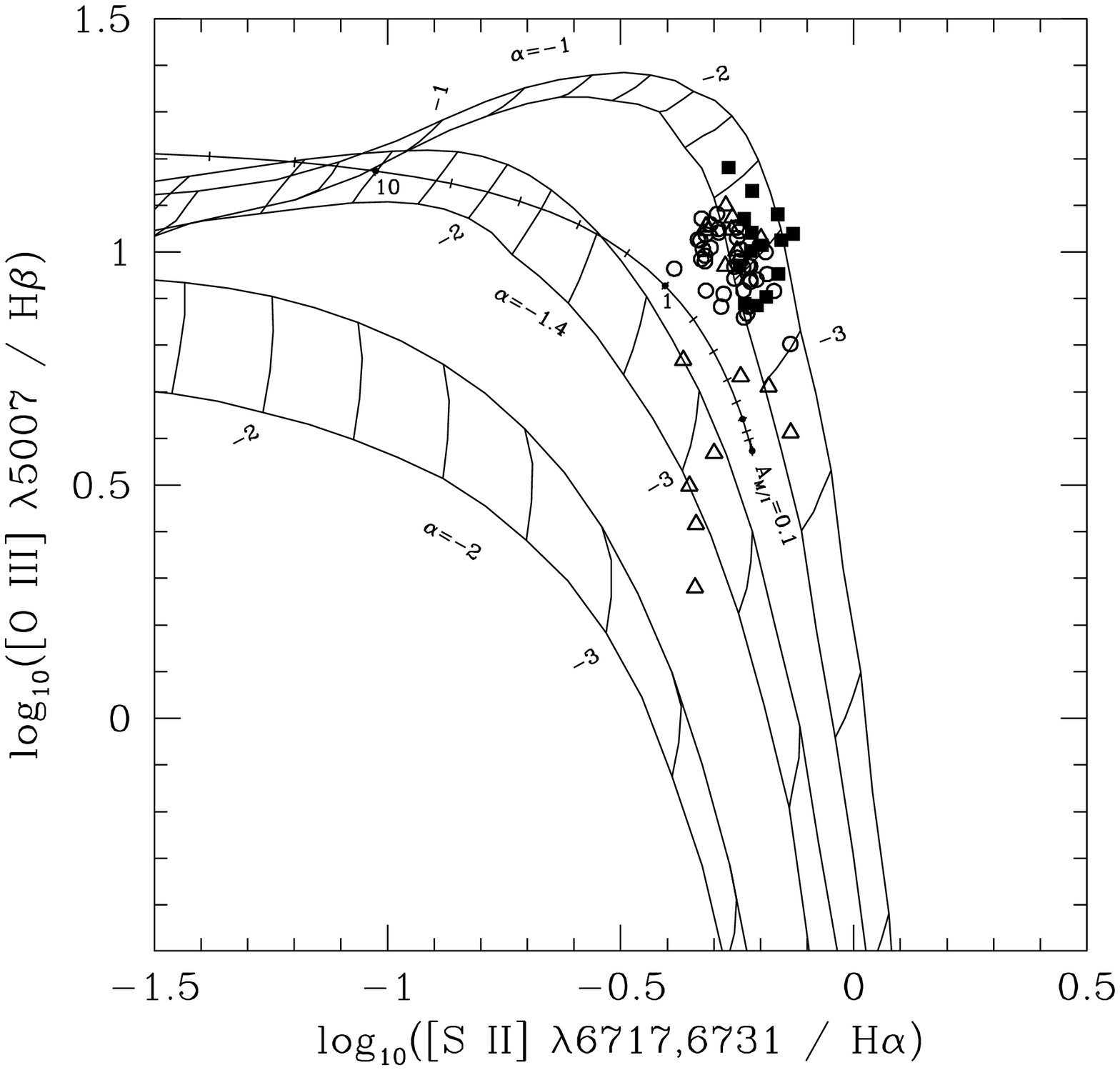]{[\ion{O}{3}]~$\lambda$5007/H$\beta$ versus
[\ion{S}{2}]~$\lambda\lambda$6717,6731/H$\alpha$. The data points and 
model grids are as described for Figure~10.}

\figcaption[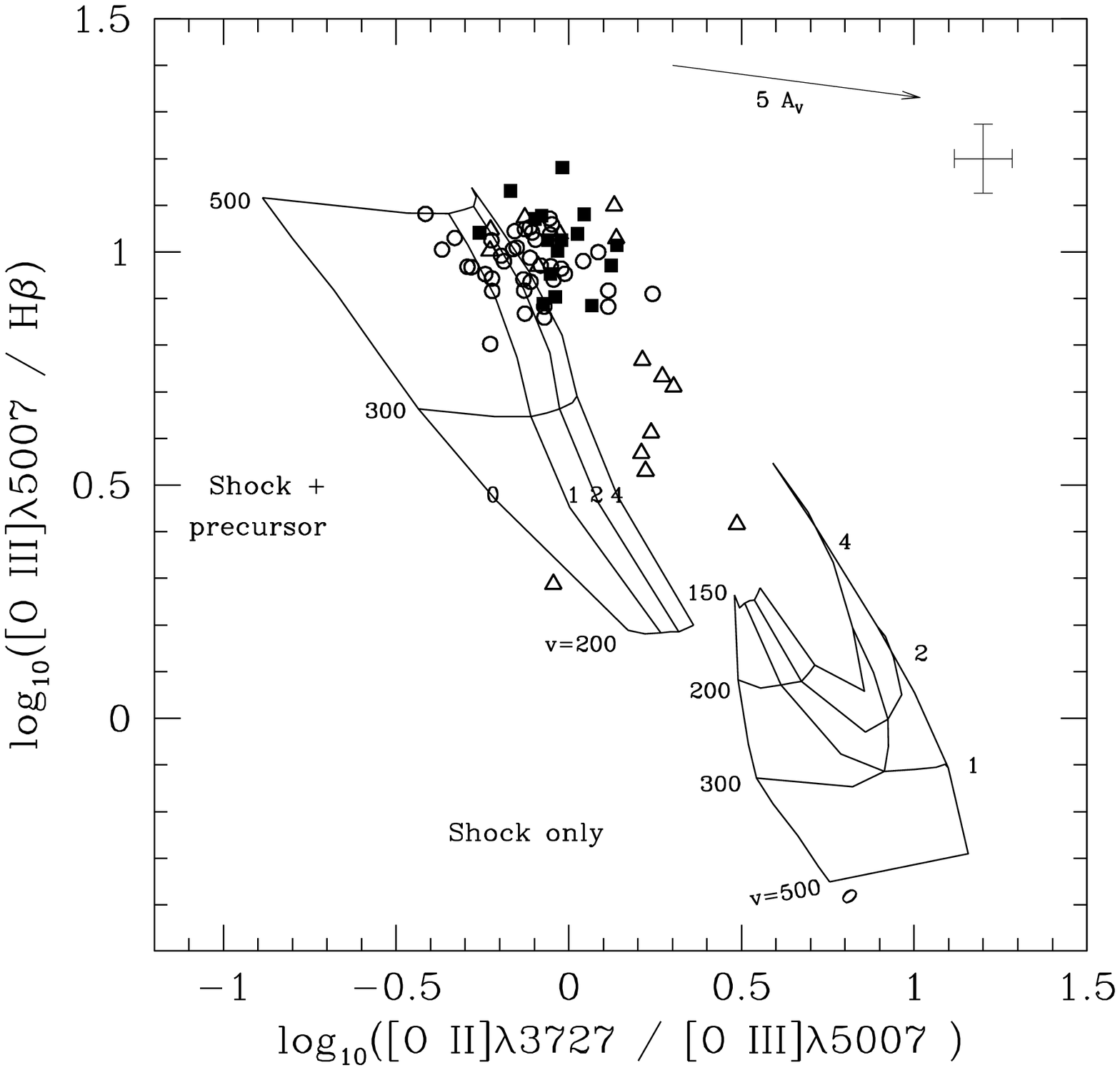,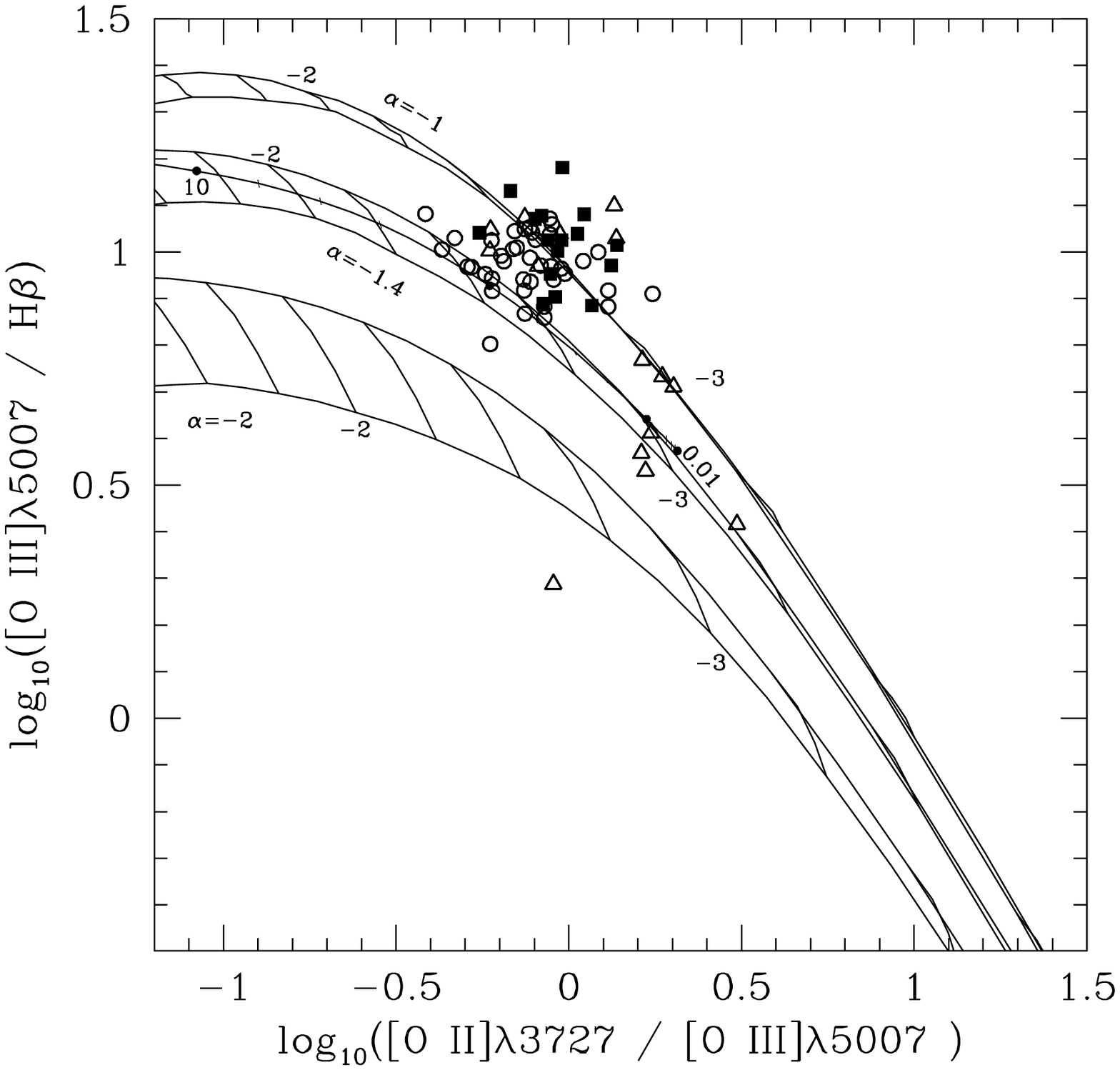]{[\ion{O}{3}]~$\lambda$5007/H$\beta$ versus
[\ion{O}{2}]~$\lambda$3727/[\ion{O}{3}]~$\lambda$5007. The data points and 
model grids are as described for Figure~10. }

\figcaption[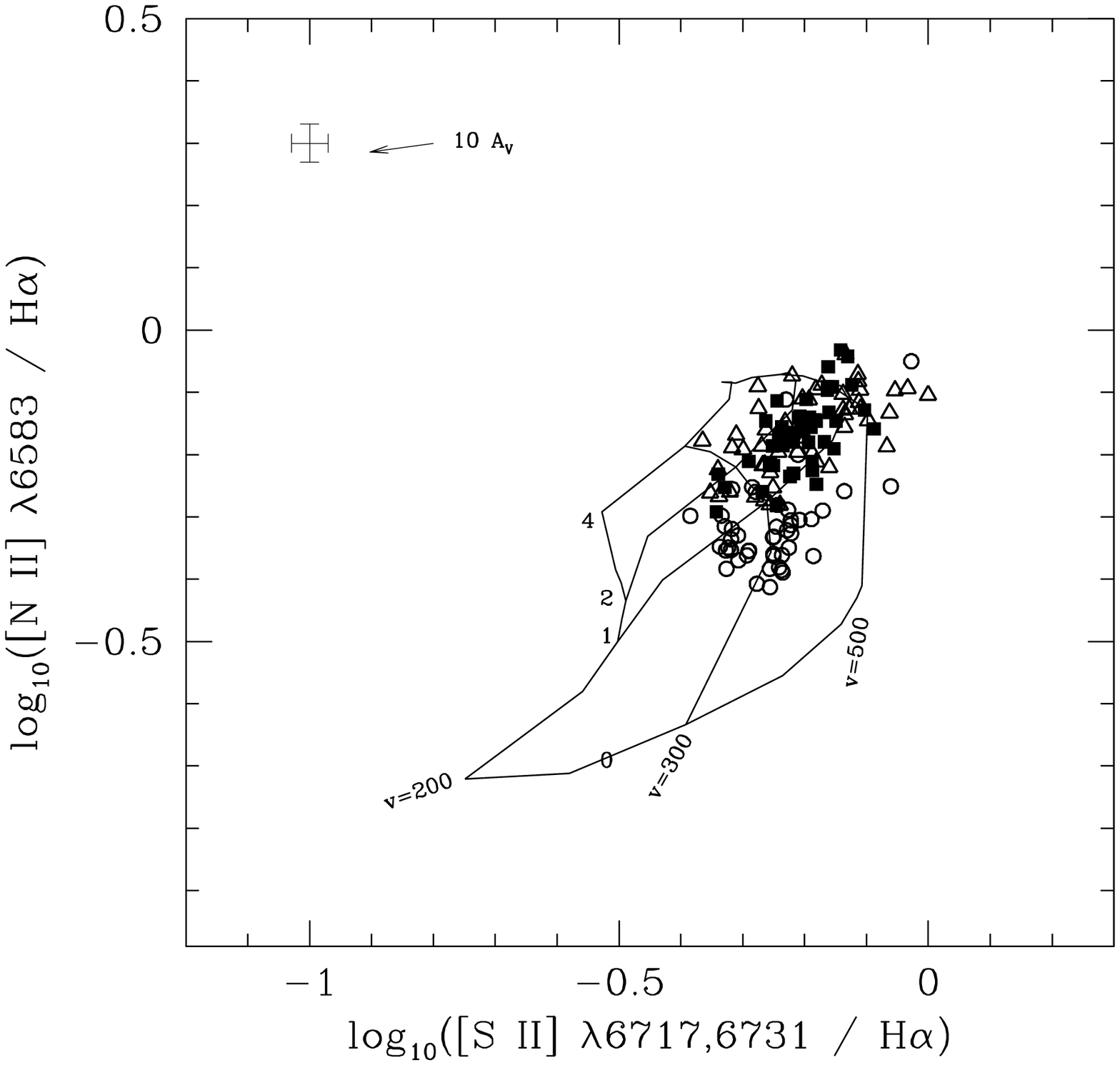,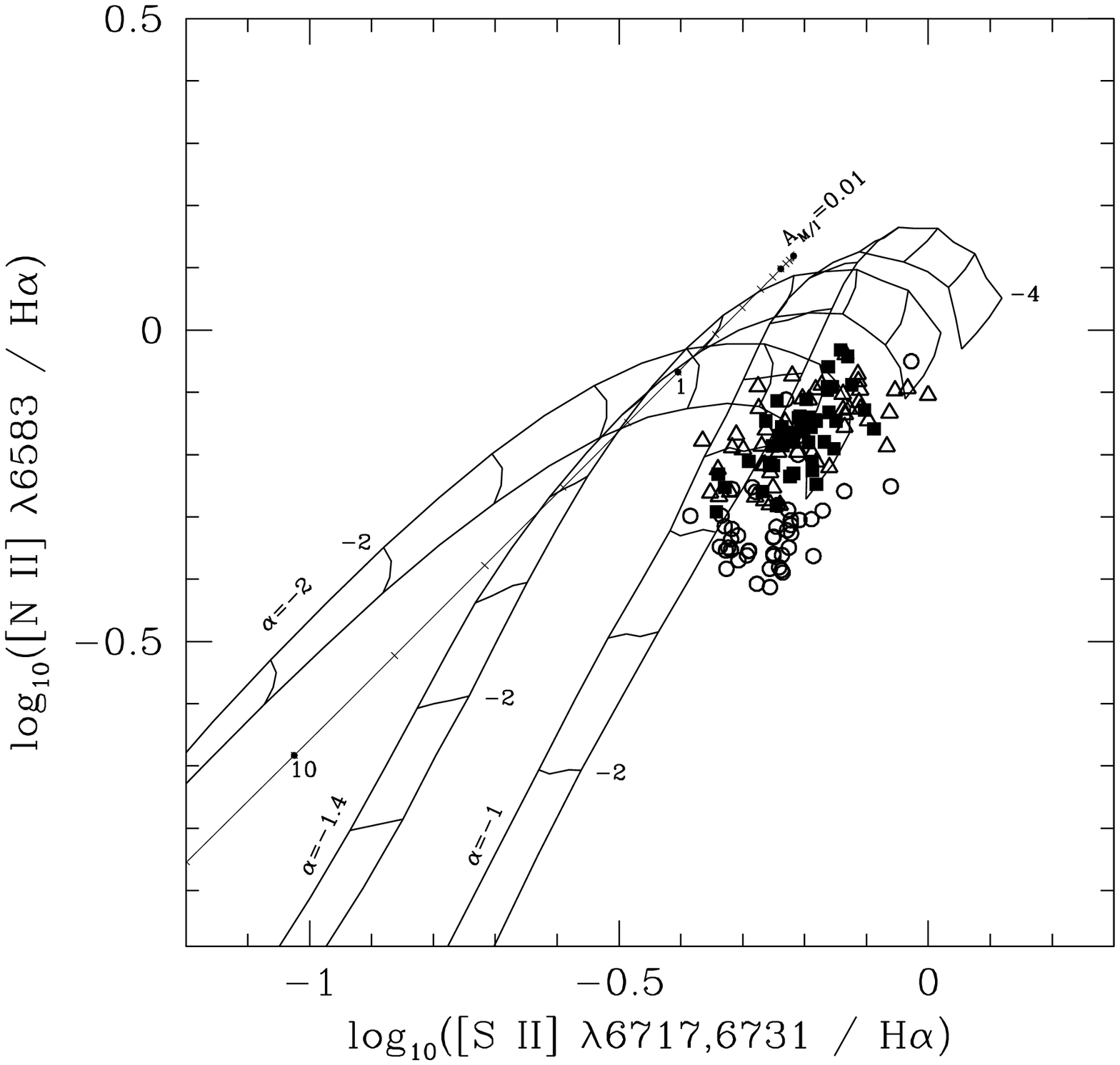]{[\ion{N}{2}]~$\lambda$6583/H$\alpha$ versus
[\ion{S}{2}]~$\lambda\lambda$6717,6731/H$\alpha$. The data points and 
model grids are as described for Figure~10. }

\figcaption[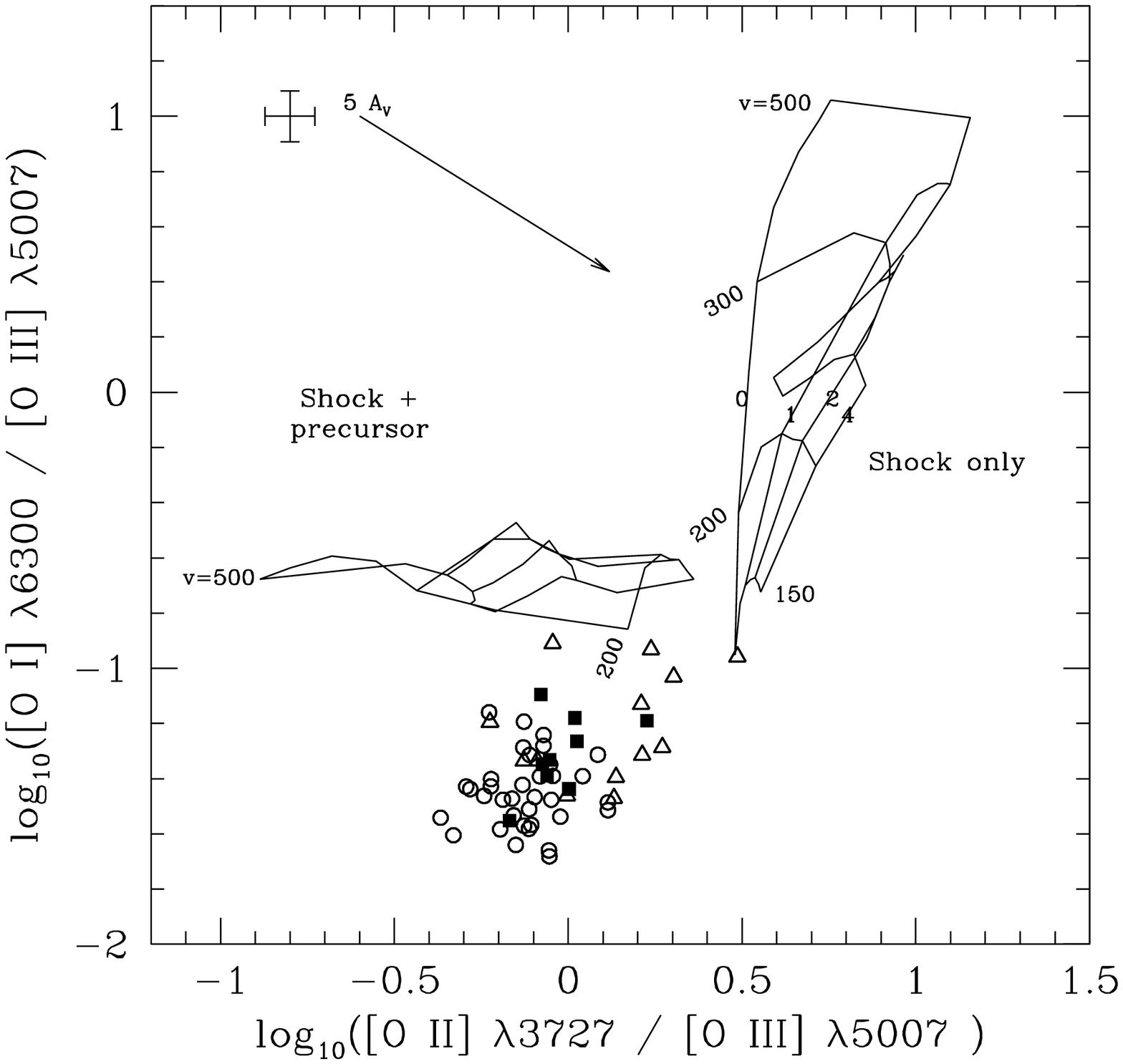,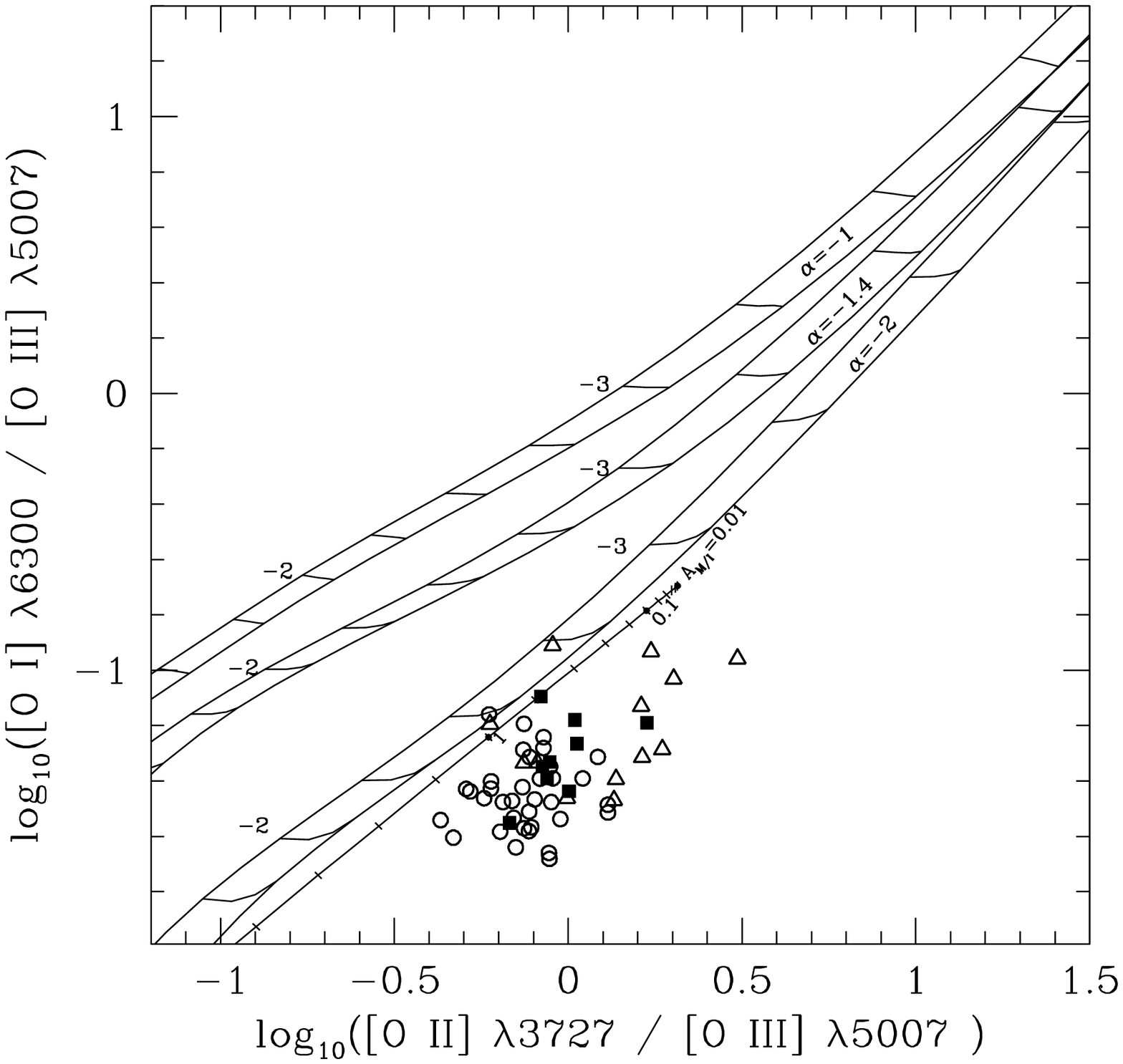]{
[\ion{O}{1}]~$\lambda$6300/[\ion{O}{3}]~$\lambda$5007 versus
[\ion{O}{2}]~$\lambda$3727/[\ion{O}{3}]~$\lambda$5007. The data points and 
model grids are as described for Figure~10. }

\figcaption[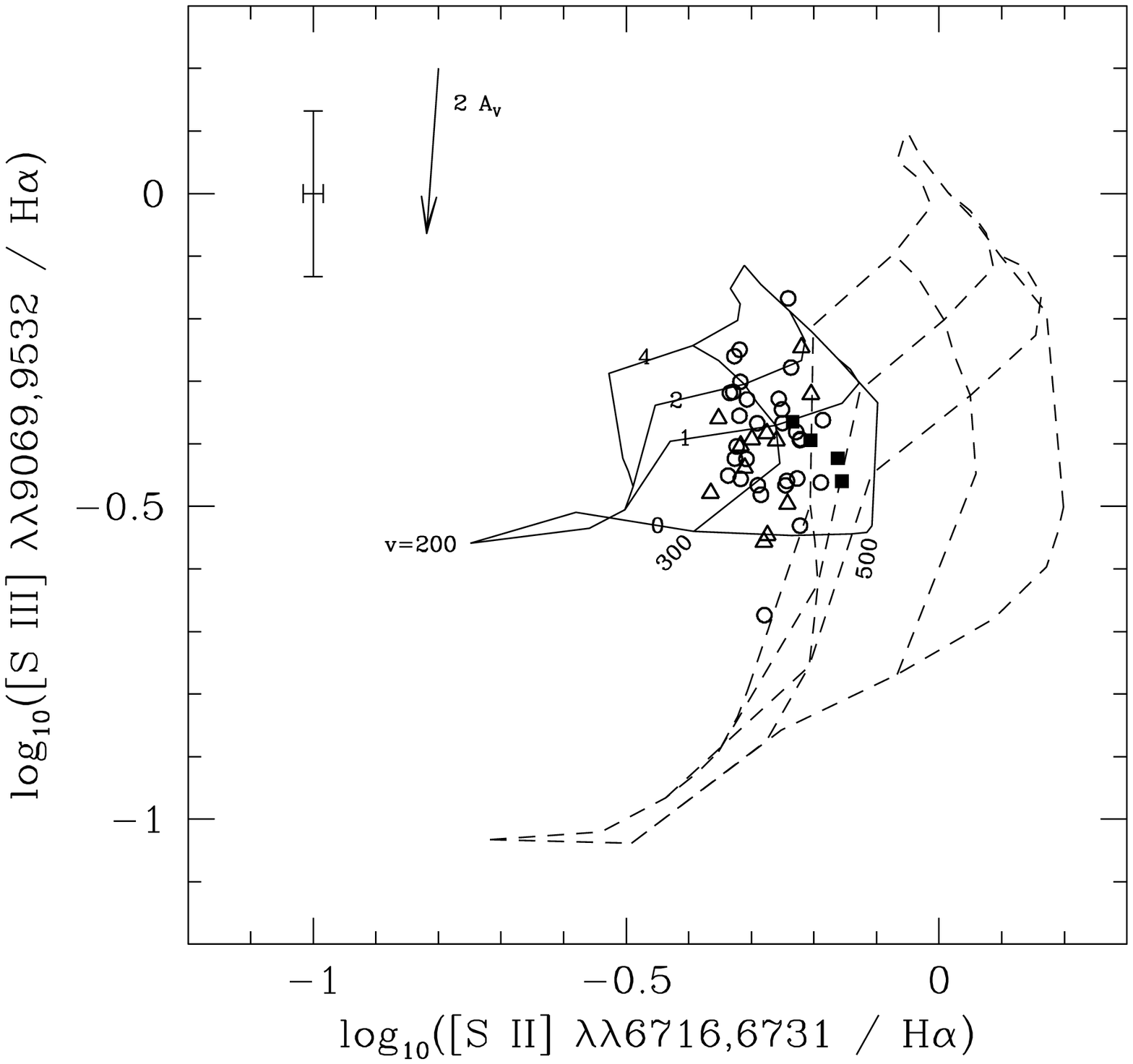,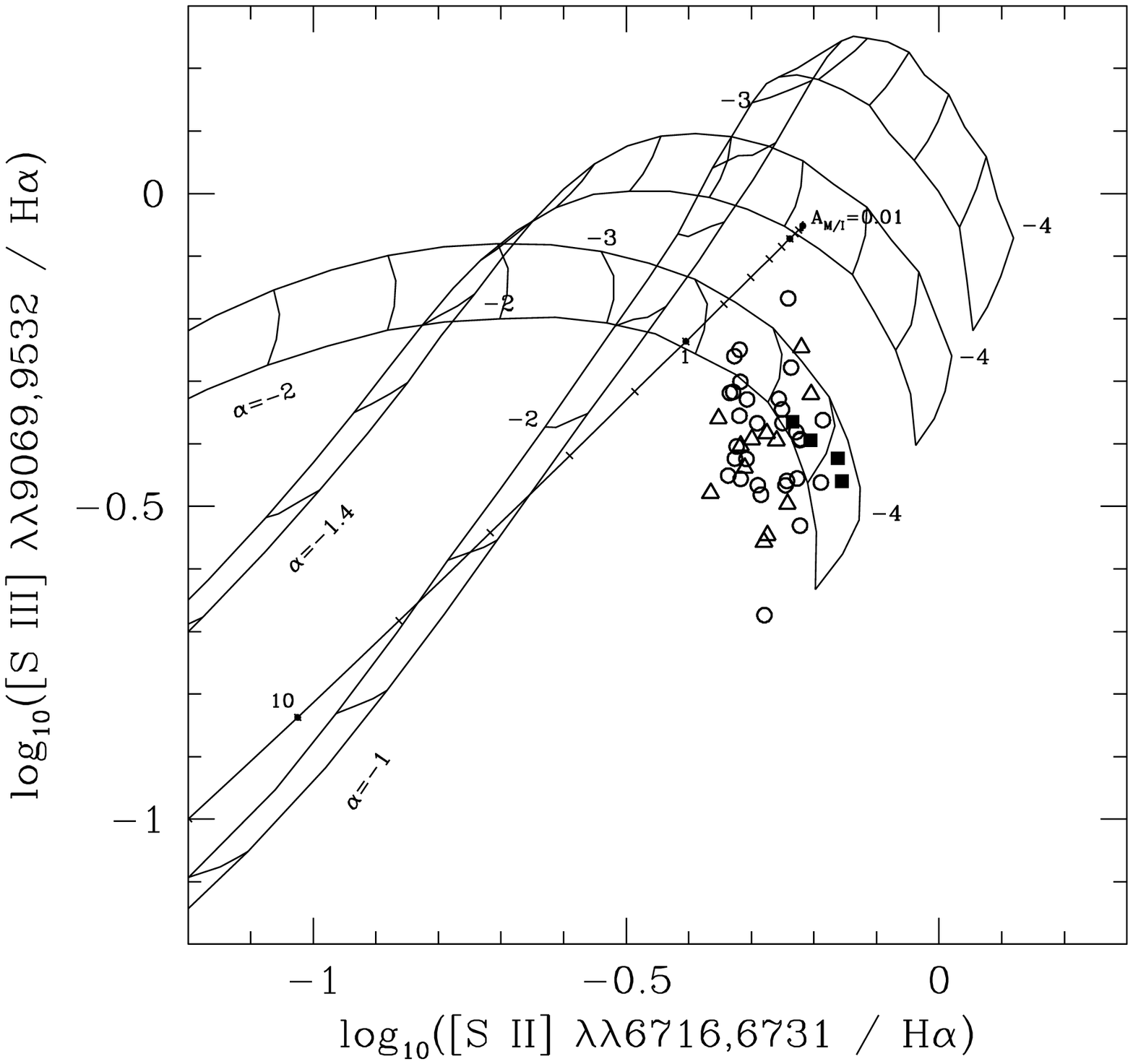]{
[\ion{S}{3}]~$\lambda\lambda$9069,9532/H$\alpha$ versus
[\ion{S}{2}]~$\lambda\lambda$6717,6731/H$\alpha$. In panel (a)
the shock-only grid is shown with the dashed line. The data points and 
model grids are otherwise as described for Figure~10. }

\figcaption[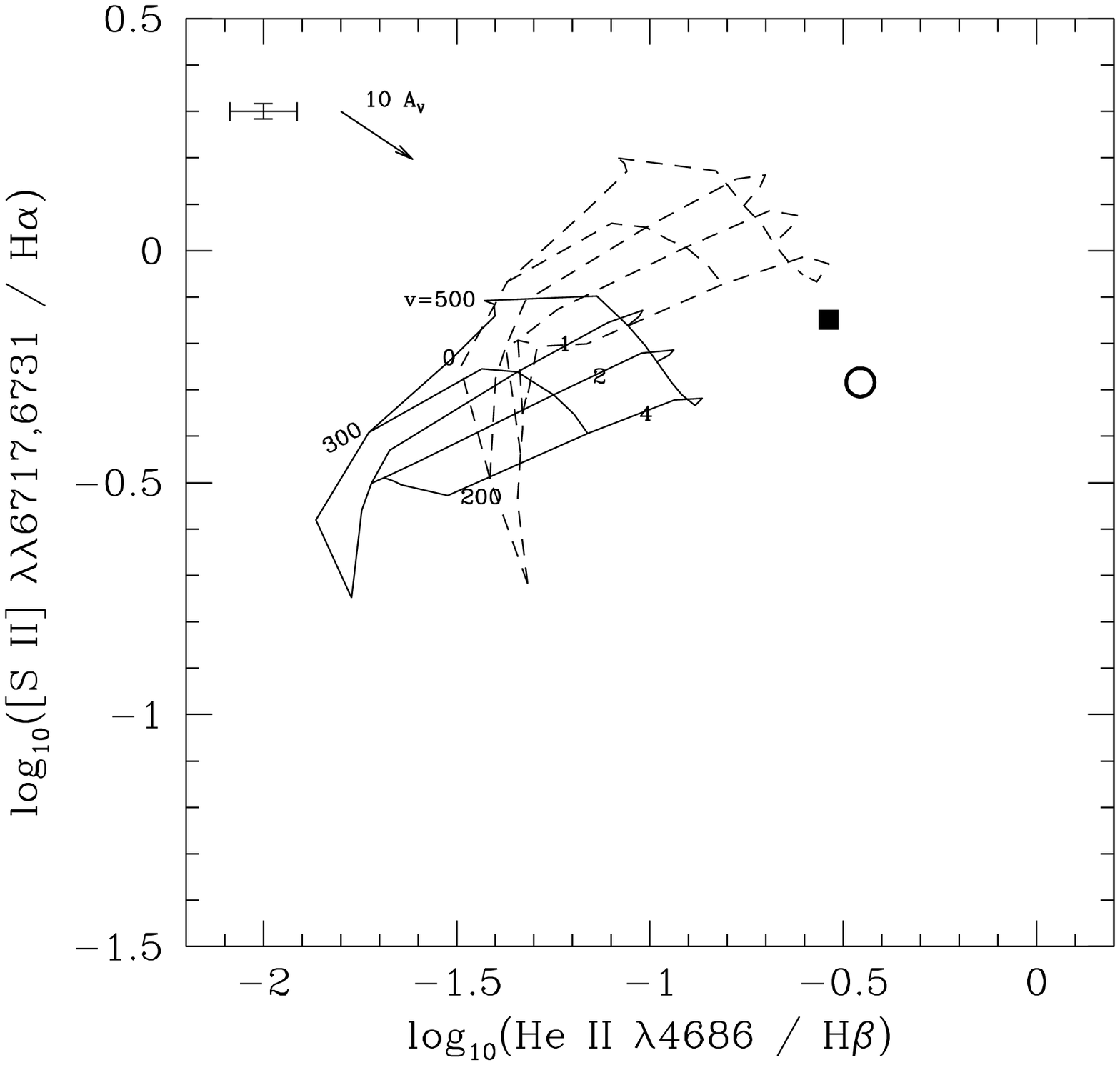,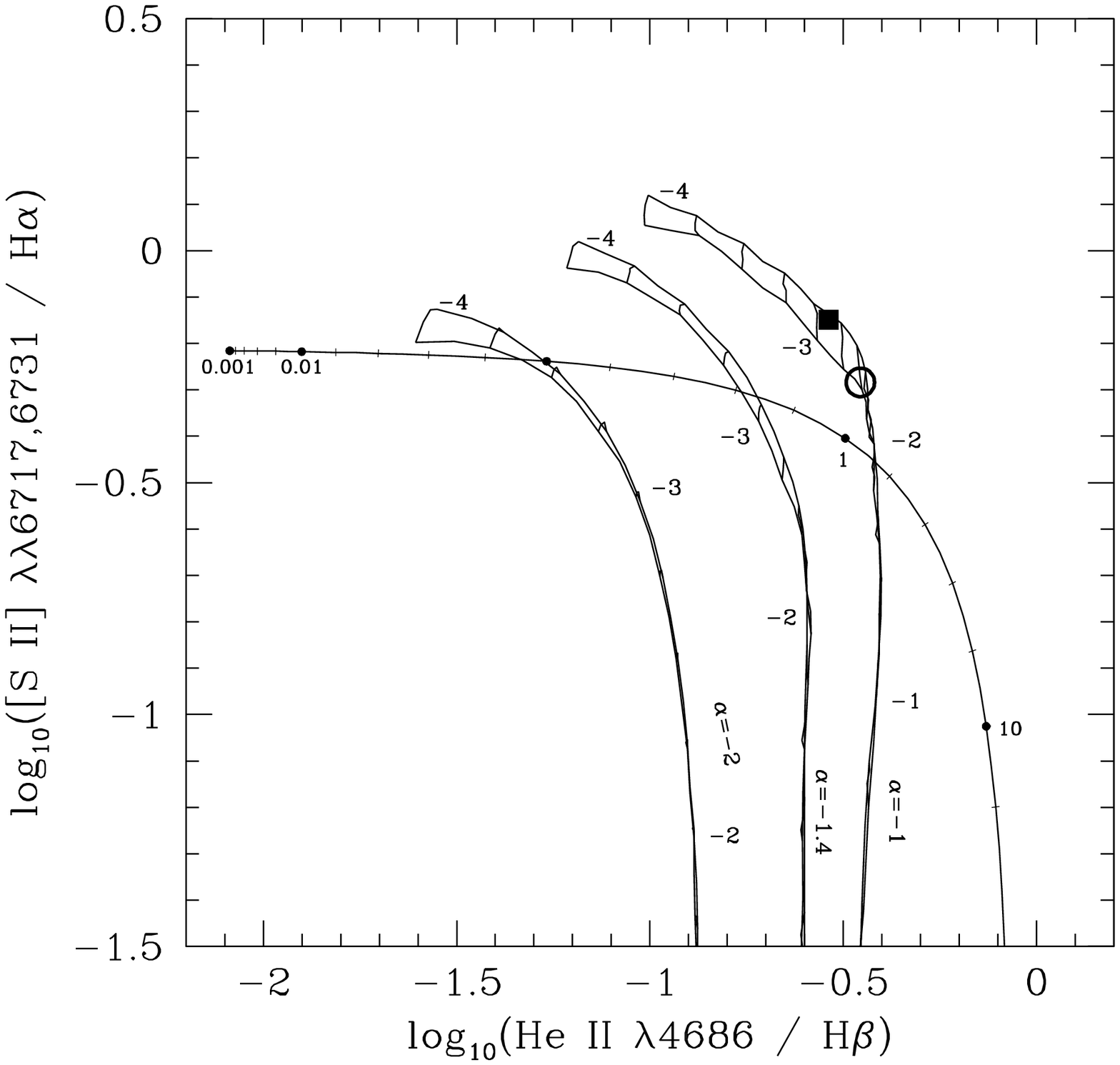]{
[\ion{S}{2}]~$\lambda\lambda$6717,6731/H$\alpha$ versus
\ion{He}{2}~$\lambda$4686/H$\beta$. In panel (a)
the shock-only grid is shown with the dashed line. The line ratios 
for the average spectra of the NW and SE cones are shown as the 
open circle, and the solid square respectively. The model grids
are as described for Figure~10.}

\figcaption[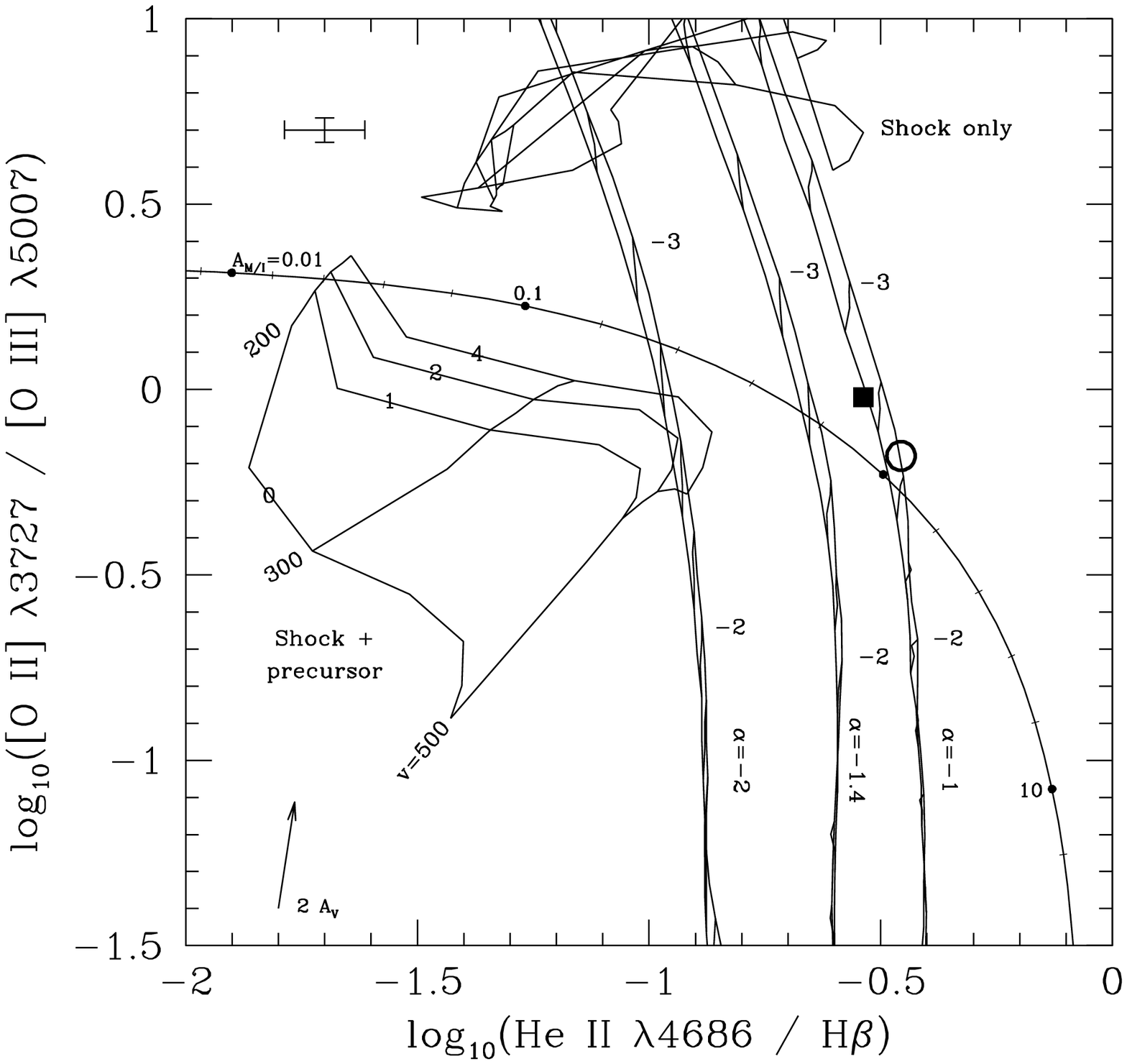]{\ion{He}{2}~$\lambda$4686/H$\beta$ versus
[\ion{O}{2}]~$\lambda$3727/[\ion{O}{3}]~$\lambda$5007. 
The line ratios for the average spectra of the NW and SE cones are
shown as the open circle, and the solid square respectively.  The
shock, shock+precursor, $U$-sequence and $A_{M/I}$ sequence models are shown
and are labelled as described in Figure~10.}

\figcaption[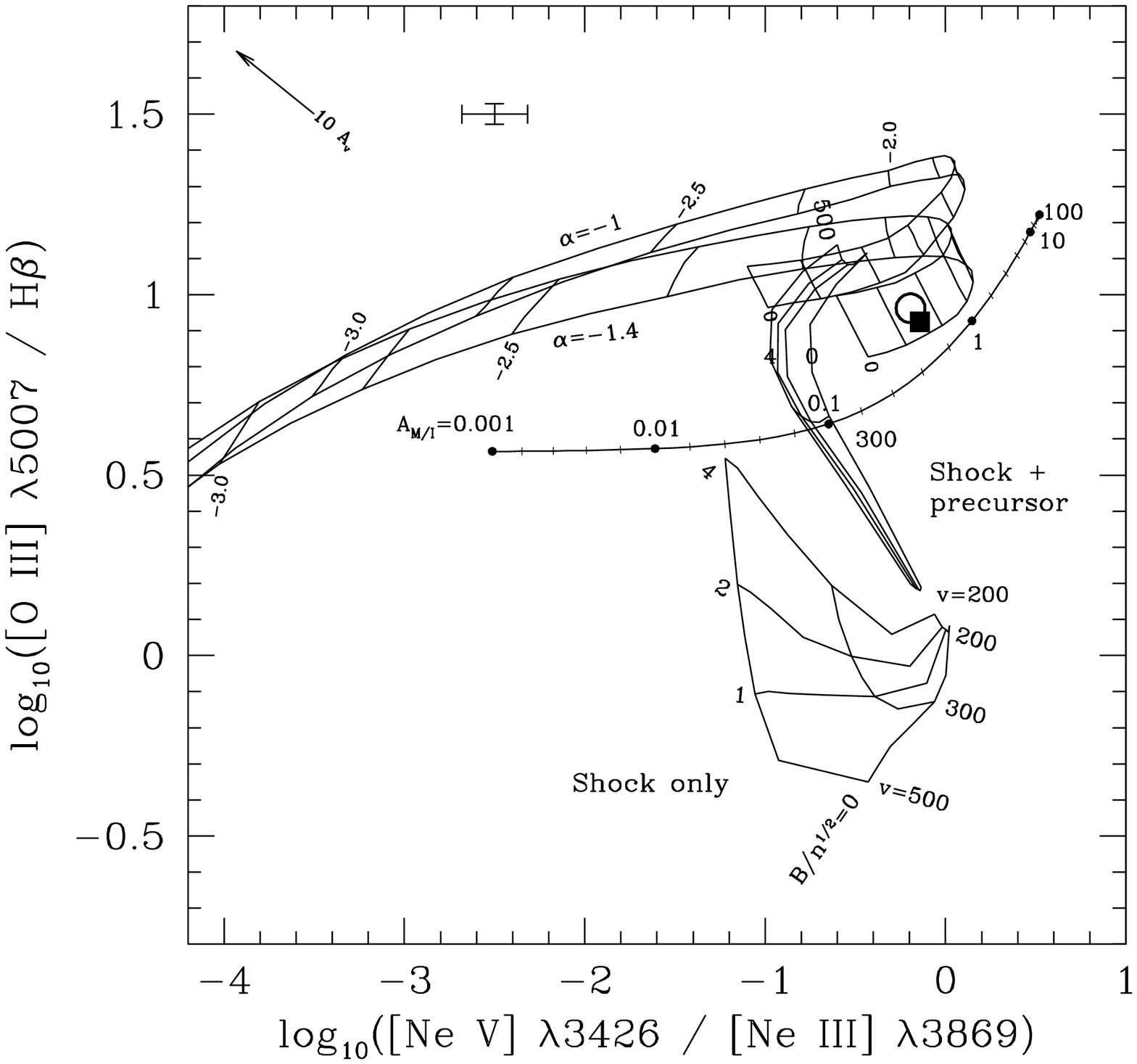]{
[\ion{Ne}{5}]~$\lambda$3426/[\ion{Ne}{3}]~$\lambda$3869 versus
[\ion{O}{3}]~$\lambda $5007/H$\beta$ ratio. The line ratios for the 
average spectra of the NW and SE cones are shown with the open circle, 
and the solid square respectively.  The
shock, shock+precursor, $U$-sequence and $A_{M/I}$ sequence models are shown
and are as described in Figure~10.}

\figcaption[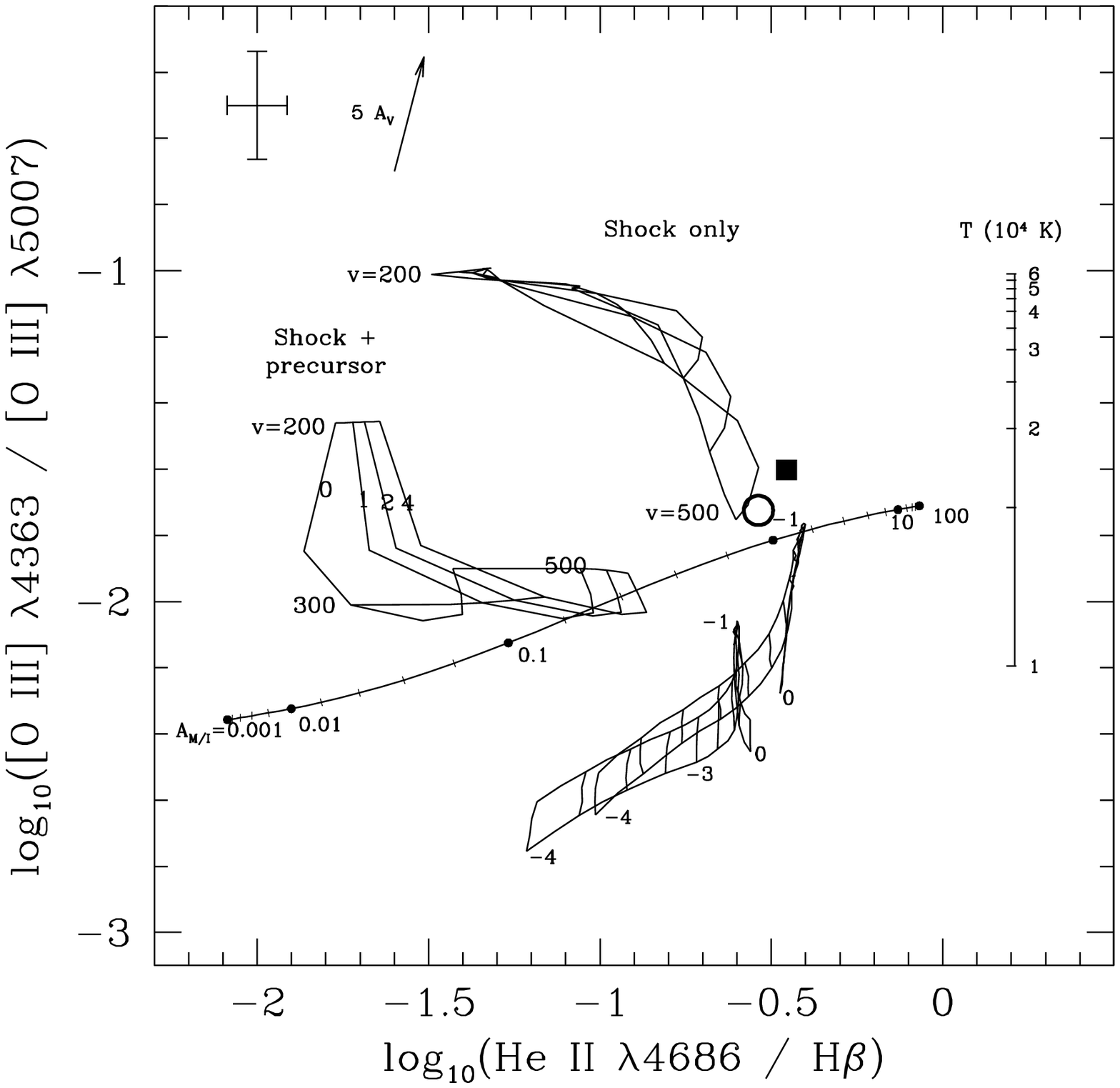]{Temperature sensitive diagnostic diagram:
[\ion{O}{3}]~$\lambda$4363/[\ion{O}{3}]~$\lambda$5007 versus
\ion{He}{2} $\lambda$4686/H$\beta$. The R$_{OIII}$ temperature scale 
is shown on the left side of the diagram. The line ratios for the 
average spectra of the NW and SE cones are shown with the open circle, 
and the solid square respectively. The shock, shock+precursor, 
$U$-sequence and $A_{M/I}$ sequence models are shown
and are as described in Figure~10.} 

\figcaption[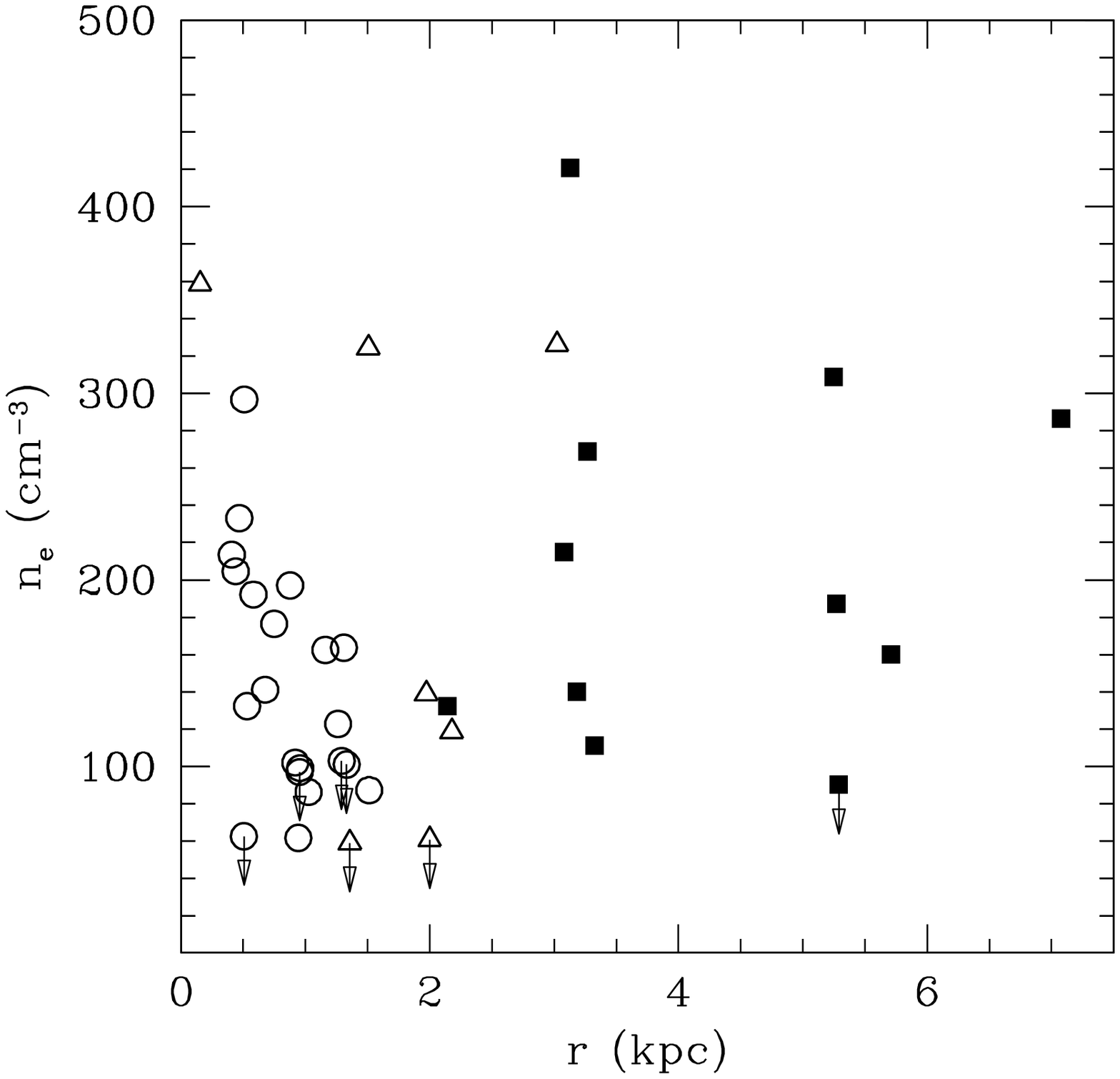,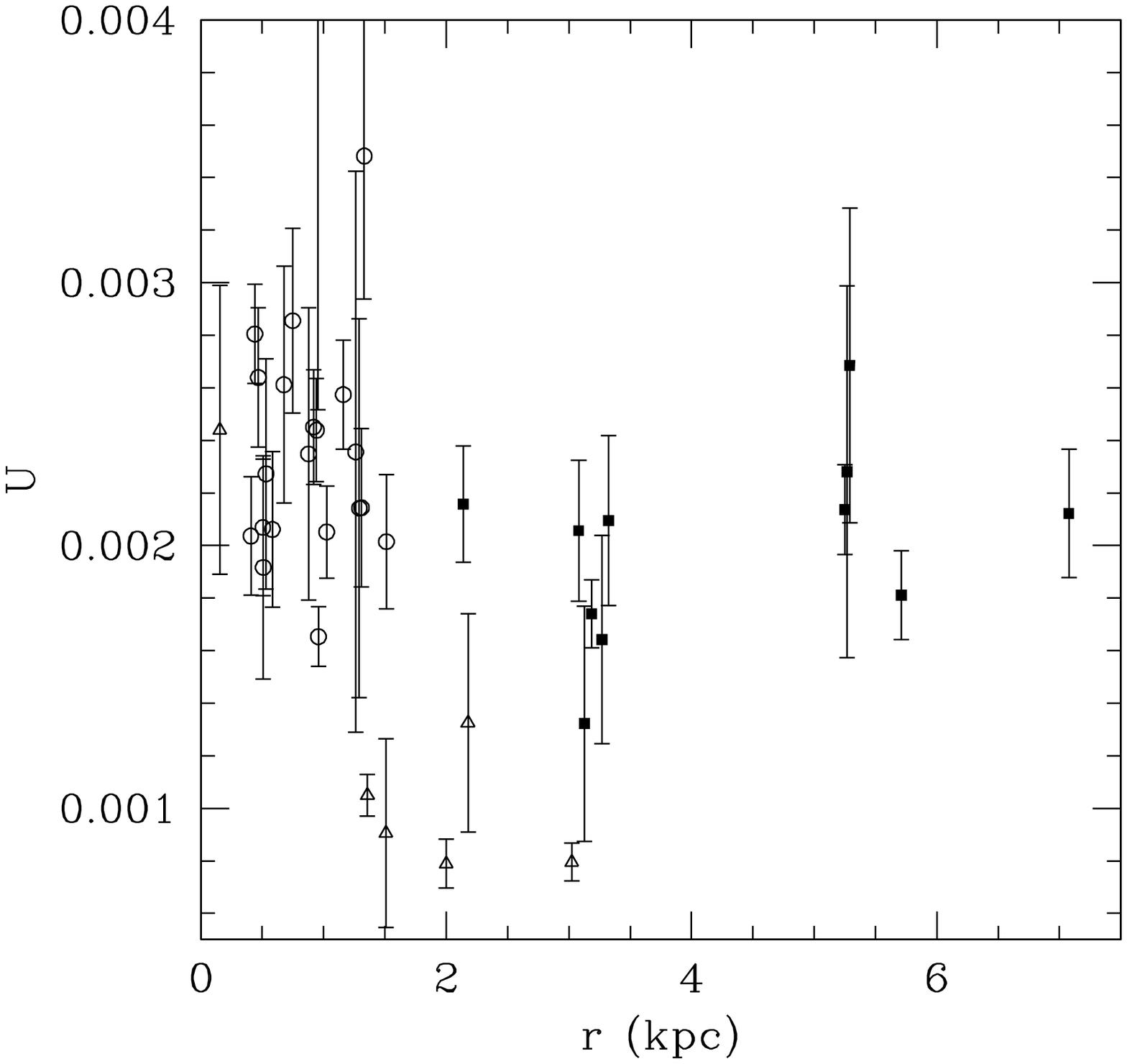]{ Densities (a), and 
Ionization parameters (b) for points within
the NW and SE cone regions that were used in the $q$-test calculations.}

\figcaption[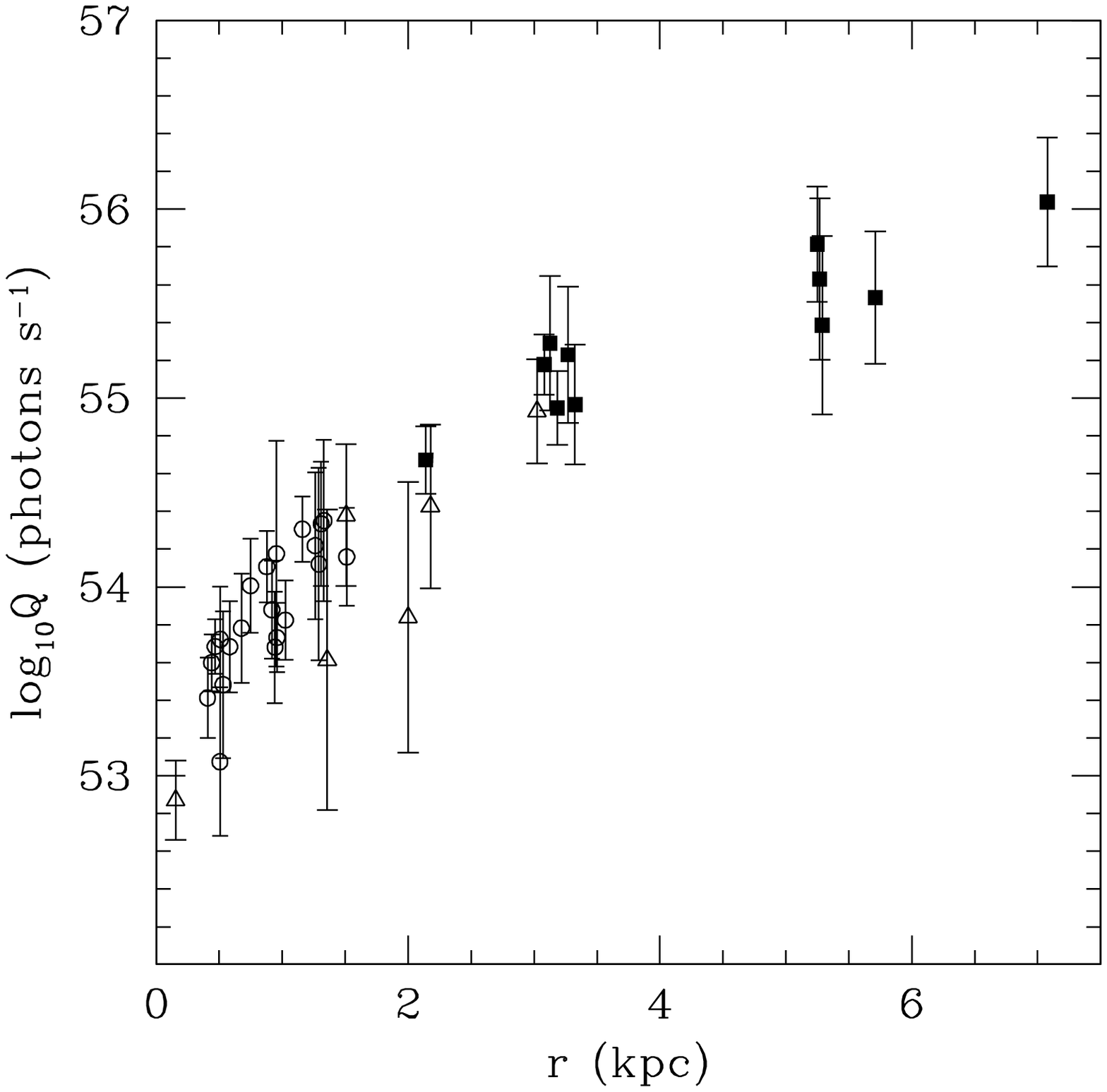,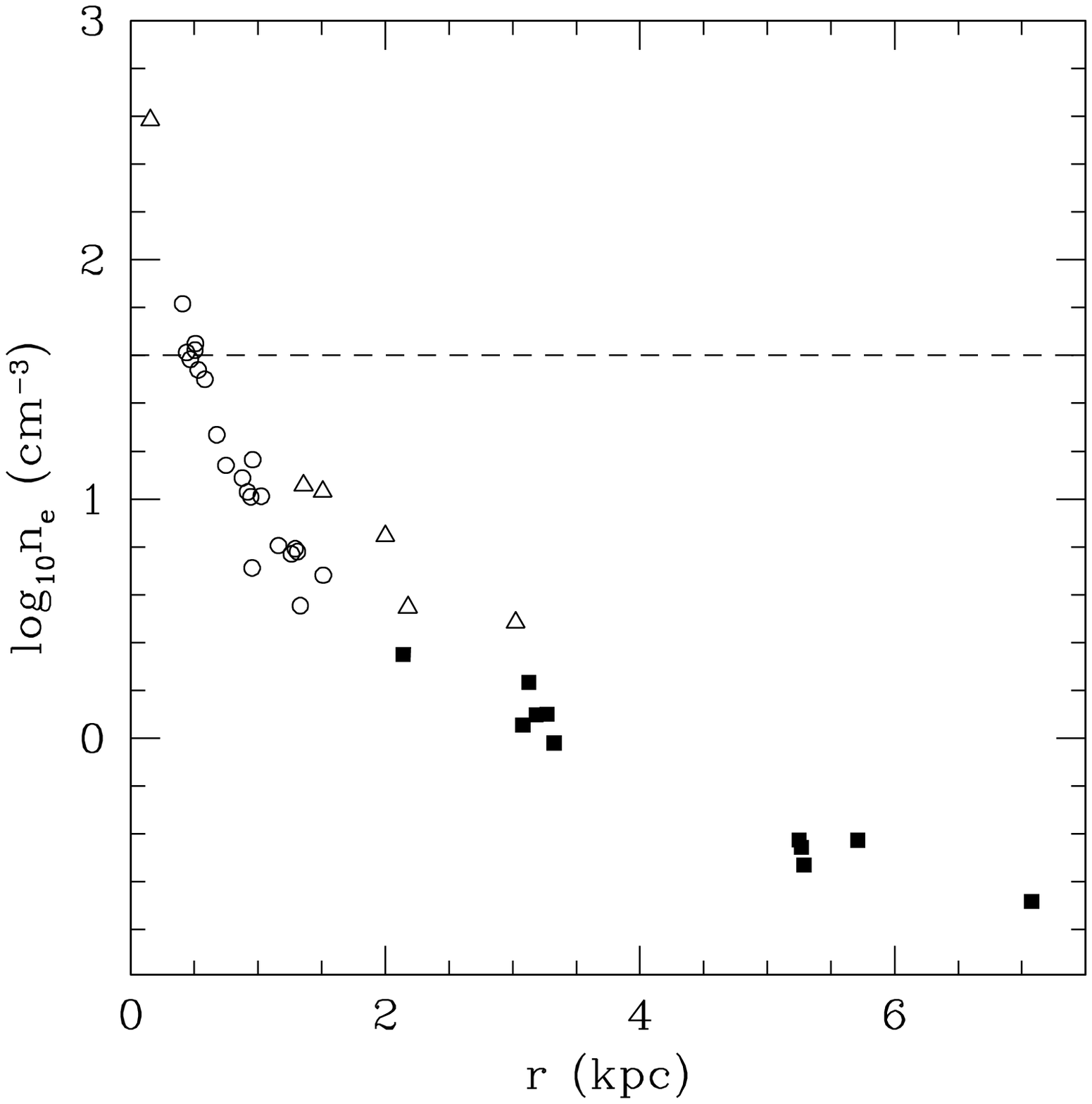,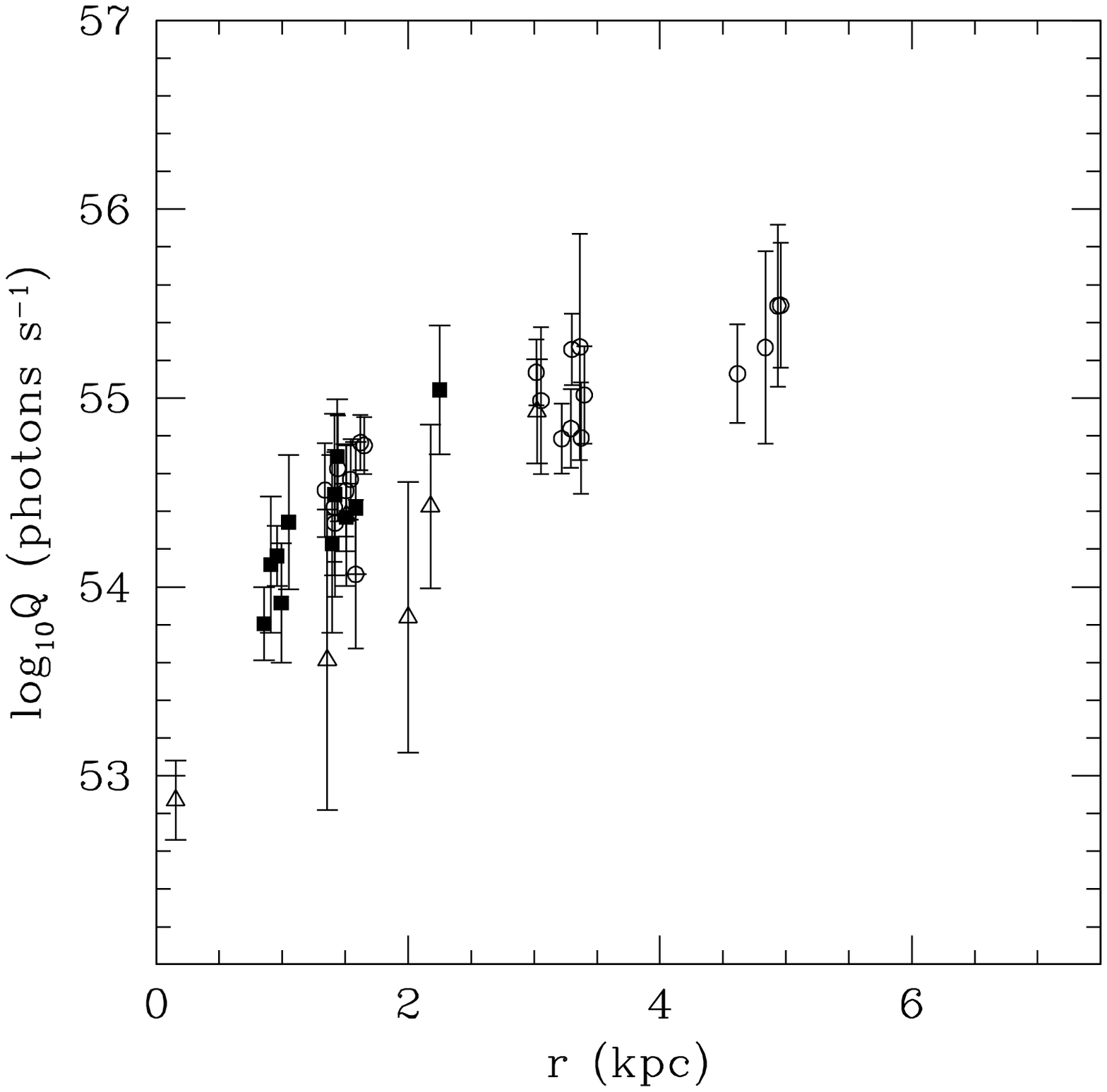,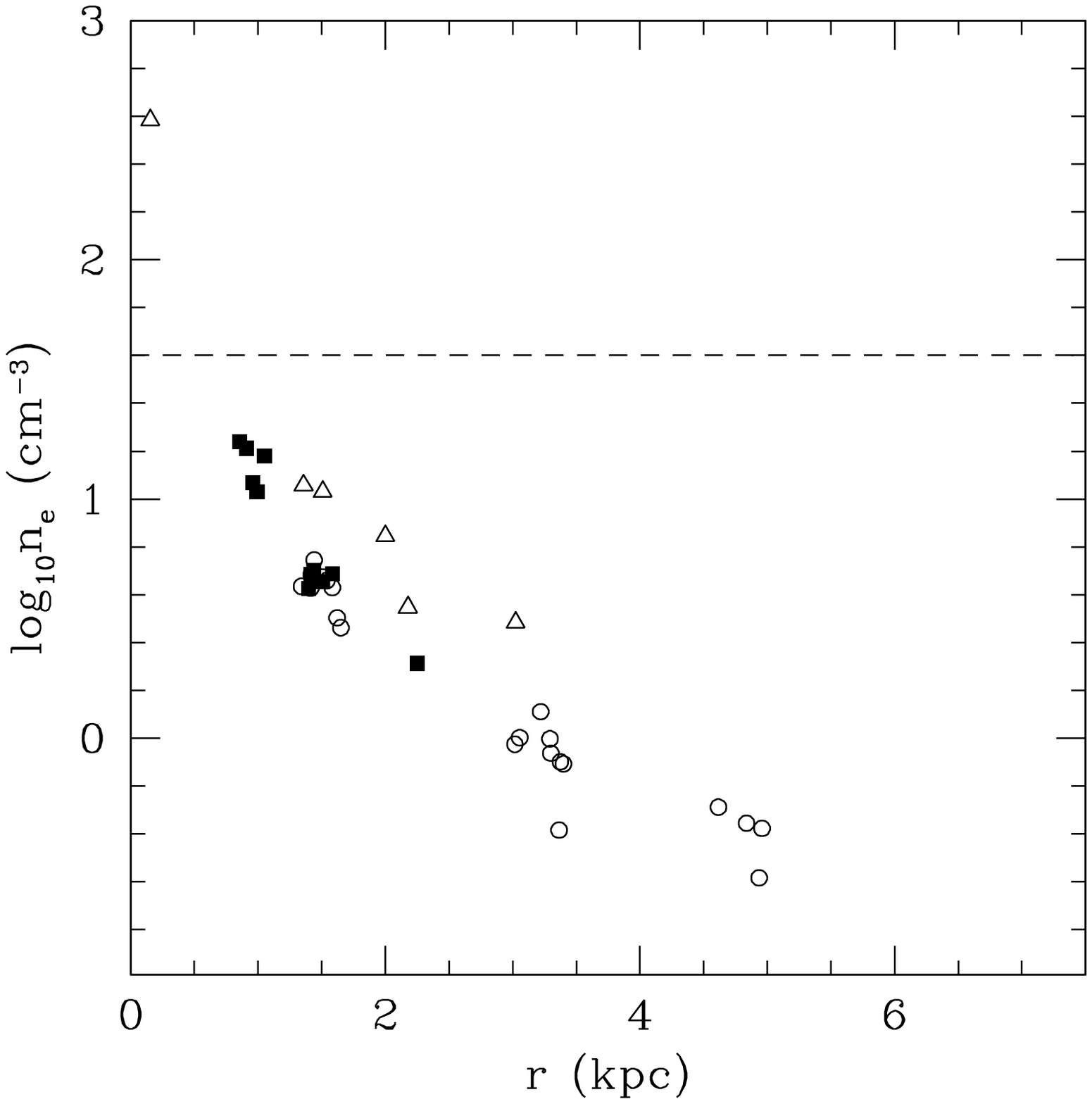,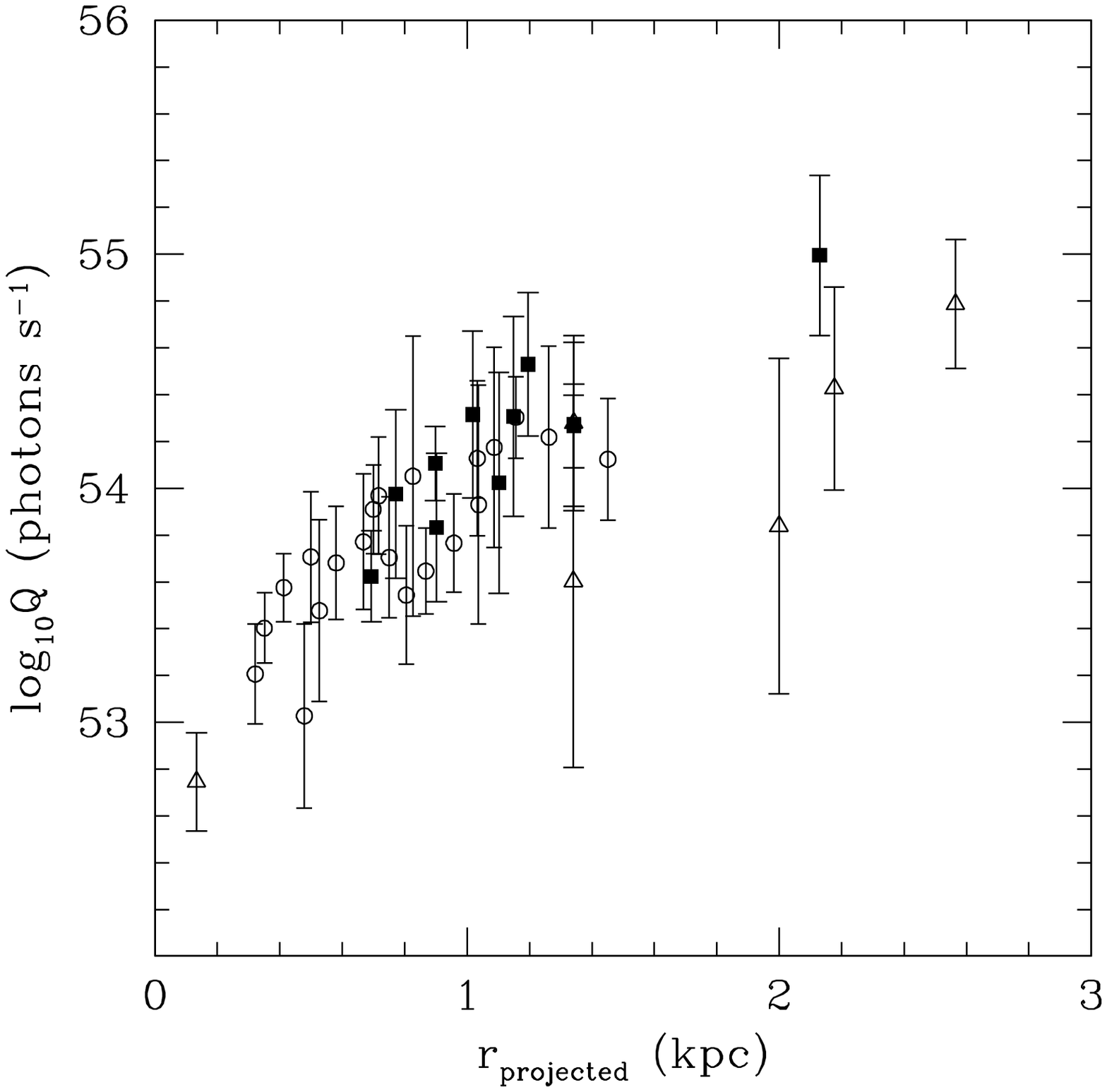,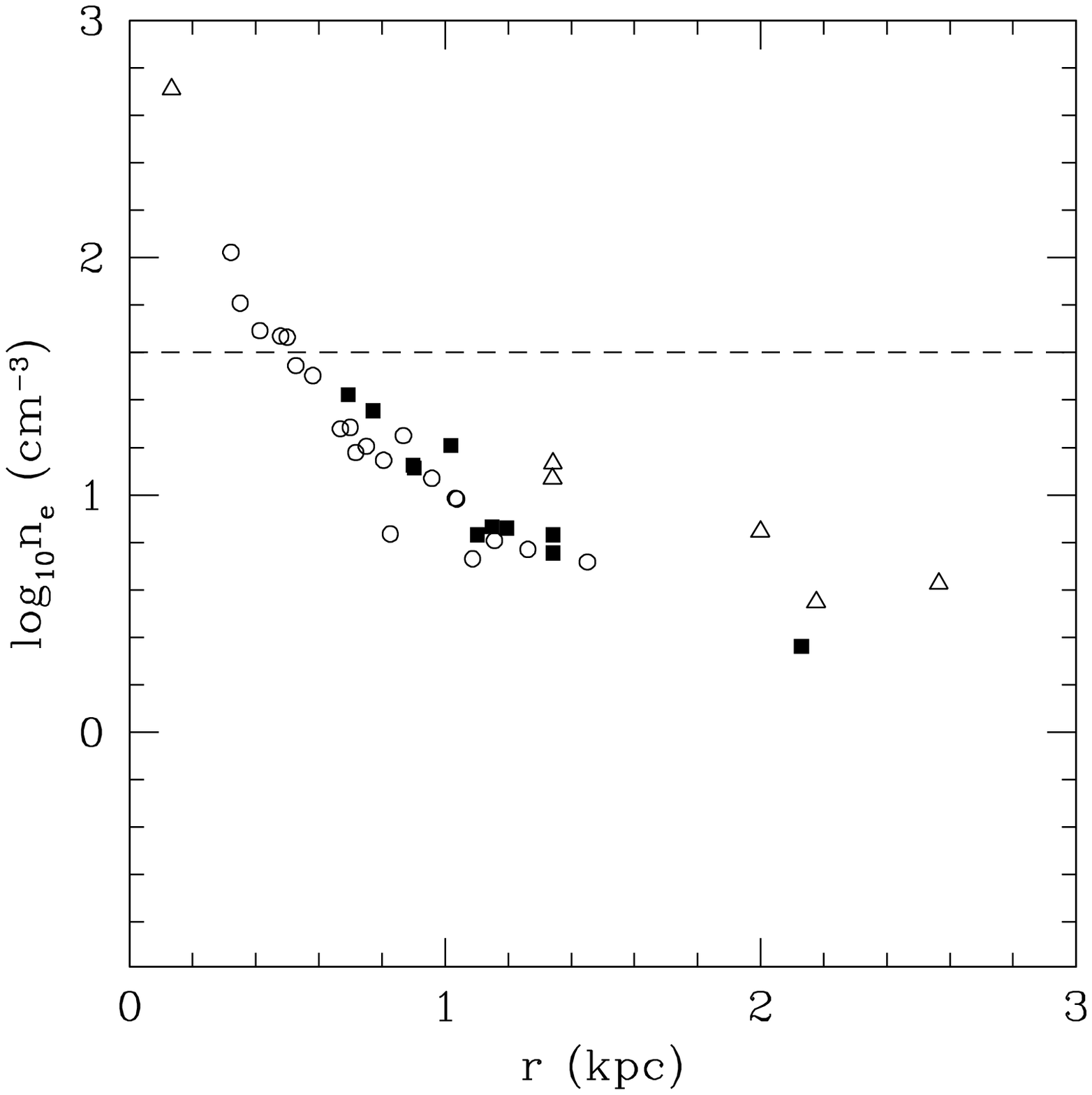]{
Q versus radius, and $n_{e,Q}$ versus radius for three different
de-projections of the points within the ionization cones of NGC~2992.
The diagrams in the left panels show Q versus radius
for de-projection onto the near side for both cones (top), 
de-projection onto the far side of the cones (middle)
and with no de-projection (bottom). The diagrams in the right panels
show the corresponding plots of $n_{e,Q}$ 
versus radius. The plotting symbols represent the different regions of the 
ENLR as shown in Figure~6. The lower useful limit for the 
[\ion{S}{2}] density diagnostic is shown with the dashed line in the
diagrams in the right panels.
}

\figcaption[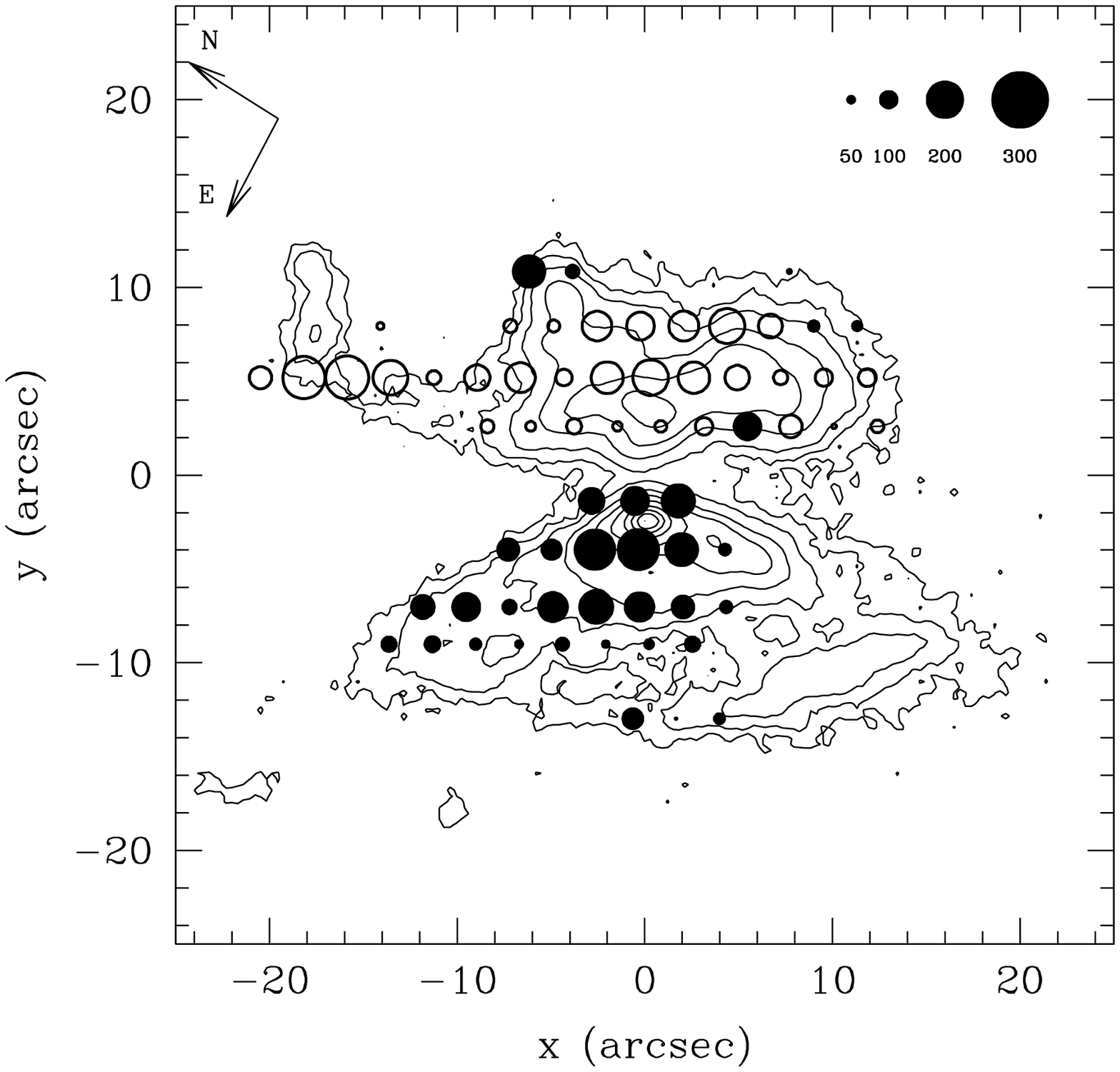,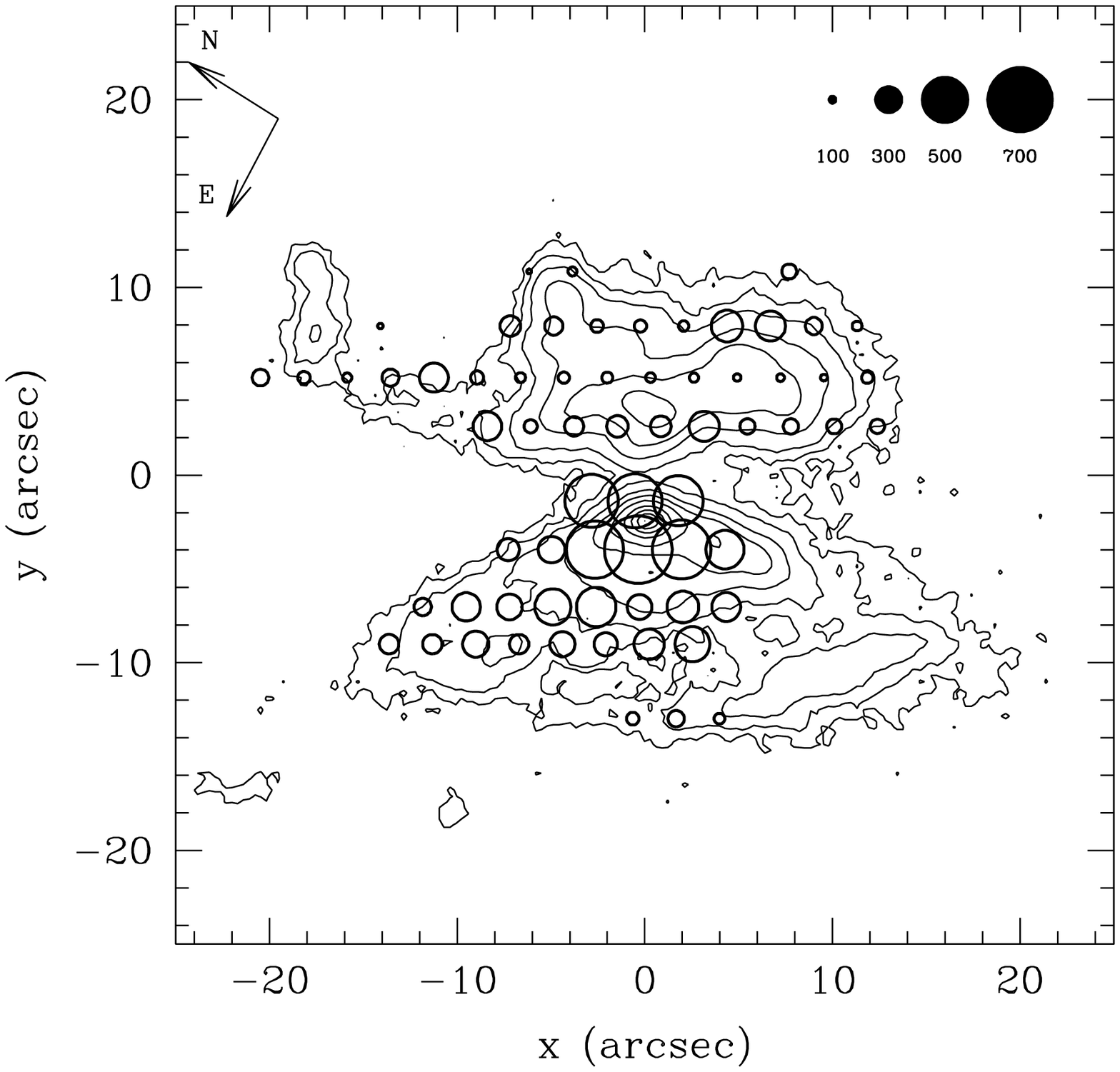]{a) Outflow velocity field
overplotted on the [\ion{O}{3}]~$\lambda$5007 contour map. The size
of the circle at each point indicates the difference in velocity of
the wind component relative to the rotational component.  The key in
the top right hand corner indicates the magnitude of this velocity in
km~s$^{-1}$.  Solid points correspond to velocities that are blue
shifted relative to the rotation, and open circles indicate red
shifted velocities.  b) FWHM map of the major outflow component.
The image contours are the same as in Figure~2.}

\figcaption[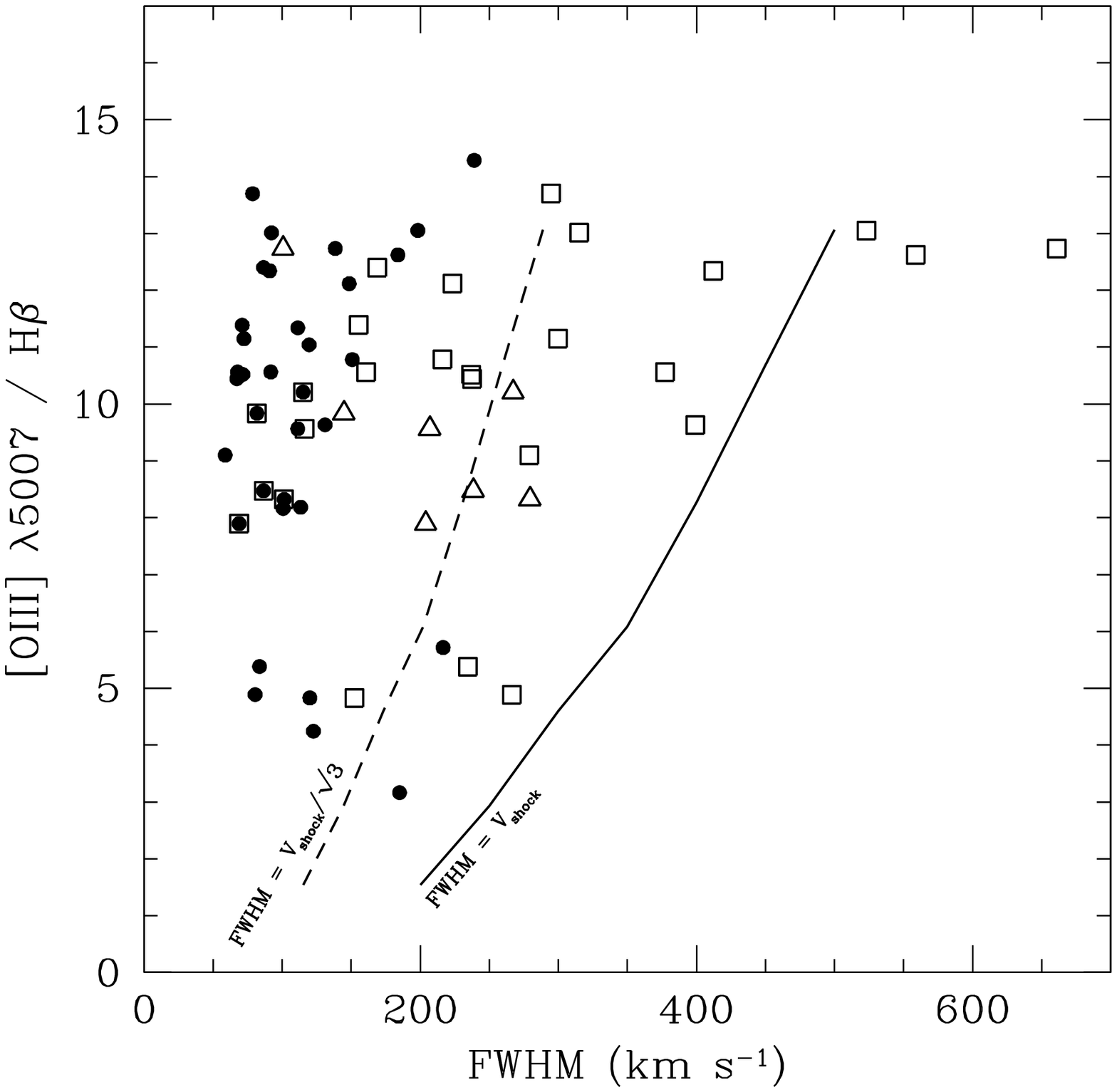,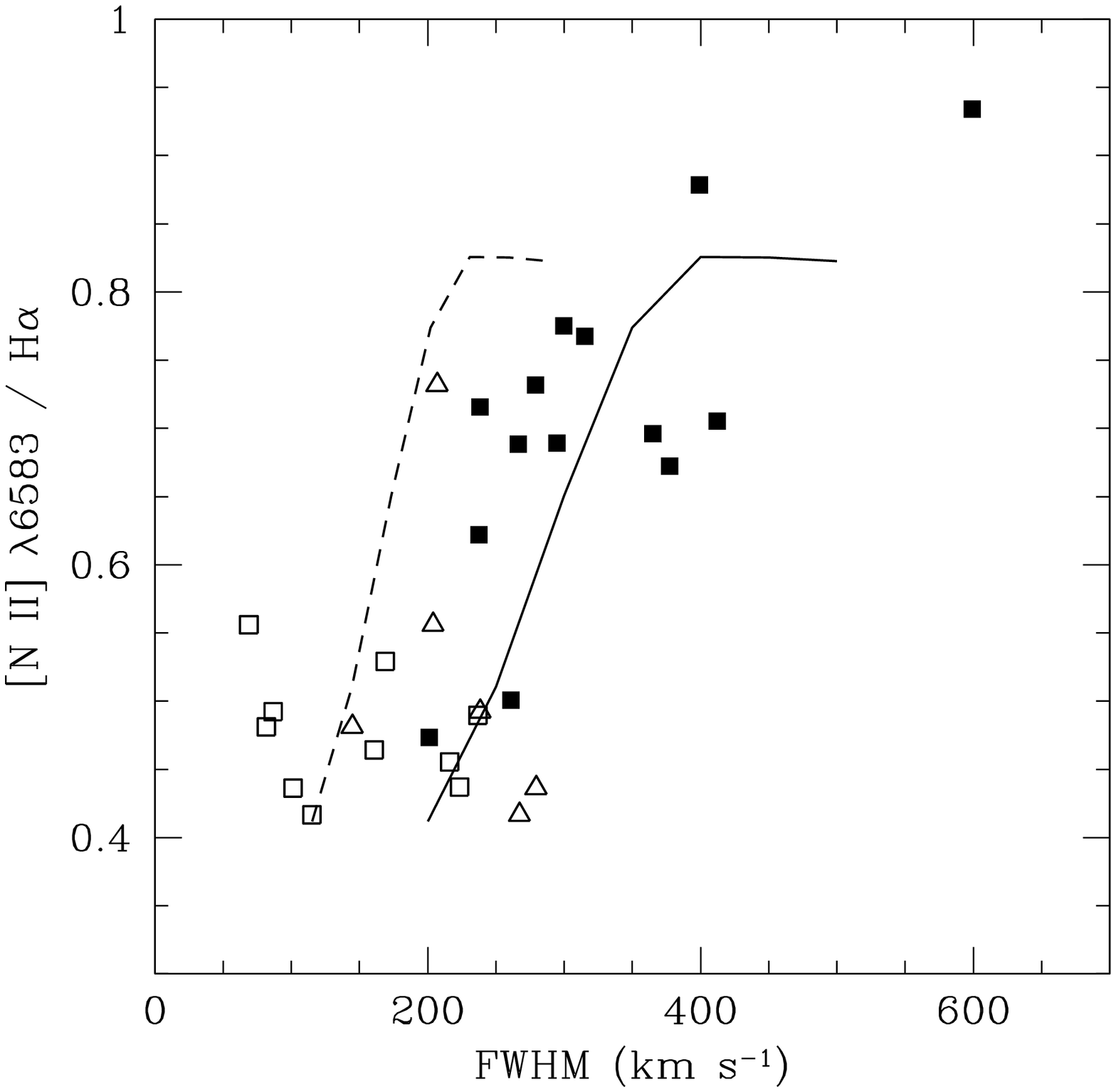]{Excitation versus line width diagrams.
(a): [\ion{O}{3}]~$\lambda$5007/H$\beta$ versus the FWHM of
the kinematical components. The kinematical component identified with
the galaxy rotation is plotted as the solid circles, the main outflow
components are plotted as open squares and the extra component  
(detected at a few locations) is plotted as open triangles. 
(b): [\ion{N}{2}]~$\lambda$6583/H$\alpha$ versus FWHM. 
As in (a), the square points indicate the main outflow component, and the 
triangles indicate the extra component. In (b) we make the distinction between 
the cones by using filled symbols for the SE cone and open symbols for the NW
cone.
In both (a) and (b) the solid curve is the shock+precursor model 
prediction where the FWHM is directly related to the
shock velocity, and the dashed curve assumes that the FWHM is
associated with the projection of randomly oriented shocks so that
$V_{shock}\simeq$FWHM/$\sqrt{3}$.
The uncertainty in the FWHM is generally less than 100~km~s$^{-1}$.
}

\clearpage




\centerline{\hbox{
\psfig{figure=fig2.ps,height=3.5in}}}
\vspace{1cm}
{\bf \Large \hspace{7cm} Figure 2 }

\clearpage

\centerline{\hbox{
\psfig{figure=fig3a.ps,width=7in}}}
\vspace{1cm}
{\bf \Large \hspace{7.0cm} Figure 3a }

\clearpage 

\centerline{\hbox{
\psfig{figure=fig3b.ps,width=7in}}}
\vspace{1cm}
{\bf \Large \hspace{7.0cm} Figure 3b }

\clearpage

\centerline{\hbox{
\psfig{figure=fig3c.ps,width=7in}}}
\vspace{1cm}
{\bf \Large \hspace{7.0cm} Figure 3c }

\clearpage

\centerline{\hbox{
\psfig{figure=fig4a.ps,width=3.5in}}}
\vspace{0.2cm}
{\bf \Large \hspace{7.0cm} Figure 4a }
\par

\vspace{2cm}
\centerline{\hbox{
\psfig{figure=fig4b.ps,width=3.5in}}}
\vspace{0.2cm}
{\bf \Large \hspace{7.0cm} Figure 4b }
\par

\clearpage 

\centerline{\hbox{
\psfig{figure=fig5a.ps,height=3.5in}}}
\vspace{0.2cm}
{\bf \Large \hspace{7.0cm} Figure 5a }
\par
\centerline{\hbox{
\psfig{figure=fig5b.ps,height=3.5in}}}
\vspace{0.2cm}
{\bf \Large \hspace{7.0cm} Figure 5b }
\par

\clearpage

\centerline{\hbox{
\psfig{figure=fig6.ps,height=3.5in}}}
{\bf \Large \hspace{7.0cm} Figure 6 }
\clearpage 

\centerline{\hbox{
\psfig{figure=fig7.ps,width=3.5in} }}
\vspace{1cm}
{\bf \Large \hspace{8.0cm} Figure 7 }

\clearpage

\centerline{\hbox{
\psfig{figure=fig8.ps,height=3.5in}}}
{\bf \Large \hspace{7.0cm} Figure 8 }

\par
\centerline{\hbox{
\psfig{figure=fig9.ps,height=3.5in}}}
{\bf \Large \hspace{7.0cm} Figure 9 }

\clearpage
\centerline{\hbox{
\psfig{figure=fig10a.ps,height=3.5in,width=3.5in}
\psfig{figure=fig10b.ps,height=3.5in,width=3.5in}
}}
{\bf \Large \hspace{3.0cm} Figure 10a  \hspace{6.5cm} Figure 10b}
\par
\vspace{1cm}
\centerline{\hbox{
\psfig{figure=fig11a.ps,height=3.5in,width=3.5in}
\psfig{figure=fig11b.ps,height=3.5in,width=3.5in}
}}
{\bf \Large \hspace{3.0cm} Figure 11a  \hspace{6.5cm} Figure 11b}
\par

\clearpage
\centerline{\hbox{
\psfig{figure=fig12a.ps,height=3.5in,width=3.5in}
\psfig{figure=fig12b.ps,height=3.5in,width=3.5in}
}}
{\bf \Large \hspace{3.0cm} Figure 12a  \hspace{6.5cm} Figure 12b}
\par
\centerline{\hbox{
\psfig{figure=fig13a.ps,height=3.5in,width=3.5in}
\psfig{figure=fig13b.ps,height=3.5in,width=3.5in}
}}
{\bf \Large \hspace{3.0cm} Figure 13a  \hspace{6.5cm} Figure 13b}

\clearpage
\centerline{\hbox{
\psfig{figure=fig14a.ps,height=3.5in,width=3.5in}
\psfig{figure=fig14b.ps,height=3.5in,width=3.5in}
}}
{\bf \Large \hspace{3.0cm} Figure 14a  \hspace{6.5cm} Figure 14b}
\par
\centerline{\hbox{
\psfig{figure=fig15a.ps,height=3.5in,width=3.5in}
\psfig{figure=fig15b.ps,height=3.5in,width=3.5in}
}}
{\bf \Large \hspace{3.0cm} Figure 15a  \hspace{6.5cm} Figure 15b}

\clearpage

\centerline{\hbox{
\psfig{figure=fig16a.ps,height=3.5in,width=3.5in}
\psfig{figure=fig16b.ps,height=3.5in,width=3.5in}
}}
{\bf \Large \hspace{3.0cm} Figure 16a  \hspace{6.5cm} Figure 16b}
\par
\centerline{\hbox{
\psfig{figure=fig17a.ps,height=3.5in,width=3.5in}
\psfig{figure=fig17b.ps,height=3.5in,width=3.5in}
}}
{\bf \Large \hspace{3.0cm} Figure 17a  \hspace{6.5cm} Figure 17b}

\clearpage
\centerline{\hbox{
\psfig{figure=fig18.ps,height=3.5in,width=3.5in}
\psfig{figure=fig19.ps,height=3.5in,width=3.5in}
}}
{\bf \Large \hspace{3.0cm} Figure 18   \hspace{6.5cm} Figure 19}

\clearpage

\centerline{\hbox{
\psfig{figure=fig20.ps,height=6in,width=6in}
}}
\vspace{1cm}
{\bf \Large \hspace{7.0cm} Figure 20 }

\clearpage

\centerline{\hbox{
\psfig{figure=fig21a.ps,height=3.5in,width=3.5in}
\psfig{figure=fig21b.ps,height=3.5in,width=3.5in}
}}
{\bf \Large \hspace{3.0cm} Figure 21a  \hspace{6.5cm} Figure 21b}
\par
\centerline{\hbox{
\psfig{figure=fig22a.ps,height=3.5in,width=3.5in}
\psfig{figure=fig22b.ps,height=3.5in,width=3.5in}
}}
{\bf \Large \hspace{3.0cm} Figure 22a  \hspace{6.5cm} Figure 22b}

\clearpage

\centerline{\hbox{
\psfig{figure=fig22c.ps,height=3.5in,width=3.5in}
\psfig{figure=fig22d.ps,height=3.5in,width=3.5in}
}}
{\bf \Large \hspace{3.0cm} Figure 22c  \hspace{6.5cm} Figure 22d}
\par
\centerline{\hbox{
\psfig{figure=fig22e.ps,height=3.5in,width=3.5in}
\psfig{figure=fig22f.ps,height=3.5in,width=3.5in}
}}
{\bf \Large \hspace{3.0cm} Figure 22e  \hspace{6.5cm} Figure 22f}

\clearpage

\centerline{\hbox{
\psfig{figure=fig23a.ps,height=3.5in,width=3.5in}
\psfig{figure=fig23b.ps,height=3.5in,width=3.5in}
}}
{\bf \Large \hspace{3.0cm} Figure 23a  \hspace{6.5cm} Figure 23b}
\par
\centerline{\hbox{
\psfig{figure=fig24a.ps,height=3.5in,width=3.5in}
\psfig{figure=fig24b.ps,height=3.5in,width=3.5in}
}}
{\bf \Large \hspace{3.0cm} Figure 24a  \hspace{6.5cm} Figure 24b}


\clearpage

\begin{thebibliography}{}

\bibitem[Allen, Dopita \& Tsvetanov]{all98}  Allen, M. G., Dopita, M. A.,
         \& Tsvetanov, Z. I. 1998, \apj, 493, 571

\bibitem[Antonucci \& Miller]{ant85}  Antonucci, R. R. J. \& Miller, J. S.
         1985, \apj, 297, 621

\bibitem[Axon et al.]{axo98} Axon, D. J., Marconi, A., Capetti, A.,
         Macchetto, F. D., Schreier, Robinson, A. 1998, \apj, 496, L75

\bibitem[Baldwin, Phillips \& Terlevich]{bal81}  Baldwin, J. A.,
         Phillips, M. M., \& Terlevich, R. 1981, \pasp, 93, 5

\bibitem[Bicknell et al.]{bic98} Bicknell, G. V., Dopita, M. A.,
         Tsvetanov, Z. I., Sutherland, R. S. 1998, \apj, 495, 680

\bibitem[Binette, Wilson, \& Storchi-Bergman]{bin96} Binette, L.,
        Wilson, A. S., \& Storchi-Bergmann, T. 1996, \aap, 312, 365

\bibitem[Boisson and Durret]{boi86}  Boisson, C., Durret, F. 1986, \aap, 168, 32

\bibitem[Cai \& Pradhan]{cai93} Cai, W. \& Pradhan, A. K. 1993, \apjs, 88, 329

\bibitem[Cardelli, Clayton, \& Mathis]{car89} Cardelli, J. A., Clayton, 
         G. C., \& Mathis, J. S., 1989, \apj, 345, 245

\bibitem[Colbert et al.]{col96a} Colbert, E. J. M., Baum, S. A.,
         Gallimore, J. F.,  O'Dea, C. P., Lehnert, M. D., Tsvetanov, Z. I.,
         Mulchaey, J. S., and Caganoff, S. 1996a, \apjs, 105, 75

\bibitem[Colbert et al.]{col96b} Colbert, E. J. M., Baum, S. A.,
         Gallimore, J. F.,  O'Dea, C. P., Christensen, J. A. 1996b, 
         \apj, 467, 551

\bibitem[Colbert et al.]{col98} Colbert, E. J. M., Baum, S. A.,
         O'Dea, Veilleux, S. 1998, \apj, 496, 786

\bibitem[Colina et al.]{col87}  Colina, L., Fricke, K. J.,
         Kollatschny, W., \& Perryman, M. A. C. 1987, \aap, 178, 51

\bibitem[Dopita \& Sutherland]{dop95}  Dopita, M. A. \& Sutherland, R. S.
         1995, \apj, 455, 468

\bibitem[Dopita \& Sutherland]{dop96}  Dopita, M. A. \& Sutherland, R. S.
         1996, \apjs, 102, 161

\bibitem[Dopita]{dop97} Dopita, M. A. 1997, \apss, 248, 93

\bibitem[Durret]{dur90}  Durret, F. 1990, \aap, 229, 351

\bibitem[Durret and Bergeron]{dur88}  Durret, F., \& Bergeron, J. 1988, 
         \aap, Suppl. Ser. 75, 273

\bibitem[Ferland \& Netzer]{fer83}  Ferland, G. J., \& Netzer, H. 1983,
         \apj, 264, 105

\bibitem[Glass]{gla97} Glass, I. S. 1997, \mnras, 292, L50

\bibitem[Hummel et al.]{hum83}  Hummel, E., van Gorkom, J. H., \&
         Kotanyi, C. G. 1983, \apjl, 267, L5

\bibitem[Hutchmeier]{hut82}  Hutchmeier, W. K. 1982, \aap, 110, 121

\bibitem[Koski]{kos78}  Koski, A. T. 1978, \apj, 223, 56

\bibitem[Lawrence \& Elvis 1982]{law82}  Lawrence, A. \& Elvis, M. 1982, 
         \apj, 256, 410


\bibitem[Malkan]{mal83}  Malkan, M. A. 1983, \apjl, 264, L1

\bibitem[M\'{a}rquez et al.]{mar98} M\'{a}rquez, I., Boisson, C., 
         Durret, F., and Petitjean, P. 1998, \aap, 333, 459

\bibitem[Mendoza]{men83} Mendoza, C. 1983, in ``Planetary Nebulae, I.A.U. 
         Symposium No. 103'' ed. D. R. Flower, (Dordrecht: D. Reidel), 143

\bibitem[Metz et al.]{met96} Metz, S. K., Tadhunter, C. N., Robinson. A., Axon, D. J.
        1996, in Emission Lines in Active
        Galaxies: New Methods and Techniques, ed. B. M. Peterson, F.-Z. Cheng,
        and A. S. Wilson, Astron. Soc. Pac. Conf. Ser. Vol. 113, 390

\bibitem[Penston et al.]{pen90}  Penston, et al. 1990, \aap, 236, 53

\bibitem[Pickles]{pic85} Pickles, A. J. 1985, \apjs, 59, 33

\bibitem[Pogge]{pog96} Pogge, R. W. 1996, in Emission Lines in Active
        Galaxies: New Methods and Techniques, ed. B. M. Peterson, F.-Z. Cheng,
        and A. S. Wilson, Astron. Soc. Pac. Conf. Ser. Vol. 113, 378

\bibitem[Robinson et al.]{rob87} Robinson, A., Binette, L., Fosbury, R. A. E., 
         Tadhunter, C. N. 1987, \mnras, 227, 97 

\bibitem[Sanders and Mirabel]{san85}  Sanders, D. B., \& Mirabel, I. F.
         1985, \apjl, 298, L31

\bibitem[Stasinska]{sta84}  Stasinska, G. 1984, \aap, 135, 341

\bibitem[Storchi-Bergmann et al.]{sto96} Storchi-Bergmann, T., Wilson, A. S.,
        Mulchaey, J. S., \& Binette, L. 1996, \aap, 312, 357

\bibitem[Sutherland, Bicknell \& Dopita]{sut93}  Sutherland, R. S.,
         Bicknell, G. V., \& Dopita, M. A. 1993, \apj, 414, 510

\bibitem[Tadhunter, Robinson \& Morganti]{tad89} Tadhunter, C. N., 
        Robinson, A., \& Morganti, R. 1989, in Extranuclear Activity in
        Galaxies, ed. E. J. A. Meurs \& R. A. E. Fosbury (Garching: ESO), 293

\bibitem[Taylor, Dyson \& Axon]{tay92} Taylor, D., Dyson, J. E, \&
         Axon, D. J. 1992, \mnras, 255, 351
 
\bibitem[Terlevich et al.]{ter92} Terlevich, R., Tenorio-Tagle, G.,
         Franco, J., Melnick, J. 1992, \mnras, 255, 713

\bibitem[Tsvetanov, Dopita \& Allen]{tsv95} Tsvetanov, Z., Dopita, M.,
         Allen, M. 1995, BAAS, 27, 871

\bibitem[Ulvestad and Wilson]{ulv84}  Ulvestad, J. S., Wilson, A. S.
         1984, \apj, 285, 439

\bibitem[Unger et al.]{ung87} Unger, S. W., Pedlar, A., Axon, D. J., 
         Whittle, M., Meurs, E. J. A., \& Ward, M. 1987, \mnras, 228, 671 

\bibitem[Ward et al.]{war80}  Ward, M., Penston, M. V., Blades, J. C.,
         \& Turtle, A. J. 1980, \mnras, 193, 563

\bibitem[Ward et al.,\ 1987]{war87}  Ward, M. J., Geballe, T., Smith, M.,
         Wade, R., Williams, P. 1987, \apj, 316, 138

\bibitem[Weaver et al.]{wea96}Weaver, K et al. 1996, \apj, 458, 160

\bibitem[Wehrle \& Morris]{weh88} Wehrle, A. E., Morris, M. 1988, \aj, 95, 1689

\bibitem[Wilson, Ward, \& Haniff]{wil88} Wilson, A. S., Ward, M. J. \& 
        Haniff, C. A. 1988, \apj, 334, 121

\bibitem[Veilleux and Osterbrock]{vei87}  Veilleux, S., Osterbrock, D. E.
         1987, \apjs, 63, 295.

\bibitem[Viegas \& de Gouveia Dal Pino]{vie92} Viegas, S. M., 
         de Gouveia Dal pino, E. M. 1992, \apj, 384, 467

\end{thebibliography}
\end{document}